\documentclass[11pt]{JHEP3}

\usepackage{latexsym}
\usepackage{epsfig,amssymb,euscript}
\usepackage{amsmath,cancel,slashed}
\DeclareMathAlphabet{\mathpzc}{OT1}{pzc}{m}{it}
\usepackage{array,calc,epsfig,bbm}
\usepackage{multicol}
\usepackage{bbm}
\usepackage{fancybox}
\usepackage{tensor}

\def\be{\begin{equation}}
\def\ee{\end{equation}}

\def\bea{\begin{eqnarray}}
\def\eea{\end{eqnarray}}
\def\bseq{\begin{subequations}}
\def\eseq{\end{subequations}}
\newcommand\bbone{\ensuremath{\mathbbm{1}}}
\newcommand{\ul}{\underline}

\newcommand{\beal}{\begin{align}}

\numberwithin{equation}{section} 
\usepackage[]{graphicx}

\def\d {{\rm d}}

\def\calb         {{\cal B}}

\def\calg         {{\cal G}}

\def\cali         {{\cal I}}

\def\calk         {{\cal K}}
\def\call         {{\cal L}}

\def\caln         {{\cal N}}
\def\calo         {{\cal O}}

\def\calr         {{\cal R}}
\def\cals         {{\cal S}}

\def\calw         {{\cal W}}

\def\reals        {{\mathbb R}}

\def\tr           {\mathop{\rm Tr}}
\def\Re           {{\rm Re\hskip0.1em}}
\def\Im           {{\rm Im\hskip0.1em}}

\def\sqr#1#2{{\vcenter{\vbox{\hrule height.#2pt
 \hbox{\vrule width.#2pt height#1pt \kern#1pt \vrule width.#2pt}\hrule
 height.#2pt}}}}

\def\oh{\frac{1}{2}}
\def\a{{\alpha}}

\def\eps{{\epsilon}}

\def\sig{{\sigma}}

\def\raw{\rightarrow}

\def\ls{\ell_{\rm s}}

\def\slashchar#1{\setbox0=\hbox{$#1$}           
\dimen0=\wd0                                 
\setbox1=\hbox{/} \dimen1=\wd1               
\ifdim\dimen0>\dimen1                        
\rlap{\hbox to \dimen0{\hfil/\hfil}}      
#1                                        
\else                                        
\rlap{\hbox to \dimen1{\hfil$#1$\hfil}}   
/                                         
\fi}

\title{DWSB in heterotic flux compactifications}

\author{Johannes Held${}^{\diamondsuit}$, Dieter L\"ust${}^{\diamondsuit \clubsuit}$, Fernando Marchesano${}^{\heartsuit}$ \& Luca Martucci${}^{\clubsuit}$\\

\begin{itemize}

\item  Max-Planck-Institut f\"ur Physik\\
F\"ohringer Ring 6, 80805 M\"unchen, Germany

\item  Arnold-Sommerfeld-Center for Theoretical Physics\\
Department f\"ur Physik, Ludwig-Maximilians-Universit\"at M\"unchen\\
Theresienstra\ss e 37, 80333 M\"unchen, Germany 

\item  CERN Theory Group,\\ CH-1211 Geneva 23, Switzerland
 \end{itemize}

\bigskip
 E-mail:
\email{heldj@mppmu.mpg.de}, \email{dieter.luest@lmu.de} \& \email{luest@mppmu.mpg.de}, \email{marchesa@cern.ch}, \email{luca.martucci@physik.uni-muenchen.de}}

\received{\today}               
\revised{}
\accepted{}               

\preprint{hep-th\yymmddd}
\preprint{MPP-2010-39\\ LMU-ASC 21/10 \\ CERN-PH-TH/2010-075}

\abstract{We address the construction of non-supersymmetric vacua in heterotic compactifications with intrinsic torsion and background fluxes. In particular, we implement the approach of {\it domain-wall supersymmetry breaking} (DWSB) previously developed in the context of type II flux compactifications. This approach is based on considering backgrounds where probe NS5-branes wrapping internal three-cycles and showing up as four-dimensional domain-walls do not develop a BPS bound, while all the other BPS bounds characterizing the $\caln=1$ supersymmetric compactifications are preserved at tree-level. Via a scalar potential analysis we provide the conditions for these backgrounds to solve the ten-dimensional equations of motion including order $\alpha^{\prime}$ corrections. We also consider backgrounds where some of the NS5-domain-walls develop a BPS bound, show their relation to no-scale SUSY-breaking vacua and construct explicit examples via elliptic fibrations. Finally, we consider backgrounds with a non-trivial gaugino condensate and discuss their relation to supersymmetric and non-supersymmetric vacua in the present context.}


\begin{document}

\section{Introduction}

Supersymmetry breaking in string compactifications is one of the most important issues when making contact between strings and particle physics. As a possible solution of the hierarchy problem, we look for non-supersymmetric string vacua in which the soft supersymmetry breaking terms in the observable sector are at the TeV scale, whereas the supersymmetry breaking scale in the hidden sector of the theory can be much higher. This means that in non-supersymmetry string vacua supersymmetry has to be broken in a controllable way: no tachyonic instabilities (except the one associated to the Higgs) should arise from supersymmetry breaking. In addition, the cosmological constant should be kept zero or at least very small. It goes without saying that to date fully realistic string vacua satisfying all these requirements have not been yet found, and that this lack of success is partially due to the poor understanding of supersymmetry breaking in string theory.

Nevertheless, several promising string scenarios with broken supersymmetry have been investigated in the past. A somehow crude separation of models is given by considering either  perturbative or non-perturbative supersymmetry breaking.  In fact, it was shown already some time ago by Scherk and Schwarz  \cite{Scherk:1978ta} that supersymmetry can be broken at tree level by choosing different boundary conditions for fermions and bosons. Non-perturbative, spontaneous supersymmetry breaking was first discussed in the context of $E_8\times E_8$ heterotic strings compactified on Calabi-Yau manifolds. The most popular mechanism is to consider supersymmetry breaking by gaugino condensation on a hidden gauge group $G_{\rm hid}$ \cite{din85,drsw85}. Then, in the next step, the hidden sector SUSY-breaking is mediated to the observable sector by gravitational effects. Supersymmetry breaking by gaugino condensation is usually formulated in terms of the four-dimensional effective action where, by integrating out the condensing gauginos, a non-vanishing superpotential of the form
\begin{equation}\label{hw}
{\cal W}
\simeq c_0\, e^{-\frac{s}{b_0}}
\end{equation}
emerges, with $s$ the dilaton, $c_0$ a model-dependent quantity and $b_0$ proportional to the $\beta$-function coefficient of $G_{\rm hid}$. Supersymmetry is then spontaneously broken by the F-terms of some moduli fields, such as the overall K\"ahler modulus $t$ of the Calabi-Yau. If in addition $c_0$ depends on $t$ via a modular function, as required in a large class of models by T-duality, the $t$-field gets stabilized in the non-supersymmetric vacuum of the effective theory \cite{Font:1990nt,Ferrara:1990ei,Nilles:1990jv}. The soft term structure in this class of models has been discussed in various papers \cite{Ibanez:1992hc,Kaplunovsky:1993rd}.

%


While non-perturbative supersymmetry breaking is the most explored scenario in the realm of heterotic compactifications, the situation is somehow reversed for type II strings. There new ways for constructing non-supersymmetric vacua have been unveiled, mainly due to progress made on constructing compactifications with background fluxes (see \cite{Grana:2005jc,Douglas:2006es,Blumenhagen:2006ci} for reviews). A general result is that the backreaction of background fluxes is such that the internal space is not anymore a Calabi-Yau manifold. However, the conditions for unbroken ${\cal N}=1$ supersymmetry in four dimensions still require that the internal geometries allow for globally defined spinors, and so they can be characterized in terms of SU(3) structures \cite{chiossi02}. In particular, for the so-called SU(3)-structure manifolds the internal geometry is encoded in the SU(3)-invariant forms $J$ and $\Omega$ also present in Calabi-Yau spaces.

More precisely, the globally defined two-form $J$ and three-form $\Omega$ not only encode the internal geometry of the compactification, but also codify in a particularly useful way important information of the latter. For instance, in Calabi-Yau compactifications the conditions for $\caln=1$ supersymmetry in four dimensions can be expressed as the differential conditions $\d J = \d \Omega = 0$. In addition, $J$ and $\Omega$ measure the energy of the potential BPS objects of the compactifications, and in particular of those that upon dimensional reduction show up as 4d strings, 4d domain-walls or 4d gauge theories (i.e., fill up the 4d space-time). As shown in \cite{lucal}, this picture survives when considering flux compactifications. Indeed, as mentioned above in this more general case the internal space will not be Calabi-Yau, and so $J$ and $\Omega$ will in general not be closed. Nevertheless, supersymmetry is still equivalent to specific differential equations for $J$ and $\Omega$ \cite{gmpt05},\footnote{For SU(3)$\times$SU(3)-structure manifolds $(J, \Omega)$ should be replaced by the pure spinors/polyforms $(\Psi_1,\Psi_2)$.} and such differential equations precisely allow them to measure the tension of the BPS objects of the compactification or, in a more mathematical language, to be calibrations for probe D-branes in this background. In fact, the existence of the full set BPS bounds for 4d extended objects (i.e., the existence of calibrations) is equivalent to the conditions for 4d $\caln =1$ supersymmetry \cite{lucal}. If at least one of the above calibrations/BPS bounds is missing, 4d supersymmetry must be  broken. 

This latter observation was used in \cite{dwsb} to propose a general strategy for constructing $\caln=0$ vacua in the context of type II flux compactifications. In this approach one assumes that the ten-dimensional background allows for a globally defined spinor which, from the four-dimensional viewpoint will be the generator of an approximate supersymmetry. One can then construct $J$ and $\Omega$ from such spinor, and check whether these two objects allow to define BPS bounds for 4d strings, 4d domain-walls and 4d gauge theories. Dropping these BPS bounds will translate in 4d physics as a definite pattern of spontaneous supersymmetry breaking, more precisely as the violation of a D-term and two different kinds of F-term conditions, respectively. As shown in \cite{dwsb}, one can construct type II flux vacua with supersymmetry broken at tree level by simply dropping one of the three kinds of BPS bounds, namely the BPS bound for 4d domain-walls or, equivalently, one of the two F-term conditions in the effective action description. This way of breaking supersymmetry was dubbed {\sl domain wall supersymmetry breaking (DWSB)} in \cite{dwsb}. It was there discussed  that type II DWSB provide a controllable mechanism for supersymmetry breaking on generalized flux compactifications with vanishing tree level cosmological constant $\Lambda$. More precisely, the vanishing of $\Lambda$ has its origin in an underlying no-scale structure \cite{noscale} of the effective potential that is preserved by DWSB, just as in the $\caln=0$ flux vacua constructed in \cite{gkp}.

Compared to the somewhat rigid heterotic Scherk-Schwarz constructions, these type II DWSB constructions display a much richer landscape of tree-level SUSY-breaking vacua. Extracting the physics of these vacua beyond the supergravity approximation is however difficult, since  they contain RR background fluxes and usual worldsheet techniques do not work. A similar statement applies to gaugino condensate vacua, so heterotic tree-level SUSY-breaking seem the best option to understand stringy corrections to $\caln=0$ vacua. However, no systematic study of $\caln = 0$ heterotic flux vacua has so far been performed.

The purpose of this paper is to fill this gap and to provide a simple strategy to construct a wide class of heterotic flux vacua with supersymmetry broken at tree-level. In particular, we apply the DWSB strategy to heterotic strings on SU(3)-structure manifolds with 
non-zero $H$-flux.\footnote{Heterotic $H$-flux compactifications with torsion were already considered some time ago in
\cite{Hull,stromtorsion,Lust:1986ix}.}
The three classes of calibrated objects (4d strings, 4d domain-walls and 4d gauge theories) are now NS5-branes wrapping internal cycles. Supersymmetry will be again spontaneously broken by abandoning the domain wall calibration condition. Nevertheless, the effective potential will still be BPS-like,\footnote{See \cite{bpsaction} for previous work on the subject of heterotic BPS-like potentials.} leading to a no-scale structure as in the type II case. The corresponding effective superpotential contains the flux $H$ 
and has the form
\begin{equation}\label{hwhnonCY}
{\cal W}
\simeq \int \Omega\wedge (H-i{\rm d}J)\, 
\end{equation}
being in agreement with the microscopic heterotic flux picture that we develop.   



As we will see, along our discussion several subtleties will arise due to higher order $\alpha'$-corrections and the heterotic Bianchi identity. These subtleties can be addressed due to the better control that we have over stringy corrections in heterotic compactifications: a clear advantage when analyzing the physics of DWSB vacua. Indeed, since this construction is based on the notion of BPS bounds of a 4d $\caln=0$ theory, it is not clear whether $\a'$ or loop corrections will modify or even destroy such BPS bounds. A related question is how and at what order these corrections will violate the no-scale structure of the tree-level scalar potential, a fact of crucial importance for scenarios based on tree-level SUSY-breaking, like those proposed in \cite{kklt,LVS}. This important asset of heterotic compactifications may then be used to learn important information of dual non-supersymmetric vacua at different corners of the landscape. In particular, at the level of $\caln=1$ compactifications heterotic strings are a useful tool to understand the physics behind F-theory compactifications (see, e.g., \cite{hetF} for recent developments and references therein). It is then natural to wonder to what extent these tools could be applied for DWSB vacua.

The paper is organized as follows. In section \ref{sec:10theory} we recall some aspects of the low-energy effective action governing heterotic string theory, setting the notation and conventions used later on. In section \ref{sec:pot} we compute the effective potential for bosonic heterotic compactifications, and rewrite it in BPS-like form. In section \ref{sec:susybvacua} we make use of the potential to apply the DWSB ansatz outlined above, expressing our results in terms of the torsion classes of the compactification manifold. Section \ref{sec:calibr} discusses in detail the calibration structure underlying this class of heterotic vacua, making use of such calibrations in order to address the issue of bundle stability in them. The concept of calibration is further used in section \ref{sec:oneparSB}, in order to define a subclass of DWSB vacua, dubbed $\oh$DWSB vacua, that are then related to $\caln=0$ no-scale vacua via a four-dimensional effective description. Explicit constructions of $\oh$DWSB vacua are given in section \ref{sec:example} via elliptically fibered spaces, in particular via the standard example of $T^2$ fibered over {\rm K3}. In section \ref{sec:gaugino} we extend the class of backgrounds under study by adding non-vanishing fermion condensates, showing that the calibration/BPS bound structure used for purely bosonic backgrounds is still present there. Finally, we conclude our discussion in section \ref{sec:conclusions} with an outlook on possible applications of this work. Conventions and technical details have been relegated to the appendices.


\section{Ten-dimensional action and supersymmetry transformations}\label{sec:10theory}

In the string frame and up to order $\calo(\alpha^\prime)$, the bosonic sector of ten-dimensional $\caln=1$ heterotic supergravity is governed by the action \cite{bdr}
\be\label{10daction}
S=\frac{1}{2\kappa^2}\int\d^{10} x\sqrt{-{\rm det\, } g}\,e^{-2\phi}\big[{\cal R}_{X_{10}}+4(\d\phi)^2-\frac1{2} H^2+\frac{\alpha^\prime}4(\tr R_+^2- \tr F^2)\big]
\ee
with $2\kappa^2=(2\pi)^7\alpha^{\prime 4}$ and the square of a $p$-form $\rho$ defined as $\rho^2\equiv \rho\cdot\rho:=\frac1{p!}\rho_{M_1\ldots M_p}\rho^{M_1\ldots M_p}$.\footnote{In ten dimensions we use $M,N,\ldots$ as curved indices and $\ul{M},\ul{N},\ldots$ as flat indices.} Here $\phi$ is the dilaton and ${\cal R}_{X_{10}}$ is the Ricci scalar of the full ten-dimensional space. The two-form $F$ is the SO(32) or $E_8\times E_8$ field strength, and the Lie algebra bilinear form $\tr$ is related to the trace in the adjoint representation ${\rm tr}_{\rm adj}$ by $\tr\equiv \frac1{30} {\rm tr}_{\rm adj}$. In addition, $R^{\ul{M}}_{\pm}{}_{\ul{N}}=\frac12 R^{\ul{M}}_{\pm}{}_{\ul{N}PQ}\d x^P\wedge\d x^Q$, are curvature two-forms constructed using the connections
\bea
\omega^{\ul{M}}_\pm{}_{\ul{N} P}=\omega^{\ul{M}}{}_{\ul{N} P}\pm\frac12 H^{\ul{M}}{}_{\ul{N} P}
\eea
with $\omega^{\ul{M}}{}_{\ul{N} P}$ the ordinary torsionless spin connection. Their traces are defined as
\bea
\tr R_\pm^2:=-R^{\ul{M}}_\pm{}_{\ul{N}}\cdot R^{\ul{N}}_\pm{}_{\ul{M} }\equiv\frac12 R_{\pm MNPQ}R_\pm^{MNPQ}
\eea
Finally, $H$ stands for the Neveu-Schwarz (NS) three-form flux, whose Bianchi identity (BI) is given by
\bea\label{BI}
\d H=\frac{\alpha^\prime}4 (\tr R_+\wedge R_+-\tr F\wedge F)
\eea
where $\tr R_+\wedge R_+=-R^{\ul{M}}_+{}_{\ul{N}}\wedge R^{\ul{N}}_+{}_{\ul{M}}$.

In order to derive the equations of motion  from the action  (\ref{10daction}) a little care is needed, due to the implicit dependence  of $H$ on other elementary fields through the BI (\ref{BI}) and because of the presence of the $\alpha^\prime$-order curvature correction. The latter complication is simplified, however, by a lemma stating that the variation of the $\alpha^\prime$-order curvature correction with respect to $\omega^{\ul{M}}_+{}_{\ul{N}}$ is proportional to the leading order equations of motion \cite{bdr}. Then, for instance, the `modified' Einstein equation  can be written in the form\footnote{Given a  $p$-form $\rho$, we use the notation $\iota_M\rho\equiv \iota_{\partial_M}\rho\equiv\frac{1}{(p-1)!}\rho_{MN_1\ldots N_{p-1}}\d x^{N_1}\wedge\ldots\wedge\d x^{N_{p-1}}$.}
\be
\label{modEinst}
R_{MN}+2\nabla_M\nabla_N\phi-\frac12\iota_M H\cdot\iota_N H +\frac{\alpha^\prime}{4}\big[\tr(\iota_M R_+\cdot \iota_N R_+)  -\tr (\iota_M F\cdot\iota_N F)\big]=0
\ee 
where the dilaton equation of motion has been used to simplify the final expression.

In order to construct supersymmetric bosonic backgrounds one needs to make sure that the  supersymmetry variations of the gravitino $\psi_M$, dilatino $\lambda$ and gaugino $\chi$ vanish. At leading order in $\alpha'$ these are given by 
\bseq\label{10dsusy}
\begin{align}
\delta_\eps\psi_M&=\nabla^{-}_M\epsilon\equiv \big(\nabla_M-\frac{1}{4}\slashed{H}_M\big)\epsilon\, ,\label{gravsusy}\\
\delta_\eps\lambda&= \big(\slashed{\partial}\phi-\frac{1}{2}\,\slashed{H}\big)\epsilon\, ,\label{dilsusy}\\
\delta_\eps\chi&= \frac12\,\slashed{F}\,\epsilon\, \label{gaugesusy},
\end{align}
\eseq
where $H_M\equiv \iota_M H$ and our spinorial conventions are specified in Appendix \ref{app:fermconv}. Note that such supersymmetry transformations are not corrected at order $\alpha^\prime$ \cite{bdr}.


\section{Four-dimensional compactifications and effective potential}
\label{sec:pot}

Let us now restrict our attention to compactifications to four dimensions. We then assume that the ten-dimensional spacetime has the form $X_{10}=X_4\times M$, where $M$ is a six-dimensional compact manifold and $\d s^2_{X_{4}}$ a maximally symmetric space metric with cosmological constant $\Lambda$. We describe such geometry by the external coordinates $x^\mu$ ($\mu=0,\ldots,3$), internal coordinates $y^m$ ($m=1,\ldots,6$)\footnote{As in ten-dimensions, we underline internal flat indices.} and the metric 
\be\label{wmetric}
\d s^2_{X_{10}}=e^{2A}\d s^2_{X_{4}}+\d s^2_{M}
\ee
Moreover, in order to keep our construction as general as possible, we allow for non-trivial warping $A$, dilaton $\phi$ and fluxes $H$ and $F$ such that the 4d maximal symmetry is not broken. As a result, the ten-dimensional BI (\ref{BI}) keeps the same form, but with $R_+$ being just the curvature of the internal torsion-full connection $\omega_+^{\ul{m}}{}_{\ul{n}p}=\omega^{\ul{m}}{}_{\ul{n}p}+\frac12 H^{\ul{m}}{}_{\ul{n}p}$. In the following, $R_\pm$ will denote such internal torsion-full curvature.

The overall normalization of the non-vanishing warping is fixed in terms of the four-dimensional Planck mass $M_{\rm P}$, by the relation
\be\label{planck}
\frac{1}{\kappa^2}\int_M \text{vol}_M\, e^{2A-2\phi}=M^2_{\rm P}
\ee
with $\text{vol}_M=\d^6 y\sqrt{{\rm det\, }g}$ and $g$ the internal six-dimensional metric. The relation (\ref{planck}) is obtained by requiring that $\d s^2_{X_4}$ describes the four-dimensional Einstein frame metric, i.e.\ by requiring that the dimensional reduction of (\ref{10daction}) gives the canonical four-dimensional Einstein term $(M^2_{\rm P}/2)\int \sqrt{-g_{X_4}}\,\calr_{X_4}$. 

Note that so far we are restricting to purely bosonic heterotic configurations. One may however  add non-trivial fermionic condensates to the above bosonic background, a possibility that we will consider in section \ref{sec:gaugino}.

\subsection{Cosmological constant and warping}
\label{sec:prel}

Regarding the 4d cosmological constant and the warp factor of heterotic vacua, one can obtain some rather restricting conditions from the equation of motion (\ref{modEinst}). In particular, by choosing $M,N$ along the four-dimensional space $X_4$, one gets the equation   
\bea\label{extEinstcompl}
\nabla^m(e^{-2\phi}\nabla_m e^{4A})&=& 4 \, e^{2A-2\phi}\Lambda+\alpha^\prime\, e^{2A-2\phi}\Big\{\frac{2}3\, e^{-2A}[\Lambda-3(\d A)^2]^2\cr && +2(\nabla_m\nabla_n e^A)(\nabla^m\nabla^n e^A)+ (\iota_m H\cdot \iota_n H)\, \nabla^m e^A\nabla^n e^A\Big\}
\eea
which severely restricts the choices of $\Lambda$ and $A$. Indeed, solving this equation perturbatively, we find that at the lowest order in  $\alpha^\prime$ we have 
\bea\label{extEinstzero}
\nabla^m(e^{-2\phi}\nabla_m e^{4A})&\simeq& 4 \, e^{2A-2\phi}\Lambda
\eea
where by $\simeq$ we mean equivalence at zeroth-order in $\alpha^\prime$. By integrating this equation over the internal space, we get the lowest $\alpha^\prime$-order condition 
\bea\label{vanLambda}
\Lambda\,\int_M e^{2A-2\phi}\text{vol}_M\simeq 0\quad \Rightarrow\quad \Lambda\simeq 0
\eea
Note that multiplying (\ref{extEinstzero}) by $e^{(p-4)A}$ for any $p\neq 4$, integrating over $M$ and using (\ref{vanLambda}) one gets
\bea
\int_M \text{vol}_M\, e^{pA-2\phi}\,(\d A)^2\simeq 0
\eea
which implies that, at lowest order in $\alpha^\prime$, the string-frame warp factor must be constant.

Hence, at leading order in $\alpha^\prime$, the modified external Einstein equations (\ref{modEinst}) imply that $\Lambda$ vanishes and $e^A$ is constant. Plugging this back into eq.(\ref{extEinstcompl}) and expanding it in powers of $\alpha'$, one can check that the first corrections to the above result arise at order $\alpha'^3$, and can thus be ignored at the $\calo(\alpha^\prime)$-approximation we are working with.

We then conclude that, at order $\calo(\alpha^\prime)$, the external Einstein equation (\ref{extEinstcompl}) requires the four-dimensional space to be flat and the warping to be constant.  Note that this result is valid for any purely bosonic compactification, whether it is supersymmetric or not. 

From (\ref{planck}) we see that the Einstein frame condition for the four-dimensional metric requires that 
\be\label{warp_planck}
e^{2A}=\frac{g^2_{\rm s} \ls^8 M^2_{\rm P}}{4\pi\text{Vol}(M)}
\ee
where $\ls = 2\pi\sqrt{\alpha^\prime}$, $\text{Vol}(M)=\int\text{vol}_M$ and $g_{\rm s}$ is defined by
\be\label{gs}
\frac{1}{g_{\rm s}^2}=\frac{\int_Me^{-2\phi}\,\text{vol}_M}{\int_M \text{vol}_M}
\ee


\subsection{Effective potential}

As the above constraints  can be obtained by just considering the external ten-dimensional Einstein equations, one should be able to further restrict the compactification (\ref{wmetric}) by analyzing the complete set of ten-dimensional equations of motion. In order to discuss the latter, we will follow the approach of \cite{dwsb} and  consider an effective four-dimensional potential which is a functional of all the fields, and whose extremization provides all the equations of motion. 

Such a potential can be obtained by restricting the action (\ref{10daction}) to the compactification ansatz described at the beginning of  this section, imposing a flat four-dimensional metric $\d s^2_{X_4} = \d s^2_{\mathbb{R}^{1,3}}$ and setting $S=-\int_{X_4}\d^4 x\, V$. This procedure can be seen as a consistent truncation to fields preserving the four-dimensional Poincar\'e invariance and thus  as a sort of `dimensional reduction' along $X_4$. The resulting four-dimensional potential is given by
\bea\label{firstpot}
V&=&\frac{1}{2\kappa^2}\int_M \text{vol}_M\ e^{4A-2\phi}\Big[-{\cal R}+\frac12 H^2-4(\d\phi)^2+8\nabla^2 A+20(\d A)^2+\frac{\alpha^\prime}4 (\tr F^2-\tr R^2_+) \cr &&-2\alpha^\prime\, e^{-2A}(\nabla^m\nabla^n e^A) (\nabla_m\nabla_n e^A)-\alpha^\prime\, (\iota_m H\cdot \iota_n H)\, \nabla^m A\nabla^n A-6\alpha^\prime(\d A\cdot \d A)^2\Big]
\eea
where ${\cal R}$ is the scalar curvature constructed using the internal six-dimensional metric, and the second line of (\ref{firstpot}) arises from the fact that now $R_+$ denotes just the internal curvature. The ten-dimensional equations of motion for compactifications to flat space can then be obtained by extremizing (\ref{firstpot}). 

Note that the potential (\ref{firstpot}) must vanish on-shell. Indeed,  (\ref{firstpot}) must be extremized under a general variation of the warping $A$. In particular, it must be extremized by a constant shift on $A$, which demands that $V=0$ on-shell. This is indeed what is expected from pure four-dimensional arguments, since in Minkowski vacua the four-dimensional potential must vanish.

\subsection{SU(3)-structure and BPS-like potential}

So far in our analysis we have not made use of the supersymmetric structure of the heterotic effective theory and, in particular, of the equations (\ref{10dsusy}). On the other hand, one would expect such conditions to play a role in any compactification where supersymmetry is spontaneously broken. In particular, their failure to be satisfied should somehow encode the different SUSY-breaking patterns.

In this respect, the scalar potential $V$ written in the form (\ref{firstpot}) does not appear particularly useful for studying possible patterns of supersymmetry breaking. In order to improve the situation, we have to make manifest the underlying supersymmetric structure. As we are going to show, this is possible once we assume the existence of an SU(3)-structure in the internal space, which means that $M$ is an almost complex space, with an hermitian metric and a globally defined never vanishing (3,0)-form $\Omega$ and K\"ahler (or fundamental) $(1,1)$-form $J$. As reviewed in Appendix \ref{app:fermconv}, these forms can be directly related to a globally defined chiral  spinor $\eta$ in $M$. In supersymmetric compactifications $\eta$ can be seen as the internal component of the ten-dimensional Killing spinor $\epsilon$, decomposed as in (\ref{fermsplit}). Let us however stress that, at this stage, we do not require $\epsilon$ or $\eta$ to satisfy any particular requirement except being globally well-defined. In other words, the $\Omega$ and $J$ that we consider define a generic SU(3)-structure.


Now, a key point is that $\Omega$ and $J$ not only fully specify the internal spinor $\eta$ but also the six-dimensional metric $g$. Thus, in principle, one can express the scalar curvature ${\cal R}$ appearing in the potential (\ref{firstpot}) as a function of $\Omega$ and $J$. This problem has been addressed in \cite{bedulli06}, and in \cite{cassani08,dwsb} for the more general SU(3)$\times$SU(3)-structure case relevant in type II configurations. Here we use the general formula obtained in \cite{dwsb} (see eq.~(C.1) therein), from which one can derive the following identity\footnote{To obtain (\ref{curvSU(3)}) from (C.1) of \cite{dwsb}, one should set $f=1$, $A=\phi=H=0$, $\psi_1=ie^{iJ}$ and $\psi_2=\Omega$ therein.}
\bea\label{curvSU(3)}
{\cal R}=-\frac12(\d J)^2-\frac18[\d(J\wedge J)]^2-\frac12|\d \Omega|^2+\frac12|J\wedge \d\Omega|^2+\frac12 u^2-\nabla^mu_m
\eea
where for $\rho$ a complex $p$-form, $|\rho|^2\equiv \rho\cdot\bar\rho$, and\footnote{In (\ref{curvSU(3)b}),   terms of the form $\rho\lrcorner\tau$, where  $\rho$ is a $p$-form and $\tau$ is a $q$-form such that $p\leq q$, are defined as: $\rho\lrcorner\tau\equiv\frac{1}{p!}\rho^{m_1\ldots m_p}\tau_{ m_1\ldots m_p m_{p+1}\ldots m_q}\d y^{m_{p+1}}\wedge\ldots\wedge \d y^{m_q}$.} 
\bea
u=u_m\d y^m=\frac14(J\wedge J)\lrcorner \d(J\wedge J)+\frac12\Re(\bar\Omega\lrcorner\d\Omega)
\label{curvSU(3)b}
\eea
Thus, by using (\ref{curvSU(3)}), together with the BI (\ref{BI}), one can then rewrite (\ref{firstpot}) as
\bea\label{pot} 
V=V_0+V_1
\eea
with
\bseq\label{pot2}
\begin{align}
V_0=&\frac{1}{4\kappa^2}\int \text{vol}_M\ e^{4A-2\phi}\big[e^{-4A+2\phi}\d(e^{4A-2\phi} J)-*H \big]^2\cr
+ & \frac{1}{4\kappa^2}\int \text{vol}_M\ e^{4A-2\phi}\big\{ \frac14\big[e^{-2A+2\phi}\d\big(e^{2A-2\phi}J\wedge J \big)  \big]^2+ 4(\d A)^2 \big\}  \cr
+ & \frac{1}{4\kappa^2}\int \text{vol}_M\,e^{-2A+2\phi}\Big[|\d(e^{3A-2\phi}\Omega)|^2-|J\wedge \d(e^{3A-2\phi}\Omega)|^2\Big]\cr
-& \frac{1}{4\kappa^2}\int \text{vol}_M\,e^{4A-2\phi}\Big\{2\d A +\frac1{4}e^{-2A+2\phi}(J\wedge J)\lrcorner \d(e^{2A-2\phi}J\wedge J)\cr 
& \hspace{5cm}+\frac12e^{-3A+2\phi}\Re[\bar\Omega\lrcorner \d(e^{3A-2\phi}\Omega)]\Big\}^2
\label{potord0}\\
V_1=& \frac{\alpha^\prime}{8\kappa^2}\int e^{4A-2\phi}\big[\tr (F\wedge *F)+\tr(F\wedge F)\wedge J\big]  \cr
&- \frac{\alpha^\prime}{8\kappa^2}\int e^{4A-2\phi}\big[\tr (R_+\wedge *R_+)+\tr(R_+\wedge  R_+)\wedge J\big] \cr 
&-\frac{\alpha^\prime}{2\kappa^2}\int_M \text{vol}_M\ e^{4A-2\phi}\Big[2\, e^{-2A}(\nabla^m\nabla^n e^A) (\nabla_m\nabla_n e^A)\cr 
& \hspace{4cm}-\, (\iota_m H\cdot \iota_n H)\, \nabla^m A\nabla^n A-6(\d A\cdot \d A)^2\Big]
 \label{potord1}
\end{align}
\eseq
Note that the potential (\ref{pot}) depends only explicitly in the dilaton and fluxes, but both explicitly and implicitly on the metric, the latter through the SU(3)-structure tensors $J$ and $\Omega$. In order to extremize the potential, one needs to know what is the dependence of $J$ and $\Omega$ on the metric at the infinitesimal level, which is given by
\be\label{su3var}
\delta J=-\frac12\delta g^{mn}\, g_{k(m}\d y^k\wedge \iota_{n)}J\, ,\quad \delta \Omega=-\frac12\delta g^{mn}\, g_{k(m}\d y^k\wedge \iota_{n)}\, \Omega
\ee
where $\delta g^{mn}$ is a general variation of the inverse of the metric. 

In principle one could also express the curvature $R_+$ in terms of the SU(3)-structure forms $J$ and $\Omega$ and the flux $H$, but this turns out not to be necessary for our purposes. One can use the decompositions in $(p,q)$-forms induced by the almost complex structure associated to $\Omega$ to rewrite the first two lines on the r.h.s. of (\ref{potord1}) as a sum of squares
\bseq\label{squareform}
\begin{align}\label{squareformF}
& \tr (F\wedge *F)+\tr(F\wedge F)\wedge J= \text{vol}_M\,\big[2\tr |F^{(2,0)}|^2+\tr (J\lrcorner F)^2\big]\\ 
\label{squareformR}
&\tr (R_+\wedge *R_+)+\tr(R_+\wedge  R_+)\wedge J=\text{vol}_M\,\big[ 2\tr |R_+^{(2,0)}|^2+\tr (J\lrcorner R_+)^2\big]
\end{align}
\eseq

Note that by this scalar potential approach we have followed a philosophy quite similar to the one in \cite{bpsaction}, where a similar potential was constructed. Let us however point out a few differences between our potential and the one obtained there. First, we are not assuming constant warping. While this aspect will not be crucial for most of the discussions on compactifications with constant warping, allowing for a non-trivial warping makes explicit the consistency of our truncation ansatz. Second, our potential (\ref{pot})-(\ref{pot2})  is expressed in terms of the SU(3)-invariant (3,0)-form $\Omega$, and not of the associated almost complex structure as in \cite{bpsaction}. Finally, and most importantly, our potential is in a full BPS-form, namely it is a sum of squares, while the potential of \cite{bpsaction} is not, since it contains an $\calo(\alpha^{\prime 0})$-term linear in the curvature. As we will see, having a fully-BPS structure will be crucial in studying possible  mechanisms of supersymmetry breaking.

\section{Supersymmetry breaking vacua: general discussion}
\label{sec:susybvacua}

Let us now address the possible patterns of supersymmetry breaking in purely bosonic heterotic vacua. In order to do that, we will first revisit the supersymmetric case and later try to break supersymmetry. We will identify a particularly natural possibility, which we will further restrict in section \ref{sec:oneparSB} to a rather simple subfamily of constructions. As we will see in section  \ref{sec:example} this restricted class of vacua include as a subcase the supersymmetry-breaking backgrounds considered in \cite{bsb}, which were mainly motivated by duality arguments. 

\subsection{Supersymmetric vacua from the  BPS potential}
\label{sec:susy}

As explained in section \ref{sec:pot}, any vacuum must extremize the potential (\ref{pot})-(\ref{pot2}). Since this potential is a sum of squares, the simplest possibility is that each of these squares vanish separately. Let us first consider the $\calo(\alpha^{\prime 0})$ potential $V_0$ (\ref{potord0}). Imposing that all squares vanish demands that the warping should be constant, $\d A=0$, in agreement with the discussion of subsection \ref{sec:prel},  and that the following equations should be satisfied:
\bseq\label{susycond}
\begin{align}
&\d(e^{-2\phi}\Omega)=0 \label{susycond1}\\
&\d(e^{-2\phi}J\wedge J)=0 \label{susycond3}\\
&e^{2\phi}\d(e^{-2\phi}J)=*H \label{susycond2}
\end{align}
\eseq
These indeed match the conditions obtained in \cite{stromtorsion} by standard spinorial arguments (see also Appendix \ref{app:spinsb}). In particular, (\ref{susycond1}) requires $M$ to be a complex manifold with a nowhere-vanishing globally defined holomorphic $(3,0)$-form (so $c_1(M)=0$). The second condition (\ref{susycond3}) requires the internal space to be conformally balanced \cite{michelson82}. Finally, the third condition (\ref{susycond2}) imposes that, in presence of a non-vanishing three-form flux $H$ the space is {\em not} (even conformally) K\"ahler. By introducing the complexified three form $\calg:=H-ie^{2\phi}\d (e^{-2\phi}J)$, one can see (\ref{susycond2}) as an imaginary-self-duality (ISD) condition
\be\label{ISDcond}
*\calg=i\,\calg
\ee
which means that $\calg^{2,1}$ is primitive, $\calg^{3,0}=0$ and $\calg^{1,2}=\eta\wedge J$ for some (0,1)-form $\eta$. 

Note that the above supersymmetry equations can be rewritten in a slightly different form,  by introducing the three-form 
\be\label{Gthree}
G:=H-i\d J=\calg-2i\d\phi\wedge J
\ee
Indeed, one can first use  (\ref{susycond3}) to rewrite (\ref{susycond2}) as: $G^{3,0}=0=G^{1,2}$. Then, by noticing that the integrability of the almost complex structure associated to $\Omega$ implies that $(\d J)^{3,0}=0$, (\ref{susycond}) can be rewritten as
\bseq\label{susycond_2}
\begin{align}
&\d(e^{-2\phi}\Omega)=0 \label{susycond1_2}\\
&\d(e^{-2\phi}J\wedge J)=0 \label{susycond3_2}\\
&G^{1,2}=0=G^{0,3} \label{susycond2_2}
\end{align}
\eseq
As we will see in section \ref{sec:4dint}, expressing the supersymmetry conditions as in (\ref{susycond2_2}) is more natural from the viewpoint of the effective four-dimensional theory. On the other hand, (\ref{susycond2}) has a direct interpretation in terms of calibrations, to be discussed in section \ref{sec:calibr}.

Let us now consider the $\calo(\alpha^\prime)$ potential (\ref{potord1}). Again, we require squared terms to separately vanish. The terms in the last line automatically vanish since we have already imposed that $A$ is constant. On the other hand, from the first line in (\ref{potord1}) and via (\ref{squareformF}) one gets the conditions
\be\label{HYM}
F^{0,2}=0\, ,\quad J\lrcorner F=0
\ee
This means that the gauge bundle should be holomorphic and with a primitive field-strength or, in other words, the gauge bundle must be Hermitian-Yang-Mills (HYM). Finally, from the second line of (\ref{potord1}) or (\ref{squareformR}) one gets 
\be
R_+^{0,2}=0\, ,\quad J\lrcorner R_+=0
\label{RHYM}
\ee
Conditions which, up to higher order $\a'$ corrections, are automatically implied by the supersymmetry conditions \cite{sentorsion}. 

\subsection{Torsion induced SUSY-breaking vacua}
\label{sec:torsusyb}

Let us now turn to non-supersymmetric configurations. The strategy can be divided in two steps. First we can look for a supersymmetry breaking ansatz such that it violates the supersymmetry conditions of subsection \ref{sec:susy} but still leads to a vanishing potential $V$. As a second step, we need to consider whether $V$ can be extremized within this ansatz, and which further constraints such extremization may impose.

Focusing on the $\calo(\alpha^{\prime 0})$ piece of the potential, one sees that  $V_0$ is the sum of positive and negative definite terms, and that violation of the supersymmetry conditions implies that some of the positive definite terms  do not vanish. Hence a $V_0=0$ can only be attained if there is an exact cancellation between the positive and negative definite terms of (\ref{potord0}).

Now, from the general remarks of subsection \ref{sec:prel}, we know that the warping should be constant up to order  $\alpha'^2$, and so we can already set $\d A=0$. While we are still left with a large number of terms in $V$, a drastic simplification is obtained by imposing that the conditions (\ref{susycond2}) and (\ref{susycond3}) are not violated in the non-supersymmetric vacuum. As discussed in section \ref{sec:calibr}, this  guarantees the  geometrical structure behind the stability of the gauge bundle and space-time filling NS5-branes, and so it  appears particularly natural in the context of compactification with stable gauge sectors. To summarize, we impose
\bseq\label{sbcond0}
\begin{align}
&\d(e^{-2\phi}J\wedge J)=0\label{sbcond3} \\
&e^{2\phi}\d(e^{-2\phi}J)=*H\quad (\Leftrightarrow\ *\calg=i\,\calg) \label{sbcond2}
\end{align}
\eseq
but we allow for
\be\label{violdw}
\d(e^{-2\phi}\Omega)\neq 0
\ee
This choice makes all the terms of $V_0$ containing derivatives of $J$ vanish and encodes the origin of the supersymmetry breaking in the violation of (\ref{susycond1}). If we further simplify the potential by imposing that 
\be\label{addcond}
\bar\Omega\lrcorner \d(e^{-2\phi}\Omega)=0\
\ee 
then the last line on the r.h.s.\ of (\ref{potord0}) also vanishes. We are thus left with the following non-vanishing contributions to the potential
\be
V_0^\prime=\frac{1}{4\kappa^2}\int \text{vol}_M\,e^{4A-2\phi}\big[|e^{-3A+2\phi}\d(e^{3A-2\phi}\Omega)|^2-|J\wedge \d \Omega|^2\big]
\label{potpiece}
\ee
which, taking into account that $\d A=0$, vanishes if and only if \footnote{Notice that, in fact,  we could have vanishing potential even by violating the condition (\ref{sbcond3}), due to a non-trivial cancellation of the terms containing $\d(e^{2A-2\phi}J\wedge J)$ in (\ref{potord0}). However, in this case the extremization of these terms in (\ref{pot2}) is not straightforward and needs to be checked separately. This kind of SUSY-breaking can be thought as driven by D-terms and removes part of the integrable geometrical structure which could be crucial to study the stability of the bundle,  cf.~sections \ref{sec:calibr} and \ref{sec:4dint} below.}
\be\label{sbcond}
|e^{2\phi}\d(e^{-2\phi}\Omega)|^2=|J\wedge \d\Omega|^2
\ee
Hence, this supersymmetry-breaking pattern originates from the r.h.s.\ of (\ref{sbcond}): 
\bea\label{sbcond1}
\text{SUSY-breaking}\quad\Leftrightarrow\quad (J\wedge J)\lrcorner\d(e^{-2\phi}\Omega)\neq 0
\eea 
An important implication of this condition is that this  supersymmetry breaking mechanism  is possible only if the complex structure defined by the SU(3)-structure {\em is not} integrable. 

This way of breaking supersymmetry  can be seen as the heterotic counterpart of the type II  supersymmetry breaking pattern discussed in \cite{dwsb}, which generalizes the flux-induced  SUSY-breaking pattern of type IIB warped Calabi-Yau/F-theory backgrounds \cite{gkp}. In \cite{dwsb}, this mechanism was named `domain-wall supersymmetry breaking' (DWSB for short) because of its interpretation in terms of calibrations. As discussed in section \ref{sec:calibr}, this interpretation is possible in the heterotic case as well, and so the present vacua will be dubbed in the same manner.

In order to make contact with the flux literature, it is useful to translate the above conditions to the language of torsion classes (see Appendix \ref{app:fermconv}).\footnote{See e.g. \cite{cardoso02} for a discussion of these supersymmetry conditions in the language of SU(3)-torsion classes.} First, (\ref{sbcond3}) and (\ref{addcond}) are equivalent to fixing $W_4$ and $W_5$ in terms of the dilaton
\be\label{45tor}
W_4\,=\, \d\phi\quad \quad \quad W_5\,=\, 2\d\phi
\ee
Second, (\ref{addcond}) and (\ref{sbcond}) are  equivalent to impose that
\be\label{sbtor}
e^{2\phi}\d(e^{-2\phi}\Omega)=W_1\, J\wedge J+W_2\wedge J\qquad \text{with}\qquad |W_2|^2=24|W_1|^2
\ee
where we recall that $W_2$ is a primitive $(1,1)$-form. Finally, (\ref{sbcond2}) can be rewritten as
\bseq\label{fluxtorsion}
\begin{align}
&H^{3,0}=-\frac34\,\bar{W}_1\,\Omega \label{fluxtor1}\\
&H^{2,1}=-i(\d\phi)^{1,0}\wedge J-iW_3^{2,1} \label{fluxtor2}
\end{align}
\eseq
Note that in this language the supersymmetry breaking can be associated to a non-vanishing $W_1$, that is, a non-vanishing $(\d J)^{3,0}$. Because of eq.(\ref{sbcond2}) this is directly related to a non-vanishing $H^{3,0}$, as eq.(\ref{fluxtor1}) shows. We can thus characterize this supersymmetry breaking mechanism in terms of the three-form $G$ defined in (\ref{Gthree}) as follows. By using (\ref{sbcond3}), we can write (\ref{sbcond2}) as 
\be\label{Fflat}
G^{3,0}=G^{1,2}=0
\ee as in the supersymmetric case. Then, by comparing with (\ref{susycond2_2}), we can identify the origin of supersymmetry breaking with the non-vanishing of $G^{0,3}=-\frac32\, W_1\,\bar\Omega$: 
\bea\label{sb2}
\text{SUSY-breaking}\quad\Leftrightarrow\quad G^{0,3}\neq 0
\eea   
Formally, this condition is identical to the SUSY-breaking condition in type IIB warped Calabi-Yau/F-theory backgrounds \cite{gkp}, where $G$ is constructed as $G=F_{\rm RR}+ie^{-\phi}H$ with $F_{\rm RR}$ the Ramond-Ramond three-form flux. 

This supersymmetry breaking mechanism can also be described in the more standard language of Killing spinor equations. By using the results of Appendix \ref{app:spinsb}, one can see that the gravitino and dilatino Killing spinor equations are violated as follows
\bseq\label{spinviol}
\begin{align}
\delta\psi_\mu&=0 \label{spinviol1}\\
\delta\psi_m&=-\frac{i}4\cals_{mn}\,\zeta\otimes\gamma^n\eta^*+\text{c.c.}\label{spinviol2}\\
\delta\lambda&=\,3\,W_1\,\zeta\otimes \eta^*+\text{c.c.}\label{spinviol3}
\end{align}
\eseq
where we have introduced the two-form
\be\label{cals}
\cals:=W_2+4W_1 J 
\ee

In order to conclude our $\calo(\alpha^{\prime 0})$ discussion, it remains to impose the extremization of $V_0$, which unlike in the supersymmetric case is not automatic. However, it is sufficient to impose the extremization of $V_0^\prime$ given in (\ref{potpiece}), since all other terms are automatically extremized, being quadratic in vanishing terms. In particular, it is easy to see that the only non-trivial contribution comes from the extremization of $V_0^\prime$ under variations of the metric. By using eq.(\ref{su3var}), the resulting residual equations of motion are given by
\bea\label{eomtor}
\Im\big[\iota_{(m}\bar\Omega\cdot\iota_{n)}\d\cals\big]&=&
8g_{mn}|W_1|^2-2\Re[\bar W_1(\iota_m W_2\cdot \iota_n J)]-\Re[\iota_{m}W_2\,\cdot\,\iota_{n}\bar{W}_2]\cr
&=& |W_1|^2\Big\{9g_{mn}-\Re\Big[ \iota_m\Big(\frac{W_2}{W_1}+J\Big)\cdot\iota_n\Big(\frac{\bar W_2}{\bar W_1}+J\Big) \Big]\Big\}
\eea
Note that only the (3,0) and primitive (2,1) components of $\d\cals$ can contribute to the l.h.s.\ of (\ref{eomtor}). Since the r.h.s.\ is a (1,1)-tensor, it can only be matched in the l.h.s.\, by a (3,0) component of $\d\cals$, and this implies that the primitive (2,1) components of $\d\cals$ must vanish
\be\label{eomtor1}
(\d\cals)^{(2,1)}_{\rm P}=0
\ee
In fact, the $(3,0)$ component of $\d\cals$ also vanishes, as can be seen by using (\ref{cals}) and (\ref{sbtor}).\footnote{Indeed, such component is proportional to $\d\cals\wedge\bar\Omega=-\cals\wedge \d\bar{\Omega}$, which vanishes upon imposing (\ref{sbtor}).} We conclude that the r.h.s.\ of (\ref{eomtor}) must vanish identically. By introducing the matrix 
\be\label{Udef}
U^m{}_n=\frac13\Big(\frac{W_2}{W_1}+J\Big)^m{}_n=\frac13\, g^{mk}\Big(\frac{W_2}{W_1}+J\Big)_{kn}
\ee
we arrive at the following matrix equation
 \be\label{eomtor2}
 \Re(U\cdot U^\dagger)=\bbone
 \ee
where $(U^\dagger)^m{}_n=(U_n{}^m)^*=g_{nk}g^{mr}(U^k{}_r)^*$.
To sum up, the equations of motion (\ref{eomtor}) boil down to the conditions (\ref{eomtor1}) and (\ref{eomtor2}). Furthermore, the primitivity of $W_2$ implies that
\be\label{traceU}
\tr (IU)=-2
\ee
where $I^m{}_n=g^{mk}J_{kn}$ is the almost complex structure associated to $\Omega$.
While at this point these conditions look rather mysterious, we will provide a simple geometrical interpretation for them in section \ref{sec:oneparSB}.


At order $\alpha^\prime$, we also need to impose the extremization of the term $V_1$ in (\ref{potord1}). Since $\d A=0$, terms containing the warping do not provide further constraints. On the other hand, the terms containing the gauge bundle field-strength $F$ are extremized if 
\be\label{almostHYM}
F^{0,2}=0\, ,\quad J\lrcorner F=0
\ee
as in the supersymmetric case. Recall that  the almost-complex structure of $M$ is not integrable. Following \cite{bryant}, we can say that (\ref{almostHYM}) requires the bundle  to be {\em pseudo-HYM} and in particular that  the condition $F^{0,2}=0$ requires the bundle to be {\em pseudo-holomorphic}. Clearly, one cannot use the standard theory of bundles on K\"ahler spaces. However, as we will discuss in more detail in section \ref{sec:calibr}, the conditions (\ref{sbcond2}) and (\ref{sbcond3}) still allow to define a sort of stability of the gauge bundle, analogous to the one for bundles on K\"ahler spaces. 

Finally, from the second line of (\ref{potord1}) one gets 
\be\label{curvHYM}
R_+^{0,2}=0\, ,\quad J\lrcorner R_+=0
\ee
Unlike in the supersymmetric case, these equations are no longer  automatically satisfied. However, imposing that the supersymmetry breaking is mild enough compared to the compactification scale, the violation of (\ref{curvHYM}) is expected to be mild as well, and possibly negligible at $\calo(\alpha^\prime)$. We will come back to this point in subsection \ref{sec:higherorder}.

\subsection{Gravitino and gaugino mass}
\label{sec:gravitinomass}

While eqs.(\ref{sbcond1}) and (\ref{sb2}) translate into geometry the fact that supersymmetry is broken, in principle one would like to provide a more physical measure of the amount of SUSY-breaking by computing the 4d gravitino and gaugini masses. Note that in general a simple consistent truncation ansatz does not necessarily exists for these backgrounds and, as a result, there is no precise definition of the four-dimensional gravitino. Nevertheless, one can introduce a sort of four-dimensional gravitino $\psi^{\rm 4D}_\mu$ defined by the fermionic decomposition
\be\label{gravdec}
\psi_\mu+\frac12\Gamma_\mu(\Gamma^m\psi_m-\lambda)=\psi^{\rm 4D}_\mu\otimes\eta^*+\text{c.c.}
\ee
Note that defined in this way $\psi^{\rm 4D}_\mu$ may depend on the internal coordinates and thus cannot be considered as a properly defined four-dimensional gravitino. Nevertheless, the combination of the ten-dimensional gravitino and dilatino appearing on the l.h.s.~of (\ref{gravdec}) is such that the four-dimensional kinetic term for  $\psi^{\rm 4D}_\mu$, resulting from the ten-dimensional action of \cite{bdr},\footnote{Our conventions and the ones used in \cite{bdr} are related by: $\phi^{\rm there}=\exp(2\phi^{\rm here}/3)$, $H^{\rm there}=H^{\rm here}/3\sqrt{2}$, $\psi^{\rm there}_M=\psi^{\rm here}_M$, $\lambda^{\rm there}=-\lambda^{\rm here}/2\sqrt{2}$ and $\chi^{\rm there}=-\chi^{\rm here}$.} has a canonical form and does not mix with other fermions. 

One can now introduce a function $\mathfrak{m}_{3/2}$ which plays the role of the gravitino mass but generically also depends on the internal coordinates. Indeed, let us define $\mathfrak{m}_{3/2}$ as follows
\be\label{defgravmass}
\delta\psi^{\rm 4D}_\mu=\frac12\, \mathfrak{m}_{3/2}\,\hat\gamma_\mu\zeta~
\ee
i.e., by plugging the usual 4d SUSY-breaking formula which relates the variation of the gravitino to the gravitino mass.\footnote{In order to identify the four-dimensional spinor $\zeta$ in (\ref{fermsplit}) with the generator of the four-dimensional supersymmetry, it is convenient to choose the normalization $\eta^\dagger\eta=e^A$ for the internal spinor $\eta$.}

Applied to the SUSY-breaking backgrounds described in subsection \ref{sec:torsusyb}, by (\ref{spinviol}) we see that the above definitions provide the expression
\be\label{warptor}
\mathfrak{m}_{3/2}=3\, e^A\, W_1~
\ee
So $\mathfrak{m}_{3/2}$ can be related to the scale set by the (4d normalized) torsion class $W_1$. Note again that this scale depends on the internal coordinates of the internal manifold.

In order to make contact with four-dimensional effective theory, one would however like to have a more standard expression for the gravitino mass $m_{3/2}$. This can be obtained by imposing $\psi^{\rm 4D}_\mu$ to be constant in the internal space and averaging $\mathfrak{m}_{3/2}$ with an appropriate dilaton-factor 
\be\label{gravmass}
m_{3/2}=\langle\mathfrak{m}_{3/2}\rangle:=\frac{\int_M \text{vol}_M\,e^{-2\phi}\mathfrak{m}_{3/2}}{\int_M \text{vol}_M\,e^{-2\phi}}=\frac{i e^A\int_M\, e^{-2\phi}\,\Omega\wedge G}{4{\int_M \text{vol}_M\,e^{-2\phi}}}
\ee
where in the last step we have used (\ref{sbtor}) and the condition $G^{3,0}=0$ for the three-form $G$ defined in (\ref{Gthree}). One can then fix the four-dimensional Einstein frame by setting $e^A$ as in (\ref{warp_planck}) and gets
\be\label{gmasstor}
m_{3/2}=\frac{i\, g^3_{\rm s}\ls^4\, M_{\rm P}\int_M\,e^{-2\phi}\,\Omega\wedge G}{8\sqrt{\pi}\,\text{Vol}(M)^{3/2}}
\ee

Let us now turn to the gaugino mass. The four-dimensional gaugino $\chi^{\rm 4D}$ is related to the ten-dimensional gaugino by the KK decomposition
\be\label{chidec}
\chi=e^{-2A}\chi^{\rm 4D}\otimes \eta+\, \text{c.c.}
\ee
The relevant terms in the ten dimensional action \cite{bdr} are given by
\be\label{chiaction}
-\frac{\alpha^\prime}{4\kappa^2}\int \d^{10}x\sqrt{-g}\, e^{-2\phi}\big(\tr\bar\chi\slashed\nabla\chi-\frac14\tr\bar\chi\slashed{H}\chi\big)
\ee
Plugging (\ref{chidec}) into (\ref{chiaction}) and integrating over the internal space by keeping $\chi^{\rm 4D}$ constant on it, one obtains the following value of the gaugino mass
\be
m_{1/2}=\frac{i\, e^A\int_M\, e^{-2\phi}\Omega\wedge H}{2\int_M \text{vol}_M\,e^{-2\phi}}=\frac{i e^A\int_M\, e^{-2\phi}\,\Omega\wedge G}{4{\int_M \text{vol}_M\,e^{-2\phi}}}
\ee 
where, in the last step, we have again used the condition $G^{3,0}=0$. We thus see that the gaugino mass equals the gravitino mass (\ref{gravmass})
\be\label{mm}
m_{1/2}\equiv m_{3/2}
\ee 
As we will see in section \ref{sec:4dint}, this result has a very simple four-dimensional interpretation. 

\subsection{Conditions on the curvature}
\label{sec:higherorder}

Let us now discuss the conditions on the curvature  (\ref{curvHYM}) that arise at order $\a'$ from the minimization of the potential piece (\ref{potord1}). First of all, it is easy to see that
\be
R_{+mnpq}=R_{-pqmn}-(\d H)_{mnpq}
\ee
so, by using the BI (\ref{BI})
\be\label{Rplusminus}
R_{+mnpq}=R_{-pqmn}+\calo(\alpha^\prime)
\ee
Hence, up to $\calo(\alpha^{\prime 2})$ terms in the scalar potential, (\ref{curvHYM}) can be rewritten as
\be\label{curvHYM2}
\Omega_{kmn}R^{mn}_-=0\quad  \quad J_{mn}R^{mn}_-=0
\ee
These conditions can be rephrased by saying that the internal spinor $\eta$ specifying the SU(3)-structure is covariantly constant with respect to the torsion-full covariant derivative $\nabla^-_m$.  From (\ref{gravsusy}) and (\ref{spinviol2}), we know that this is not satisfied in the torsional SUSY-breaking backgrounds of subsection \ref{sec:torsusyb}. However, as already mentioned, we are actually assuming a mild SUSY-breaking.  So let us assume that $\nabla_m^-\eta\sim\calo(\alpha^{\prime\beta})$, with $0 < \beta \leq 1$. Roughly speaking, this would mean that both equations  in  (\ref{curvHYM})  are violated at $\calo(\alpha^{\prime\, \beta})$. In particular, by using (\ref{squareformR}), the curvature squared term in (\ref{potord1}) would be of $\calo(\alpha^{\prime 2\beta})$, and so negligible in our approximation for $\beta\geq 1/2$. Under this condition, the full potential would be extremized at the order of approximation that we are assuming.   

We can make this argument more concrete. From  (\ref{gravsusy}) and (\ref{spinviol2}) we have that
\be\label{comm-}
\nabla^-_m\eta=-\frac{i}{4}\cals_{mn}\gamma^n\eta^*
\ee
Thus, taking into account (\ref{cals}) and the condition $|W_2|^2=24|W_1|^2$, we qualitatively have that $\nabla^-_m\eta\sim W_1\gamma_m\eta^*$. Let us now introduce a dimensionless scale $L_{\rm SB}$, measured in string units, which characterizes  the geometrical origin of the SUSY-breaking. The torsion class $W_1$ has the dimension of mass and defines a dimensionless  SUSY-breaking length scale $L_{\rm SB}$ through
\be\label{torscale}
W_1\sim (\ls L_{\rm SB})^{-1}
\ee
Then, taking $g_{mn} \sim \ls^2 L^2_{\rm KK}$, with $L_{\rm KK}$ being the KK length measured in string units we have $\nabla_m^-\eta\sim\,L_{\rm KK}\,L^{-1}_{\rm SB}$, and so in order to have a SUSY-breaking of $\calo(\alpha^{\prime\beta})$ one must have $\nabla_m^-\eta\sim L^{-2\beta}_{\rm KK}$ and then 
\be\label{sbkk}
L_{\rm SB}\sim L^{2\beta+1}_{\rm KK}
\ee  

Furthermore, by introducing the four-dimensional KK-scale $M_{\rm KK}=e^A/(\ls L_{\rm KK})$ and recalling (\ref{warptor}), we can restate (\ref{torscale}) in a more physical way
\be
\mathfrak{m}_{3/2}\sim M_{\rm KK}\,L_{\rm KK}\,L^{-1}_{\rm SB}
\ee
Then we have a mild SUSY-breaking,  which can be seen as spontaneous from the four-dimensional point of view, if $\mathfrak{m}_{3/2}\ll M_{\rm KK}$ and this condition corresponds to 
\be\label{lowsb}
\frac{L_{\rm KK}}{L_{\rm SB}}\,\ll\, 1
\ee
 If for example $\beta=1/2$ in (\ref{sbkk}) we can identify $L_{\rm SB}$ with $L^2_{\rm KK}$, and the condition (\ref{lowsb}) is guaranteed if $L_{\rm KK}\gg 1$, which is required in the regime of validity of the supergravity approximation.  

Let us now consider in more detail curvature terms which enter (\ref{curvHYM2}).
By using (\ref{comm-}) and the formula
\be
[\nabla^-_p,\nabla^-_q]\eta=\frac14 R_{-mnpq}\gamma^{mn}\eta
\ee
one obtains
\bseq\label{curv2tor}
\begin{align}
J_{mn}R^{mn}_-{}_{pq}&=2\,P^{rs}\cals_{r[p}\,\cals^*_{q]s}\\
\Omega_{kmn}R^{mn}_-{}_{pq}&=\-4iP^r{}_k\nabla^-_{[p}\cals_{q]r}+\frac12\Omega_k{}^{rs}\,\cals_{r[p}\,\cals^*_{q]s}
\end{align}
\eseq
where $P^m{}_n=(1/2)(\delta^m{}_n-iI^m{}_n)$ projects onto the flat holomorphic indices of the almost complex structure $I^m{}_n\equiv g^{mk}J_{kn}$. Then, we have the following curvature squared terms contributing to (\ref{potord1}) 
\bseq\label{curv2tor_est}
\begin{align}
  |J_{mn}R^{mn}_-|^2&\sim\, |W_1|^4\\
|\Omega_{kmn}R^{mn}_-|^2 &\sim\, |\partial W_1|^2+|W_1^2\partial W_1| +|W_1|^4\nonumber\\
&\sim\, L^{-2}_{\rm KK}|W_1|^2+L^{-1}_{\rm KK}|W_1|^3 + |W_1|^4\label{jr}
\end{align}
\eseq
By using (\ref{torscale}), the $\calo(\alpha^\prime)$ corrections to the equations of motions  associated to the curvature terms in  (\ref{potord1}) -- cf.\ equation (\ref{modEinst}) -- are of order
\bea\label{firstordcorr}
\text{(EoM)}_{\calo(\alpha^\prime)}&\sim&L^2_{KK}(|\Omega_{kmn}R^{mn}_-|^2+|J_{mn}R^{mn}_-|^2)\cr
&\sim& \frac{1}{L^2_{\rm KK}}\Big[\Big(\frac{L_{\rm KK}}{L_{\rm SB}}\Big)^2+\Big(\frac{L_{\rm KK}}{L_{\rm SB}}\Big)^3+\Big(\frac{L_{\rm KK}}{L_{\rm SB}}\Big)^4\Big]
\eea
The overall factor $L^{-2}_{\rm KK}$ is associated to the leading $\calo(\alpha^\prime)$ behavior, while the terms in squared brackets provides  a further supression factor because of (\ref{lowsb}). In order to have a correction (\ref{firstordcorr}) which goes like $L^{-4}_{\rm KK}$ and is thus $\calo(\alpha^{\prime 2})$ and negligible at our $\calo(\alpha^\prime)$ approximation, we must have $L_{\rm SB}=L^2_{\rm KK}$, and thus $\beta=1/2$. On the other hand, we could further relax this condition,  depending on the details of the background. If for example in (\ref{jr}) we have $|\partial W_1|\lesssim|W_1|^2$, then it is enough to take $L_{\rm SB}=L^{3/2}_{\rm KK}$, i.e.\ $\beta=1/4$.


\section{NS5-branes, calibrations and bundle stability}
\label{sec:calibr}

As already discussed in the literature (see e.g. \cite{gauntlett1}) the supersymmetry conditions (\ref{susycond}) admit a clear interpretation in terms of the so-called $p$-form calibrations \cite{lawson,papa0}, which are nothing but $p$-forms that measure the energy of BPS extended objects of the theory. Such family of BPS objects are most conveniently classified in terms of their four-dimensional appearance, as illustrated in figure \ref{califig}, since a different calibration exists for each 4d BPS object. In particular, the two-form $e^{-2\phi}J$ is a calibration for an NS5-brane that wraps an internal two-cycle of $M$ and fills the four-dimensional space-time $X_4$. The four-form  $e^{-2\phi}J\wedge J$, on the other hand, is a calibration for NS5-branes wrapping an internal four-cycle and filling two directions in $X_4$ (i.e., showing up as 4d strings upon dimensional reduction). Finally, for any constant phase $e^{i\theta}$, $e^{-2\phi}\Im(e^{i\theta}\Omega)$ calibrates NS5-branes wrapped on three internal and three external directions, thus appearing as a domain-wall in four dimensions. More schematically, we have the following dictionary between calibrations and BPS objects of the compactification\footnote{By the discussion below, gauge bundles could also be added to this dictionary on the same footing as space-time filling NS5-branes.}

\EPSFIGURE{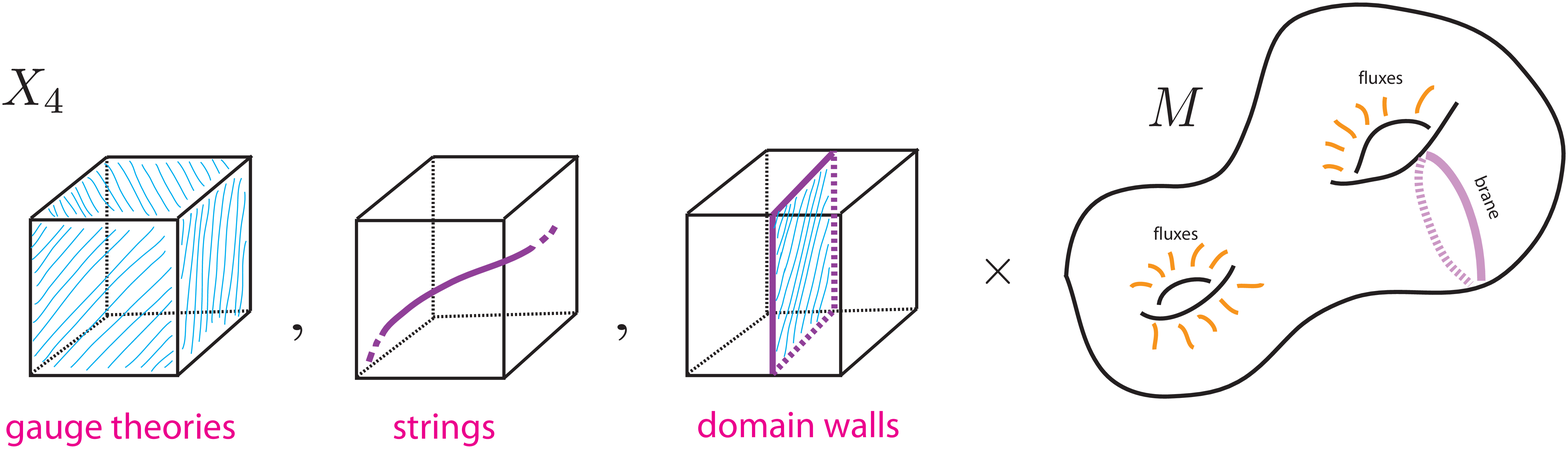, width=6in}
{\label{califig} BPS objects of the theory, classified in terms of their 4d appearance.}

%
\TABLE{\renewcommand{\arraystretch}{1.5}
\begin{tabular}{ccccc} 
Calibration & \quad &10d BPS object & \quad &4d BPS object\\ \hline
$e^{-2\phi}J$ & \quad & NS5 on $X_4 \times \Pi_2$ & \quad & gauge theory \\
$e^{-2\phi}\Omega$ & \quad & NS5 on $X_3 \times \Pi_3$ & \quad & domain wall \\
$e^{-2\phi} J \wedge J$ & \quad & NS5 on $X_2 \times \Pi_4$ & \quad & string \\
\end{tabular} }
\noindent
where $\Pi_p$ is a $p$-dimensional submanifold of $M$, and $X_d$ is a $d$-dimensional slice of $X_4$. More precisely, the statement is that all these $p$-forms can be defined as calibrations {\em only if} the corresponding differential conditions (\ref{susycond}) are satisfied. 

Now, recall that our SUSY-breaking pattern is characterized by (\ref{sbcond0}) and (\ref{violdw}). Hence  we see that, even if supersymmetry is broken,  $e^{-2\phi}J$ and $e^{-2\phi}J\wedge J$ can still be identified as calibrations, while $e^{-2\phi}\Im(e^{i\theta}\Omega)$ cannot be considered as such. For this reason, we may dub this pattern as `domain-wall supersymmetry breaking' (DWSB) as done in \cite{dwsb} for the analogous pattern in the context of type II flux compactifications, which used the interpretation in terms of calibration provided by \cite{lucal,lucal2,km07}.  In order to understand better the implications of this observation, let us recall the main properties of a calibration.


In general, a calibration structure provides a natural BPS bound for certain branes in our 10d theory. Let us, for instance, consider NS5-branes filling $X_4$ and wrapping an internal two-cycle $\Pi_2 \subset M$. These branes couple magnetically to the three-form flux $H$ and thus modify the (internal) BI as
\be\label{BIfin}
\d H=\frac{\alpha^\prime}4 (\tr R_+\wedge R_+-\tr F\wedge F)+2\kappa^2\tau_{\rm NS5}\,\delta^4(\Pi_2)
\ee
where $\tau_{\rm NS5}=(2\pi)^{-5}(\alpha^\prime)^{-3}$.  It is clear that $H$ and $\Pi_2$ cannot be considered as independent and for this reason it is convenient to go to the dual description, where the NS5-brane couples electrically to the seven-form flux $\hat H=\text{vol}_M\wedge (e^{-2\phi}*H)$.\footnote{In order to simplify the notation, in the remainder of this section we will set $e^A=1$.} We can then write $\hat H={\rm vol}_M \wedge \tilde H$, where $\tilde H=e^{-2\phi}*H$ is a three-form of $M$. The dualization procedure in absence of NS5-branes is reviewed in Appendix \ref{app:dual} and leads to a dual formulation of the supergravity potential $V$ introduced in section \ref{sec:susybvacua}. The only modification due to the addition of a NS5-brane is the addition of the NS5-potential to $V$
\be
\label{NS5pot}
V_{\rm NS5}=\tau_{\rm NS5}\int_{\Pi_2} (e^{-2\phi}\sqrt{\det g|_{\Pi_2}}\,\d^2\sigma-\tilde B)
\ee 
where $\tilde B$ is the potential two-form of $\tilde H$, $\tilde H=\d \tilde B$, and $\d^2\sigma=\d\sigma^1\wedge \d\sigma^2$ is the usual volume density induced by the world-volume coordinates $(\sigma^1,\sigma^2)$ on $\Pi_2$.

As stated above, to such brane corresponds the calibration $e^{-2\phi}J$. Indeed, this two-form provides the following algebraic inequality
\be\label{algineq}
e^{-2\phi}\sqrt{\det g|_{\Pi_2}}\,\d^2\sigma\geq e^{-2\phi}J|_{\Pi_2}
\ee
for any (appropriately oriented) $\Pi_2$. When the above inequality is saturated at every point of $\Pi_2$, then one says that the cycle $\Pi_2$ is calibrated
\be\label{calcycle}
\Pi_2\ \text{calibrated}\quad\Leftrightarrow\quad  e^{-2\phi}\sqrt{\det g|_{\Pi_2}}\,\d^2\sigma= \, e^{-2\phi}J|_{\Pi_2}
\ee
Now, the differential condition ({\ref{susycond2}})/(\ref{sbcond2}) can be rewritten as  
\be\label{diffconddual}
(\text{\ref{susycond2})/(\ref{sbcond2}})\quad \Leftrightarrow\quad \d(e^{-2\phi}J)=\tilde H
 \ee 
and allows to prove the following important statement: an NS5-brane wrapping a calibrated two-cycle $\Pi_2$ globally minimizes its potential energy (\ref{NS5pot}) under continuous deformations. More precisely, considering any other two-cycle $\Pi_2^\prime$ connected to $\Pi_2$ by a three chain $\Gamma$, $\partial\Gamma=\Pi_2^\prime-\Pi_2$, one gets $V_{\rm NS5}(\Pi_2^\prime)\geq V_{\rm NS5}(\Pi_2)$. Indeed
\be
V_{\rm NS5}(\Pi_2^\prime)\geq \tau_{\rm NS5}\int_{\Pi_2^\prime} (e^{-2\phi}J-\,\tilde B)=  \tau_{\rm NS5}\int_{\Sigma} (e^{-2\phi}J-\,\tilde B)= V_{\rm NS5}(\Pi_2)
\ee
where  we have used  (\ref{algineq}) in the first step, the differential condition (\ref{diffconddual}) in the second one, and the definition of calibrated cycle (\ref{calcycle}) in the last step. The same arguments equally apply for the calibration $e^{-2\phi}J\wedge J$.

Thus, we see that in DWSB compactifications we have a  natural notion of BPSness and stability for space-filling and string-like NS5-branes. These structures are typically associated to supersymmetric settings and we can then see them as a distinguished property of the SUSY-breaking pattern considered here, in analogy with the type II setting of \cite{dwsb}.  

In fact, the calibration structures provided by $e^{-2\phi}J$ and $e^{-2\phi}J\wedge J$ have also implications on the notion of gauge-bundle stability in this non-supersymmetric context. Since the gauge bundle is associated to an induced NS5-brane charge density proportional to $-\tr(F\wedge F)$ (which is at the origin of the BI identity (\ref{BIfin})), it is again convenient to work in the dual formulation reviewed in Appendix \ref{app:dual}. In this formulation, one can isolate the following contribution of the gauge bundle to the total potential
\be
V_{\rm bundle}=\frac{1}{8\kappa^2}\int_M \big[ e^{-2\phi}\tr(F\wedge * F) +\tilde B\wedge \tr (F\wedge F)\big]
\ee
Now, the bundle-analog of the inequality (\ref{algineq}) for NS5-branes is
\be\label{bundleineq}
\tr(F\wedge * F) \geq-\, \tr(F \wedge F)\wedge J
\ee
and the analog of the calibration condition (\ref{calcycle}) is the pseudo-HYM \cite{bryant} condition\footnote{The prefix `pseudo' comes from the non-integrability of the almost-complex structure of $M$.}
\be\label{dualHYM}
\tr(F\wedge * F) =-\, \tr(F \wedge F)\wedge J\quad \Leftrightarrow\quad \left\{ \begin{array}{ll} F^{0,2}=0 &\quad  \text{(F-flatness)}\\  J\wedge J\wedge \tr F=0&\quad \text{(D-flatness)}\end{array}\right.
\ee 
Now, as for NS5-branes, one can easily show that the pseudo-HYM gauge bundles are absolute minima of $V_{\rm bundle}$ under continuous deformations, and again the differential condition (\ref{diffconddual}) is crucial for such result. Indeed, suppose that $F$ is a pseudo-HYM  field strength and $F^\prime$ is any other field-strength which is cohomologous  to $F$, so that there is a three-form $\alpha$ such that  $\tr(F^\prime\wedge F^\prime)=\tr(F\wedge F)+\d\alpha$. Then, we have $V_{\rm bundle}(F^\prime)\geq V_{\rm bundle}(F)$, since
\bea
V_{\rm bundle}(F^\prime)&\geq&-\frac{1}{8\kappa^2}\int_M \tr (F^\prime \wedge F^\prime)\wedge ( e^{-2\phi}  J-\tilde B)\cr
&=&-\frac{1}{8\kappa^2}\int_M \tr (F \wedge F)\wedge ( e^{-2\phi}  J-\tilde B)=V_{\rm bundle}(F)
\eea
where now we have used  (\ref{bundleineq}) in the first step, the differential condition (\ref{diffconddual}) in the second one,   and the pseudo-HYM condition (\ref{dualHYM}) in the last one. 
 
As written in (\ref{dualHYM}), the pseudo-HYM condition can be split into  two parts, which can be seen as F-flatness and D-flatness conditions, along the lines of \cite{witten86}. The first, $F^{0,2}=0$, demands the bundle to be pseudo-holomorphic \cite{bryant} and can be seen as an F-flatness condition, while the second one can be seen as a D-flatness condition.

Now, suppose that we can solve the F-flatness condition $F^{0,2}=0$. Then, clearly the D-flatness in (\ref{dualHYM}) admits a solution only if
\be
\int_M e^{-2\phi} J\wedge J\wedge \tr F=0
\ee  
This necessary condition is quasi-topological since it depends only on the first Chern class of the bundle, but changes under the deformations of $e^{-2\phi} J\wedge J$.  A natural question is then if one can extend this necessary condition in order to get a quasi-topological necessary and {\em sufficient} condition. In other words, one can ask whether a notion of quasi-topological stability exist for pseudo-holomorphic bundles in our backgrounds, which are only almost-complex spaces but nevertheless admit the calibration structures described above.   

In the complex (supersymmetric) case, the existence of a solution of the D-flatness equation for a holomorphic gauge bundle is equivalent to the so-called $\mu$-stability by the theorems of Donaldson-Uhlenbeck-Yau \cite{DUY}, which are valid for K\"ahler spaces, and their generalizations to non-K\"ahler hermitian spaces \cite{liyau}. In particular, in the non-K\"ahler case of interest for supersymmetric heterotic flux compactifications,  the $\mu$-stability of a  bundle $E$ is defined in terms of the $\mu$-slope of $E$
\be
\mu(E)=\frac{1}{\text{rank}E}\int_M e^{-2\phi} J\wedge J\wedge \tr F
\ee
Then a bundle $E$ is $\mu$-stable if $\mu(E^\prime)< \mu(E)$, for all coherent subsheaves $E^\prime$ of $E$. In the heterotic  case one has furthermore to impose $\mu(E)=0$ and this leads to considering the semi-stability condition  $\mu(E^\prime)\leq0$. 

 A key ingredient for obtaining all the above results is the closure of the four-form $e^{-2\phi}J\wedge J$, i.e., the fact that the internal non-K\"ahler space is balanced. Intriguingly, such property is preserved also in the non-supersymmetric almost-complex backgrounds we are discussing about and this suggests a possible extension of the above notion of stability to our non-supersymmetric setting. This extension would  then ultimately originate from the existence of the calibration structures characterizing our backgrounds -- see \cite{HL2} for a recent study of the properties of calibrated geometries. However, a proper analysis of this possibility is beyond the scope of the present paper.


\section{$\frac12$ Domain-Wall supersymmetry breaking}
\label{sec:oneparSB}

Via the above dictionary relating supersymmetry conditions and calibrations, it is possible to identify  a subclass of the SUSY-breaking configurations discussed in section \ref{sec:susybvacua} with particularly interesting properties. Such subclass, dubbed $\frac12$DWSB vacua in the following, 
present a rather constrained internal geometry and SUSY breaking pattern with respect to the general DWSB case. More precisely, via an effective four-dimensional interpretation we will identify these compactifications as particular realizations of 4d no-scale vacua with broken supersymmetry. The latter result is not so surprising after one realizes that, when translating the definition of $\frac{1}{2}$DWSB vacua to the type II context, one finds that the $\caln=0$ vacua described in \cite{gkp} fall into this subclass.\footnote{The same applies to the so-called one-parameter DWSB vacua constructed in \cite{dwsb}.} We would therefore expect that, upon the usual chain of dualities, any of the $\caln=0$ vacua of \cite{gkp} are mapped within the class of heterotic backgrounds described in this section. 

\subsection{$\frac{1}{2}$DWSB vacua}
\label{1/2DWSB}

Let us first define our subclass of backgrounds by those that in addition to eqs.(\ref{sbcond0}) satisfy the condition
\be
\Im\Big[ e^{i\theta}\d \big( e^{-2\phi} \Omega \big)\Big] \, =\, 0
\label{susycond1/2}
\ee
for some phase $e^{i\theta}$. This condition is clearly weaker than (\ref{susycond1}), and thus trivially satisfied on any supersymmetric background. In fact, in the case where the phase $e^{i\theta}=e^{i\theta_0}$ is constant, one can think of eq.(\ref{susycond1/2}) as half-imposing the supersymmetry equation (\ref{susycond1}), in the sense that $\Omega$ does not satisfy (\ref{susycond1}) but $\Im ( e^{i\theta_0} \Omega)$ does. Note that this notion of half-imposing a supersymmetry condition can be given a well-defined physical meaning by interpreting (\ref{susycond1}) and (\ref{susycond1/2}) as domain-wall calibrations conditions (see section \ref{sec:calibr}), from which we are led to dub the present subclass of backgrounds as $\oh$DWSB backgrounds. Finally, note that in terms of differential conditions satisfied by our background, imposing eqs.(\ref{susycond2}), (\ref{susycond3}) and (\ref{susycond1/2}) seems as close as we may get to a supersymmetric background, since by additionally imposing (\ref{susycond1/2}) with the choice of $\theta' \neq \theta\, {\rm mod\, } \pi$ the whole set of SUSY conditions (\ref{susycond}) follow.

Decomposing (\ref{susycond1/2}) in terms of $J$ and $\Omega$ it is easy to convince oneself that it is equivalent to require the condition (\ref{addcond}), as well as $\Im (e^{i\theta} W_1) = 0$ and $\Im (W_2/W_1) = 0$. The latter condition alone is equivalent to impose that the matrix $U$ defined in (\ref{Udef}) is real. This in turn implies that $U^\dagger=-U$ and so the constraint (\ref{eomtor2}) reads
\be
U^2=-\bbone\quad\Leftrightarrow\quad (IU)^2=\bbone
\ee
where we have also used that $[U,I]=0$. It is then natural to introduce the matrices
\be
P_N=\frac12(\bbone+IU)\quad \quad P_N^\perp=\frac12(\bbone - IU)
\ee
satisfying the following properties
\be
P_N^\dagger=P_N\quad\quad P_N^2=P_N\quad \quad [P_N, I]=0
\ee 
and similarly for $P_N^\perp$. These properties imply that $P_N$ and $P_N^\perp$ are projector operators that split the tangent bundle into two orthogonal sub-bundles 
\be\label{tangsplit}
T=T_N\oplus T_N^\perp
\ee
which preserve the almost complex structure in the sense that $IT_N\subset T_N$ and $IT_N^\perp\subset T^\perp_N$, and so we can write $I=I_{T_N}+I_{T_N^\perp}$, with $I_{T_N}$ and $I_{T_N^\perp}$ almost-complex structures on $T_N$ and $T^\perp_N$ respectively. On the other hand, (\ref{traceU}) implies that 
\be
\tr P_N=2
\ee
and so $T_N$ is a two-dimensional vector space. Finally, by using (\ref{Udef}) one can see that 
\be
W_2=2W_1\,J(P^\perp_N-2P_N)
\ee
which together with (\ref{addcond}) gives
\be
e^{2\phi}\d(e^{-2\phi}\Omega)=3W_1(JP^\perp_N)\wedge (JP^\perp_N)
\label{splitsb0}
\ee
By Frobenius' theorem, this implies that the subbundle $T_N$ can be integrated into two dimensional submanifolds. We are thus led to consider a fibered space of the form 
\be\label{fibration}
N\  \hookrightarrow\  M\ \xrightarrow{\pi}\ \calb
\ee
with a two-dimensional fiber $N$ and a four-dimensional base $\calb$. Note that $I_{T_N}$ defines a preferred (integrable) complex structure $I_N$ on $N$ at each point of $\calb$. This fibration structure induces a dual decomposition of the cotangent bundle $T^*=T^*_\calb\oplus T^{*\perp}_\calb$, with $T^*_\calb\equiv (T^\perp_N)^*$ and $T^{*\perp}_\calb\equiv (T_N)^*$, and so we can then decompose the SU(3)-structure accordingly
\be\label{JOmega1par}
J=J_{\calb}+j\, \quad\quad \Omega=\Omega_\calb\wedge \Theta
\ee
where $j=JP_N=-\frac{i}{2}\Theta\wedge \bar\Theta$, $J_\calb = JP_N^\perp$, etc. Using this notation, we can rewrite (\ref{splitsb0}) as
\be
e^{2\phi}\d(e^{-2\phi}\Omega)=3W_1J_\calb\wedge J_\calb
\label{splitsb}
\ee

An important implication of (\ref{splitsb}) is that the fibration (\ref{fibration}) is equipped with a transverse complex structure, i.e.\ an integrable complex structure along the base. To see this, let us introduce some (non-canonical) local coordinates $x^a, y^m$, where $x^a$ are along the fiber $N$ and  $y^m$ are along the base $\calb$, in the sense that the fibers are described by $y^m=\text{const}$. We can then write $\Omega_\calb=\frac12(\Omega_\calb)_{mn}\d y^m\wedge \d y^n$ and, by using the coordinate co-frame $\d x^a$ and $\d y^m$ for  $T^*_M$, we can (non-canonically) split $\Theta=\Theta_\calb+\Theta_N$ (with $\Theta_N\neq 0$) and $\d=\d_\calb+\d_N$, with obvious notation. Now, (\ref{splitsb}) implies that 
\be\label{splitdwsb}
\d_\calb\Omega_\calb\wedge\Theta_N+\Omega_\calb\wedge\d_\calb\Theta_N+\Omega_\calb\wedge\d_N\Theta_\calb=0\quad\quad \d_N\Omega_\calb\wedge\Theta_N+\Omega_\calb\wedge\d_N\Theta_N=0
\ee
The first condition in (\ref{splitdwsb}) is telling us that $(\d_\calb\Omega_\calb-\alpha_\calb\wedge\Omega_\calb)|_{x^a=\text{const}^a}=0$, for some one-form $\alpha_\calb=(\alpha_\calb)_m\d y^m$, and this in turn means that $\Omega_\calb$ defines an  {\em integrable} complex structure on the four-dimensional slice $x^a=\text{const}$.  The second equation  implies that $\partial_a\Omega_\calb\propto \Omega_\calb$ and thus that such complex structure is in fact constant along the fiber coordinates $x^a$. Note that, even if the choice of coordinates $x^a,y^m$ is not canonical, as one may go to a different coordinate system 
\be
x^a\rightarrow \tilde{x}^a(x,y)\quad \quad y^m\rightarrow \tilde{y}^m(y)  
\ee
the above conclusions are clearly covariant under such a change of coordinates.
Thus, we see that $\Omega_\calb$ defines an {\em integrable} complex structure on the base or, in other words, that one can introduce complex coordinates $(z^1,z^2)$ on the base such that $\Omega_\calb=\frac12(\Omega_\calb)_{ij}\d z^i\wedge \d z^j$ and $\partial_a(\Omega_\calb)_{ij}\propto (\Omega_\calb)_{ij}$.

On the other hand, since $N$ is two-dimensional the almost complex structure $I_N$ of the fiber is always integrable.  We thus see that our background is characterized by two integrable complex structures on the base and the fiber, which however do not combine into an integrable complex structure on the whole space. For this to be possible, we need a non-trivial fibration. Explicit examples of such will be discussed in section \ref{sec:example}.

To summarize, for $\oh$DWSB backgrounds the compactification manifold $M$ can be described by a fibration of a two-dimensional fiber $N$ over a four-dimensional base $\calb$, both $N$ and $\calb$ being complex manifolds. Moreover, the torsion classes of $M$ are given by (\ref{45tor}) and
\be
W_2 \, =\, 2 W_1 \left(J - 3 j\right) \quad \quad {\rm with} \quad \quad W_1\,  =\, f\, e^{-i\theta}
\label{dwsb.52}
\ee
with $f$ a real function and $j$ a real (1,1)-form such that $j \cdot J\, =\, j \cdot j\, =\, 1$ (in fact, $j = J P_N$ and $J-j = J_{\calb}$).  The remaining torsion class $W_3$ is constrained via the presence of the flux $H$ and eq.(\ref{sbcond2}).

Note that in order to obtain vacua  one also needs to impose the condition (\ref{eomtor1}) on $\cals$, which here reads
\be
\label{cals1/2}
\cals\, =\,  6\, W_1 J_\calb
\ee
The two-form $\cals$  appears in the internal gravitino variation (\ref{spinviol2}) 
and so the violation of the gravitino and dilatino Killing spinor equations takes the form
\bseq\label{spinviol1par}
\begin{align}
\delta\psi_\mu&=0 \label{spinviol1_1par}\\
\delta\psi_m&=-\frac{3i}2\,W_1 \, (J_\calb)_{mn}\,\zeta\otimes\gamma^n\eta^*+\text{c.c.} \label{spinviol2_1par}\\
\delta\lambda&=\frac32\,W_1\otimes \eta^*+\text{c.c.} \label{spinviol3_1par}
\end{align}
\eseq
As we have discussed in section \ref{sec:gravitinomass}, $W_1$ is directly related to the gravitino mass. $J_{\calb}$ is in turn related to the source of SUSY-breaking, and more precisely to the chiral fields that develop non-vanishing F-terms, as the 4d analysis of the next subsection shows.

\subsection{Four-dimensional interpretation}
\label{sec:4dint}

In the following we would like to show that $\frac12$DWSB vacua can be interpreted as the ten-dimensional realization of a no-scale supersymmetry breaking in four dimensions \cite{noscale}.
In order to see this, we need an effective four-dimensional supergravity describing these kinds of flux compactifications. Unfortunately, such a theory is not available. However, the problem of identifying the four-dimensional theory governing quite general heterotic flux compactifications has been investigated in several papers (see e.g. \cite{gurrieri07,louis08} and references therein) which, under suitable simplifying assumptions, arrived at precise expressions for the effective four-dimensional theory. One of these assumption is that  all the scalar quantities in the internal spaces are assumed to be constant. Then, in the following we will implicitly approximate $W_1$ and the dilaton to be constant, which by (\ref{45tor}) implies that $W_4= W_5=0$. Note also that $W_1$ constant implies that $\theta = \theta_0$ is constant, and so without loss of generality we can set it to zero. This implies that our `smeared' compactification manifold $\tilde{M}$ satisfies $\Im W_1 = \Im W_2 = W_4 = W_5 = 0$, or in other words that in such smeared limit $\frac12$DWSB heterotic vacua reduce to compactifications in half-flat manifolds, like those studied in \cite{halfflat,gurrieri07}.

In terms of the SU(3)-structure described by $J$ and $\Omega$, the K\"ahler potential reads\footnote{Here, for simplicity, we do not consider the gauge bundle contribution to the K\"ahler potential.}
\be\label{kaehler}
\calk=-\log (s+\bar s)-\log \Big(\frac1{3!\,\ls^6}\int_M J\wedge J\wedge J\Big)-\log\Big(-\frac{i}{8\, \ls^6}\int_M\Omega\wedge \bar\Omega\Big)
\ee
where $\ls=2\pi\sqrt{\alpha^\prime}$ and the real part of $s$ is given by
\be
\Re s=\frac{1}{2\pi\ls^6}\int \text{vol}_M\, e^{-2\phi}=\frac{\text{Vol}(M)}{2\pi\ls^6g^2_{\rm s}}
\ee

 On the other hand, the superpotential \cite{beckersup,bpsaction} has the standard Gukov-Vafa-Witten \cite{gvw} form\footnote{The overall factor in (\ref{gvwsup}) has been fixed by reproducing (\ref{kaehler}) and (\ref{gvwsup}) following the approach of \cite{tentofour}, which combines the domain-wall arguments, analogous to the ones originally used in \cite{gvw},  and the use of superconformal supergravity in four dimensions.}
\be\label{gvwsup}
\calw=\frac{i\,M^3_P}{8\pi\, \ls^5}\int_M \Omega\wedge (H-i\d J)\equiv \frac{i\,M^3_P}{8\pi\, \ls^5}\int_M \Omega\wedge G
\ee 
We can then use these expressions to compute the gravitino mass from the standard four-dimensional supergravity formula
\be\label{4dgravmassform}
m_{3/2}=\frac{1}{M^2_{\rm P}}\,e^{\calk/2}\calw
\ee
In order to do this it is enough to compute $\calw$ and $\calk$ by using $\Omega$ and $J$ restricted by the conditions provided in section \ref{sec:torsusyb},\footnote{Notice that, in fact, the $(3,0)$-form $\Omega$ appearing in (\ref{kaehler}) and (\ref{gvwsup}) has no fixed normalization and only matches the $\Omega$ used in the rest of the paper (normalized as $\Omega\wedge\bar\Omega=i8\text{vol}_M$) up to a overall constant. Such a change of normalization corresponds to a K\"ahler transformation in the four-dimensional theory and thus does not affect physical quantities like $|m_{3/2}|$, as it is clear from (\ref{4dgravmassform}), (\ref{kaehler}) and (\ref{gvwsup}).} approximating dilaton and $W_1$ to constants. The result is indeed in agreement with  (\ref{gmasstor}). 

In order to show the no-scale structure \cite{noscale}, of $\frac12$DWSB vacua let us  introduce an expansion of $J=J_\calb+j$ as follows
\be
J_\calb=l^2_{\rm s}(\Re t^a)\,\omega_a\, ,\quad j=l^2_{\rm s}(\Re u)\, j^{(0)}
\ee
where $\omega_a$ is some basis of two-forms on the base $\calb$ and $j^{(0)}$ represents a fixed reference two-form orthogonal to the base. Then, $t^a$ and $u$ may be considered as pseudo-K\"ahler moduli of the base and the fiber respectively, complexified into chiral four-dimensional fields by the coefficients appearing in the expansion of the internal two-form $B$ in the same basis. Analogously, the chiral field $s$ appearing in (\ref{kaehler}) must be considered as the chiral field obtained by complexifying $\Re s$ by the axion dual to the external $B_{\mu\nu}$. By assuming off-shell the condition $\d\Omega\wedge J_\calb=0$ which follows from (\ref{splitsb}), it is easy to be convinced that the superpotential (\ref{gvwsup}) depends only on the fiber pseudo-K\"ahler modus $u$ and  pseudo-complex structure moduli $z^i$ encoded in $\Omega$:
\be\label{supnoscale}
\frac{\partial \calw}{\partial s}\equiv 0\, ,\quad \frac{\partial \calw}{\partial t^a}\equiv 0
\ee
On the other hand, we have that
\bea\label{kpot}
\calk&=&-\log (\Re s)-\log [ h_{ab}\,(\Re t^a)\, (\Re t^b )]-\log (\Re u)-\log\Big(-\frac{i}{8\,l^6_{\rm s}}\int_M\Omega\wedge \bar\Omega\Big)
\eea
where
\be
h_{ab}=\frac12\int_M \omega_a\wedge \omega_b\wedge j^{(0)}
\ee
Introducing common indices $\alpha,\beta,\ldots$ for $(s,t^a)$, it is easy to see that
\be\label{norm3}
\calk^{\alpha\beta}\partial_\alpha\calk\partial_\beta\calk=3
\ee
where $\calk^{\alpha\beta}$ is the inverse of the matrix $\partial_\alpha\partial_\beta\calk$. The conditions (\ref{supnoscale}) and (\ref{norm3}) are typical of no-scale models and indeed are sufficient to give a semi-positive  definite potential
 \be\label{4dpotnoscale}
 V_{\text{4D-SUGRA}}\equiv e^\calk \calk^{I\bar J}D_I\calw D_{\bar J}\bar\calw+(\text{D-term})^2\geq 0
 \ee
where $D_I\calw=\partial_I\calw+(\partial_I \calk)\calw$, $\calk^{I\bar J}$ is the inverse of $\partial_I\partial_{\bar J}\calk$, and $I,J,\ldots$ are indices  collectively denoting   the chiral fields $(u,z^i)$. 

In order to extremize the potential, one needs to impose
\be\label{Fflatcondition}
D_u\calw=0\, ,\quad D_i\calw =0\, ,\quad (\text{D-term})=0
\ee
Explicitly, we have the expressions
\bseq\label{Fterm}
\begin{align}
D_u\calw&\propto \frac{1}{u+\bar u}\int_M\Omega\wedge \bar G\label{Fterm1}\\
D_i\calw&\propto \int_M\chi_i\wedge G\label{Fterm2}
\end{align}
\eseq
where
\be
\chi_i=\partial_i\Omega-\Omega\,\frac{\int_M\partial_i\Omega\wedge \bar\Omega}{\int_M\Omega\wedge \bar\Omega}
\ee
should be the basis of $(2,1)$ forms relevant for the four-dimensional description. Then, assuming that the truncated theory makes sense, imposing (\ref{Fflatcondition}) is equivalent to the conditions (\ref{Fflat}), which were obtained from our previous ten-dimensional analysis. We can then interpret (\ref{Fflat}) as the above 4d F-flatness conditions. In addition the F-terms that do not enter the scalar potential $D_s\calw$ and $D_a\calw$, are non-vanishing whenever $G^{0,3}\neq 0$ (or equivalently $W_1\neq 0$), again in agreement with the ten-dimensional analysis. The only remaining ten-dimensional condition is (\ref{sbcond3}), which becomes $J\wedge \d J=0$ in the constant dilaton approximatio, and which can be interpreted as the 4d D-flatness condition.

Finally, let us also briefly consider the gauge bundle sector. By a simple dimensional reduction of the ten dimensional action (\ref{10daction}), it is easy to see that the kinetic term for the four-dimensional gauge field is given by
\be
-\frac14\,\Re s\, \tr F^{\mu\nu}F_{\mu\nu}
\ee
Then, the holomorphic gauge coupling is given by $f(s)=s$ and from the standard formula for the gaugino mass we get
\be
m^{\rm 4D}_{1/2}=-\frac1{M^2_{\rm P}}e^{\calk/2}\,\calk^{s\bar s}D_s\calw\partial_{\bar s}\log(\Re f)=\frac{1}{M^2_{\rm P}}\,e^{\calk/2}\calw\equiv m_{3/2}
\ee
This is indeed in agreement with (\ref{mm}), which was obtained directly by dimensionally reducing the fermionic ten-dimensional action.

 
 \section{Examples via homogeneous fibrations}
\label{sec:example}

In order to illustrate the general features of $\oh$DWSB vacua, let us discuss a concrete setting in which explicit examples can be constructed. Recall that the  $\oh$DWSB ansatz implies that the compactification manifold $M$ is based on a fibration of the form
\be\label{fibration2}
N\  \hookrightarrow\  M\ \xrightarrow{\pi}\ \calb
\ee
with a two-dimensional fiber $N$ and a four-dimensional base $\calb$. It is thus natural to simplify this geometry by assuming that all geometric quantities are only base-dependent. In other words, we assume that $\phi$, $W_1$ and the forms $\Omega_\calb$ and $J_\calb$ in (\ref{JOmega1par}) can be seen as functions and forms on the base $\calb$,\footnote{Stated more precisely, we assume that $\phi$, $W_1$, $\Omega_\calb$, $J_\calb$ and $F$ can be obtained as the pull-back of corresponding functions and forms by the fibration map $\pi:M\rightarrow \calb$.} which in turn implies that $\d\phi,\ \d W_1, \d \Omega_\calb$ and $\d J_\calb$ can also be seen as forms on the base $\calb$. 

This simplifying assumption has several consequences. For instance, by (\ref{splitsb}) one can see that the pull-back of $\d\Theta$ to any fiber $N$ vanishes, $\d\Theta|_N=0$. This means that the pulled-back hermitian metric $g|_N$ on $N$ is flat, and so we are led to take a two-torus as a fiber
\be
N\simeq T^2
\ee
and so $M$ elliptically fibered. Starting with \cite{drs}, such elliptically fibered manifolds have played a key role in the literature of torsional heterotic backgrounds and in particular in those constructions motivated by duality arguments -- see e.g.\  \cite{elliptic,bs,bsb}. 
Here we see the elliptic fibration arises from imposing a rather simple pattern of torsional supersymmetry breaking. In the following we will analyze which further constraints such pattern imposes on $M$.

\subsection{Constraints on the elliptic fibration}

Since our fiber is a two-torus, the one-form $\Theta$, introduced in (\ref{JOmega1par}), takes the form
\be
\Theta=\frac{\ell_{\rm s}\, L_{T^2}}{\sqrt{\Im \tau}}\, e^C\,\theta  \quad \quad \quad  \theta=\eta^1-\tau\,\eta^2
\ee
where $\ls=2\pi\sqrt{\alpha^\prime}$ and  $\eta^a$ ($a=1,2$) are one-forms which can be locally written as
\be\label{defeta}
\eta^a=\d x^a+A^a(y)
\ee
with $x^a\simeq x^a+1$ dimensionless coordinates along the $T^2$-fiber and $A^a(y)$ one-forms along $\calb$ that only depend on the base coordinates $y^\alpha$, $\alpha=1,\ldots,4$. Clearly $\tau$ is the complex structure of the $T^2$ fiber, $L_{T^2}$ is the dimensionless $T^2$ length scale in string units and $\langle e^C\rangle \lesssim 1$ encodes the non-trivial dependence on the base coordinates, so that the  volume of the fiber is given by $\text{Vol}(T^2)=e^{2C}\, \ls^2\,L^2_{T^2}$. 

The one-forms $A^a(y)$ can be seen locally as U(1) gauge fields on $\calb$, while globally they can be further twisted by SL$(2,\mathbb{Z})$ transformations, the large diffeomorphism group of $T^2$, if the $T^2$-fibration degenerates at some points. The same applies to the associated U(1) field strengths $\omega^a=\d A^a$, which must obey an SL$(2,\mathbb{Z})$-twisted quantization condition and so define SL$(2,\mathbb{Z})$-twisted cohomology classes in $\calb$.  Note however, that $\omega^a$ are cohomologically trivial in the ambient manifold $M$, since $\d\eta^a=\omega^a$.  

Let us now see how the background quantities $J$ and $\Omega$, decomposed as in (\ref{JOmega1par}), are constrained by our $\oh$DWSB ansatz. Taking into account the eqs. (\ref{splitsb}) and (\ref{sbcond0}), one arrives to the conditions\footnote{To make our conventions compatible with those usually adopted in the literature, we take the choices
\be\nonumber
J_\calb=-e^1\wedge e^2-e^3\wedge e^4\quad\quad j=e^5\wedge e^6\quad\quad \Omega_\calb=(e^1+ie^2)\wedge (e^3+ie^4)\quad\quad\Theta=e^5-ie^6 
\ee
 where  $e^1,\ldots, e^6$ is an {\em oriented} orthonormal coframe on $M$. Indeed, note that with these choices  $J_\calb$ and $\Omega_B$ are self-dual under Hodge duality on the base: $*_\calb J_\calb= J_\calb,\ *_\calb \Omega_\calb= \Omega_\calb$:}
\bseq\label{k3susyb}
\begin{align}
\d\Big(\frac{e^{C-2\phi}}{\sqrt{\Im\tau}}\,\Omega_\calb\Big)&=0\label{k3susyb1}\\
\d(e^{2C-2\phi} J_\calb)&=0\label{k3susyb2}\\
J_\calb\wedge \chi&=0\label{k3susyb3}\\
\bar\partial\tau&=0
\end{align}
\eseq
where we have introduced the complex two-form
\be
\chi=\omega^1-\tau\omega^2=\d\theta+\d\tau\wedge \eta^2
\ee
From (\ref{k3susyb1}) and (\ref{k3susyb2}) we see that $\calb$ not only admits an integrable complex structure, but also a  K\"ahler structure, with holomorphic  (2,0)-form and K\"ahler form given by
\be
\hat\Omega_\calb=\frac{e^{2D-C}}{\sqrt{\Im\tau}}\,\Omega_\calb\, \quad \quad \hat J_\calb = e^{2D} J_\calb
\ee
and where 
\be\label{ed}
e^{D}=g_{\rm s}\,e^{C-\phi}
\ee
with $g_{\rm s}$  defined in (\ref{gs}). 
It is then natural to write  the internal metric as
\be
\d s^2_M=e^{-2D}\d\hat s^2_\calb+\ls^2\, L^2_{T^2}\,\frac{e^{2C}}{\Im\tau}\,\theta\otimes\bar\theta
\ee
where $\d\hat s_\calb^2$ is the K\"ahler metric. Note however that
\be\label{exnormcond}
\hat J_\calb\wedge \hat J_\calb=\frac12\,e^{2C}\Im \tau\,\hat\Omega_\calb\wedge\overline{\hat\Omega}_\calb
\ee
so that the metric $\d\hat s^2_\calb$ is Calabi-Yau only if $e^{2C}\Im\tau$ is constant. In addition, by imposing (\ref{sbcond2}) one obtains the following expression for the three-form $H$
\be\label{Hex}
H=\hat*_\calb\d\, e^{-2D}-\ls^2\, L^2_{T^2}\big[(\d^c e^{2C})\wedge \eta^1\wedge \eta^2+\frac{e^{2C}}{\Im \tau}\,\Re\big(\hat*_\calb\chi\wedge\bar\theta)\big]
\ee
where $\d^c:=i(\partial-\bar\partial)$. This three-form flux should satisfy the BI (\ref{BIfin}) and satisfy appropriate quantization conditions, see below.

Note that the above constraints would also apply to any supersymmetric background based on an homogeneous elliptic fibration. Indeed, the fact that our background breaks supersymmetry is purely encoded in the condition
\be\label{exdwsb}
\d(e^{-2\phi}\Omega)=g^{-2}_{\rm s}\,\ls\, L_{T^2}\,\hat\Omega_\calb\wedge\chi\neq 0
\ee
and so in order to break supersymmetry in this way one must require that  $\chi$ have non-vanishing (0,2) {\em and} (2,0) components.\footnote{Having $\chi^{0,2}\neq 0$ but $\chi^{2,0}=0$ is not sufficient, since we can change the orientation of $T^2$, basically swapping $\chi$ and $\bar\chi$, and thus getting an $\caln=1$ supersymmetry. Having $\chi^{0,2}=\chi^{2,0}=0$ means that both orientations on $T^2$ lead to preserved supersymmetry and thus we have an $\caln=2$ compactification.} Of course, by relaxing the $\oh$DWSB ansatz further ways of breaking supersymmetry arise. For instance, we see from (\ref{k3susyb3}) that $\chi$ must be primitive and, from section \ref{sec:4dint}, that a non-primitive $\chi$ corresponds to a non-vanishing D-term.

Recall now that for this class of $\caln = 0$ vacua one needs to impose the residual condition (\ref{eomtor1}) coming from the equation of motions. For our elliptic fibration, this is equivalent to require that $\partial(e^{-2D}W_1)=0$, which is solved by
\be\label{torex}
W_1=c_{\rm SB}\, e^{2D}
\ee
where $c_{\rm SB}$ is a constant which parameterizes the amount of SUSY-breaking. 

\bigskip

{\em Gravitino mass}

\smallskip

\noindent The parameter $c_{\rm SB}$ should directly enter physical quantities that measure the amount of SUSY-breaking of a compactification like, e.g., the gravitino mass. Indeed, following our discussion of section \ref{sec:gravitinomass}, let us first consider the gravitino mass density $\mathfrak{m}_{3/2}$. By comparing (\ref{torex}) and (\ref{warptor}) it reads
\be\label{gmex}
\mathfrak{m}_{3/2}=3\, e^A\, c_{\rm SB}\, e^{2D} 
\ee
Since $e^A$ is constant, we see that the SUSY-breaking is milder in the points of $\calb$ with strong conformal factor  $e^{-2D}\gg 1$. A rough estimate of the gravitino mass is obtained by the  approximation $e^C\simeq 1$ and $e^D, \tau$ constant. Then, by using (\ref{torex}), (\ref{exdwsb}) and (\ref{splitsb}) we find
\be\label{csb}
c_{\rm SB}\,\simeq\, \frac{2\ls\, L_{T^2}}{3\, \Im\tau}\times\frac{\int_\calb\hat\Omega_\calb\wedge \chi}{\int_\calb \hat\Omega_\calb\wedge \overline{\hat\Omega}_\calb}
\ee
and taking into account (\ref{gmex}), (\ref{gravmass}) and (\ref{warp_planck}), one finally gets
\be\label{apprgm}
m_{3/2}\simeq \frac{g_{\rm s}\, M_{\rm P}\, e^{4D}\| \chi^{0,2}\| }{2\,L^4_\calb\, \sqrt{\pi\,\Im\tau}}
\ee
where 
\be\label{basescale}
L^4_\calb\equiv \ls^{-4}\,\hat{\rm Vol}(\calb)
\ee
and we have introduced the quantity
\be
\|\chi^{0,2}\|:=\frac{\int_\calb\hat\Omega_\calb\wedge \chi}{\Big(\int_\calb \hat\Omega_\calb\wedge \overline{\hat\Omega}_\calb\Big)^{1/2}}
\ee
that measures the alignment of $\chi$ with $\overline{\hat \Omega}_\calb$. While generalizing  to  the above expression for non-trivial profiles of $e^\phi$, $e^C$, $e^D$ and $\tau$ is straightforward, (\ref{apprgm}) already captures most of the qualitative behavior of $m_{3/2}$ and is sufficient for the purposes of our discussion.

From (\ref{apprgm}) we see that one can suppress the SUSY-breaking scale by combining the following (possibly non-independent) conditions: $g_{\rm s}\ll 1$, $\|\chi^{0,2}\|\ll 1$, $L^4_{\calb}\gg 1$, $e^{D}\ll 1$ and $\Im\tau\gg~1$. Recall that we are implicitly assuming that $\chi^{2,0}\neq 0$, which already selects an $\caln=1$ supersymmetry in four-dimensions, eventually broken by the non-vanishing $\chi^{0,2}$. The expression (\ref{apprgm}) then refers to the gravitino which is selected by the flux $\chi^{2,0}$.

\bigskip

{\em Bianchi identity and tadpole conditions}

\smallskip

\noindent We would like now to impose the Bianchi Identity (\ref{BIfin}). From (\ref{Hex}) we obtain
\bea\label{dHex}
\d H&=&\d\hat*_\calb\d\, e^{-2D}+\ls^2\, L^2_{T^2}\Big\{(\d^c\d\, e^{2C})\wedge \eta^1\wedge \eta^2- \frac{e^{2C}}{\Im \tau}\hat*_\calb\chi\wedge \bar\chi\cr
&&-\Re\Big[ \d\Big(\frac{e^{2C}}{\Im\tau}\hat*_\calb\chi\Big)\wedge \bar\theta+ \frac{e^{2C}}{\Im\tau}\,\hat*_\calb\chi\wedge (\bar\partial\bar\tau)\wedge \eta^2\Big]+\frac{\d^c e^{2C}\wedge \Im(\chi\wedge \bar\theta)}{\Im\tau}\,\Big\}
\eea
From this expression it is clear that satisfying the BI (\ref{BIfin}) is rather complicated if we have sources along the $T^2$-fiber like, e.g., gauge bundles with $\tr (F\wedge F)$ dual to a  two-cycle in $\calb$, or  NS5-branes wrapping a two-cycle in $\calb$ and smeared along the $T^2$-fiber. For instance, an NS5-brane wrapped on a holomorphic curve $\Pi_2\subset \calb$ and smeared along $T^2$ contributes to the r.h.s.\ of (\ref{BIfin}) as $\delta^2_{\calb}(\Pi_2)\wedge \eta^1\wedge \eta^2$, and this can be compensated by (\ref{dHex}) only if $\d^c\d\,  e^{2C}\sim \delta^2_{\calb}(\Pi_2)$. This is solved by taking $e^{2C}\sim \log |f_{\Pi_2}|$, with $f_{\Pi_2}$ the section of the divisor bundle $\call_{\Pi_2}$ on $\calb$. Since $f_{\Pi_2}$ vanishes on $\Pi_2$, the $T^2$ volume degenerates on $\Pi_2$.

On the other hand, the problem simplifies somewhat drastically if we assume the absence of such bundles and NS5-branes and consistently assume that the $T^2$ fibration has constant volume, so that $e^C$ is constant along $\calb$. In particular we set 
\be
e^C\equiv 1\quad\quad\Leftrightarrow \quad\quad e^D\,=\,g_{\rm s}\, e^{-\phi}
\ee without loss of generality.  In order to solve the BI (\ref{BIfin}), we then need to take the gauge bundle $F$ on $M$ to be the pull-back of a gauge bundle $F_\calb$ on $\calb$, and $N_{\rm NS5}$ NS5-branes wrapping the $T^2$-fiber and sitting at points $p_i$ of the base. In this case,  (\ref{almostHYM}) reduces to the self-duality condition
\be\label{bundbase}
*_{\calb} F_\calb=-F_\calb
\ee
which is equivalent to $F$ being $(1,1)$ and primitive.

By (\ref{apprgm}), we can suppress the SUSY-breaking scale by considering an anisotropic fibration, with base much larger then the fiber
\be
L_\calb\gg L_{T^2}
\ee
with $L_\calb$ given by (\ref{basescale}). Moreover, by the discussion of section \ref{sec:higherorder}, such anisotropic fibration should simplify the conditions on the curvature (\ref{curvHYM}) as discussed below. Finally, it also simplifies the BI \cite{bs}. Indeed, expanding the curvature tensor $R_+$ in powers\footnote{Note that $R_+$ is invariant under an overall rescaling of the six-dimensional metric and is constructed from the torsionfull connection $\Gamma^m_{+np}=\Gamma^m{}_{nm}+\frac12 H^m{}_{np}$, which depends quadratically on $L_\calb$ and $L_{T^2}$. Hence, it only depends on even powers of $L_\calb/L_{T^2}$.} of $(L_\calb/L_{T^2})^2$ we obtain the behaviour 
\be\label{R^2hor}
\tr R_+\wedge R_+=\tr R_{\calb+}\wedge R_{\calb+}+\calo\left(\frac{L^2_{T^2}}{L^2_\calb}\right)
\ee
where $R_{\calb+}$ is the torsionfull curvature of the base $\calb$ computed using the four-dimensional metric $e^{-2D}\d \hat s^2_\calb$ and  $H_\calb=\hat*_\calb\d e^{-2D}$. Taking $L^2_{T^2}\sim \calo(10)$, as in the explicit examples of section \ref{sec:K3ex}, $\tr R_+\wedge R_+$ has only legs along the base up to $\calo(L^{-2}_\calb)$ -- i.e.\ $\calo(\alpha^\prime)$ -- corrections and from (\ref{dHex}) and the BI (\ref{BIfin}) we get
\be\label{hodgediff}
\d\left(\frac{\hat*_\calb \Re\chi}{\Im\tau}\right)=0\, ,\quad \d\left(\frac{\hat*_\calb \Re(\bar\tau\, \chi)}{\Im\tau}\right)=0
\ee
In addition, we have that 
\be
\tr R_{\calb+}\wedge R_{\calb+}=\tr \hat R_\calb\wedge \hat R_\calb\,+2\d\hat*_\calb\d\hat\nabla^2D+\d\hat *_\calb
\Big[2(\hat\nabla^2e^{-2D})\d e^{2D}+\d\big( e^{2A}\hat\nabla^2 e^{-2D}\big)\Big]
\ee
where all `hatted' quantities are computed using the metric $\d\hat s^2_\calb$.
Then, by setting $\d\hat s^2_\calb=\ls^2\,L^2_\calb\, \d\hat s^2_{{}^{(0)}}$ (with $\d\hat s^2_{{}^{(0)}}$ dimensionless and $\calo(1)$) we get
\bea\label{inteqphi}
\d\hat*_{{}^{(0)}}\d \,e^{-2D}&=&\frac{1}{16\pi^2\, L^2_\calb}\big[\tr (\hat R_\calb\wedge \hat R_\calb)-\tr (F_\calb\wedge F_\calb)\big]+\frac{1}{L^2_\calb}\sum_i\delta_\calb^4(p_i)\cr
&&+\frac{1}{16\pi^2\, L^2_\calb}\d \Delta+\frac{ L^2_{T^2}}{L^2_\calb\Im \tau}\,\hat*_{{}^{(0)}}\chi\wedge \bar\chi+\calo\left(\frac{L^2_{T^2}}{L^4_\calb}\right)
\eea
where
\be
\Delta:=2\hat*_{{}^{(0)}}\d\hat\nabla_{\!\!{}^{(0)}}^2D+\hat *_{{}^{(0)}}
\Big[2(\hat\nabla_{\!\!{}^{(0)}}^2e^{-2D})\d e^{2D}+\d\big( e^{2D}\hat\nabla_{\!\!{}^{(0)}}^2 e^{-2D}\big)\Big]
\ee

Clearly, this equation admits a solution (up to higher order corrections in $1/L_{\calb}$) only if its integrated counterpart 
\be\label{tadcalb}
N_{\rm NS5}+Q_{\rm NS5}(E)+L^2_{T^2}\int \frac{\hat*_\calb\chi\wedge \bar\chi}{\Im\tau}=Q_{\rm NS5}(\calb)
\ee
is also satisfied. Here
\be\label{pontrj}
Q_{\rm NS5}(E)=-\frac{1}{16\pi^2}\int\tr(F_\calb\wedge F_\calb)\, ,\quad Q_{\rm NS5}(\calb)=-\frac{1}{16\pi^2}\int\tr(\hat{R}_\calb\wedge \hat{R}_\calb)
\ee
give the total NS5-brane charge sourced by the gauge bundle and the curvature of the base.

Absence of anti-NS5-branes implies that $N_{\rm NS5}$ is always positive, and from (\ref{bundbase}) the same applies to $Q_{\rm NS5}(E)$. The l.h.s.\ of (\ref{tadcalb}) is then always positive, and this implies an upper bound for the number of NS5-branes and non-trivial gauge bundle that can be introduced for a fixed manifold $\calb$. 


Once the condition (\ref{tadcalb}) is satisfied, one can integrate (\ref{inteqphi}) perturbatively. More precisely, along the lines of the $\caln=2$ case discussed in \cite{bsb}, one can write (\ref{inteqphi})  in terms of the shifted conformal factor
\be
e^{-2D^\prime}=e^{-2D}-\frac{1}{8\pi^2\, L^2_\calb}\, \hat\nabla_{\!\!{}^{(0)}}^2D
\ee 
In this way (\ref{inteqphi}) takes the form of a  standard Poisson equation 
\bea\label{inteqphi2}
\d\hat*_{{}^{(0)}}\d \,e^{-2D^\prime}&=&\frac{1}{16\pi^2\, L^2_\calb}\big[\tr (\hat R_\calb\wedge \hat R_\calb)-\tr (F_\calb\wedge F_\calb)\big]+\frac{1}{L^2_\calb}\sum_i\delta_\calb^4(p_i)\cr &&+\frac{ L^2_{T^2}}{L^2_\calb\Im \tau}\,\hat*_{{}^{(0)}}\chi\wedge \bar\chi+\calo\left(\frac{L^2_{T^2}}{L^4_\calb}\right)
\eea
where on the r.h.s. of (\ref{inteqphi2}) we have omitted terms like
\be\label{cordelta}
\frac{1}{16\pi^2\, L^2_\calb}\d\hat *_{{}^{(0)}} \Big[2(\hat\nabla_{\!\!{}^{(0)}}^2e^{-2D^\prime})\d e^{2D^\prime}+\d\big( e^{2D^\prime}\hat\nabla_{\!\!{}^{(0)}}^2 e^{-2D^\prime}\big)\Big]
\ee
If (\ref{tadcalb}) is fulfilled, (\ref{inteqphi2}) can be always integrated. Hence, the possible corrections provided by (\ref{cordelta}) are of order $\calo(L^2_{T^2}/L^4_\calb)$ and can be consistently neglected.\footnote{\label{NS5breaking} Here we are implicitly ignoring the fact that, in the vicinity of NS5-branes, $\d\hat*_{{}^{(0)}}\d \,e^{-2D}$ diverges  and the tree-level supergravity approximation breaks down. However, the SUSY-breaking effects are at the $L_\calb$ scale and very close to the NS5-brane supersymmetry is restored. Thus, we expect that NS5-brane sources can be consistently incorporated.}

\bigskip

{\em The K3 case and $H$-flux quantization}

\smallskip

\noindent Let us now consider the case of constant $\tau$ in more detail. Recall that then (\ref{exnormcond}) implies that the base should be a Calabi-Yau two-fold: $\calb\equiv$ K3. Furthermore, since $\tau$ does not degenerate, the one-forms $A^a(y)$ can be seen as proper U(1) gauge fields along K3  and then the corresponding field-strengths $\omega^a=\d A^a$ are quantized as
\be
\int_{\Pi_2\subset {\rm K3}} \!\!\!\!\!\!\omega^a\ \in \mathbb{Z}
\ee 
and so the forms $\omega^a$ define  non-trivial  elements of the integral cohomology group $H^2(\text{K3},\mathbb{Z})$. In fact, from (\ref{hodgediff}) we have that $\d(\hat*_{\rm K3} \chi)=0$, and so $\chi$ must be harmonic. Finally, in order to evaluate (\ref{tadcalb}) one has to use $Q_{\rm NS5}({\rm K3})=-\frac12 p_1({\rm K3})=24$.\footnote{\label{fn:pontrj} Note that in our conventions $p_1(\calb)=\frac{1}{8\pi^2}\int\tr(\hat{R}_\calb\wedge \hat{R}_\calb)$, which apparently differs by an overall sign from the standard definition of Pontjagin classes, since we use a positive-definite $\tr=-\tr_{\rm standard}$.}

We are thus led to the setting of non-degenerate $T^2$ fibrations over K3
\be\label{T2fibr}
T^2\quad \hookrightarrow \quad M\quad\rightarrow\quad  {\rm K3} 
\ee
which is often considered in the construction of heterotic torsional backgrounds \cite{drs,elliptic,bs,bsb}. Note in particular that for this case the SUSY-breaking conditions discussed below (\ref{exdwsb}) reduce to those identified in \cite{bsb} by direct inspection of the Killing spinor equations
and of the $\calo(\alpha^{\prime0})$ equations of motion.

The K3 example allows to discuss the quantization of the $H$-flux in a rather simple way. Indeed, in  general this is a complicated problem partly because of the non-closure of $H$ due to the contributions on the r.h.s.\ (\ref{BIfin}). However, in the simplified setting (\ref{T2fibr}), the $H$-field (\ref{Hex}) reduces to
\be
H=\hat*_{\rm K3}\d\, e^{-2D}-\,\frac{\ls^2\, L^2_{T^2}}{\Im\tau}\,\Re\big(\hat*_{\rm K3}\chi\wedge\bar\theta)
\ee
The flux $H$ can then be written as $H=\pi^*(H_{\rm K3})+\pi^*(h_a)\wedge \eta^a$, where $H_{\rm K3}$ and $h_a$ are forms on K3 and $\pi^*$ is the pull-back operation induced by the projector $\pi:M\rightarrow {\rm K3}$. In particular we have that
\be
h_1=-\,\frac{\ls^2\, L^2_{T^2}}{\Im \tau}\,\Re\big(\hat*_{\rm K3}\chi)\,\quad h_2=\,\frac{\ls^2\, L^2_{T^2}}{\Im \tau}\,\Re\big(\bar\tau\hat*_{\rm K3}\chi)
\ee
are also harmonic forms. Because of (\ref{R^2hor}), this suggests that the proper quantization condition to impose is that both two-forms $\ls^{-2}\,h_a$ must be harmonic representatives of integral cohomology classes in $H^2({\rm K3};\mathbb{Z})$.\footnote{Indeed, $h_a$ could be seen as U(1) field-strengths in the eight-dimensional theory obtained by compactifying the ten-dimensional theory on $T^2$, and are thus appropriately quantized.}  
More precisely, we get the following condition on $\chi$
\be\label{Hquantcond}
\frac{L^2_{T^2}}{\Im \tau}\int_{\Pi_2}\Re\big(\hat*_{\rm K3}\chi)\in \mathbb{Z}\quad\quad\quad \frac{L^2_{T^2}}{\Im \tau}\int_{\Pi_2}\Re\big(\bar\tau\,\hat*_{\rm K3}\chi)\in\mathbb{Z} 
\ee
for any two-cycle $\Pi_2\subset$K3.

\bigskip

{\em Curvature corrections}

\smallskip

\noindent Let us finally consider the $R_+$-dependent terms in the scalar potential (\ref{potord1}) for this simplified K3 case, in order to illustrate our general discussion of section \ref{sec:higherorder}.  First of all, by using (\ref{csb}) and taking into account that $\hat\Omega_{\rm K3}$ scales as $L^2_{\rm K3}/\sqrt{\Im\tau}$, we see that $c_{\rm SB}$ scales as $L_{T^2}/L^2_{\rm K3}$ (assuming an approximately square $T^2$-fiber). In the examples of section \ref{sec:K3ex} we will see how $L_{T^2}\sim \calo(1 $-$ 10)$. Furthermore, being the $T^2$ fiber flat, the leading contribution to the curvature has its origin in K3. The dimensionless length scale $L_{\rm KK}$ introduced in section \ref{sec:higherorder} can then be identified with $e^{-D}L_{\rm K3}$ and, by comparing (\ref{torex}) and (\ref{torscale}), we see that  $L_{\rm SB}\simeq L^2_{\rm KK}$ and hence $\beta=1/2$ in (\ref{sbkk}).  

Moreover, an explicit calculation shows that the curvature terms in (\ref{curv2tor}) lead indeed to a behavior as in (\ref{curv2tor_est})
\bseq\label{curv2tor_est_ex}
\begin{align}
 |J_{mn}R^{mn}_-|^2&\sim\, e^{8D}|c_{\rm SB}|^4\\
|\Omega_{kmn}R^{mn}_-|^2 &\sim\, e^{6D}|c_{\rm SB}|^2\,|\d D|^2_{\rm K3}\,+\,e^{8D}\, |c_{\rm SB}|^2\,  \nonumber\\
&\sim\, e^{6D}\,L^{-6}_{\rm K3}\,+\,e^{8D}\, L^{-8}_{\rm K3}
\end{align}
\eseq
Then, by (\ref{firstordcorr}) we see that the $\calo(\alpha^\prime)$ correction to the equations of motion goes like
\bea
(\text{EoM})_{\calo(\alpha^\prime)}&\sim& e^{4D}|c_{\rm SB}|^2\, L^{2}_{\rm K3}\big( \,|\d D|^2_{\rm K3}\,+\,e^{2D}\, |c_{\rm SB}|^2  \,\big)\cr
&\sim& e^{4D}\,L^{-4}_{\rm K3}\,+\,e^{6D}\, L^{-6}_{\rm K3}
\eea
where, in the last step, we have used the above estimate for $c_{\rm SB}$ and the fact that from  (\ref{inteqphi}) we can  assume that $|\d D|_{\rm K3}^2\sim L^{-2}_{\rm K3}$. This already confirms the estimate made in section \ref{sec:higherorder} that the contribution of the first-order potential goes like $L^{-4}_{\rm KK}$ and is thus $\calo(\alpha^{\prime 2})$.  Actually, one could have $D$ approximately constant, e.g.\  $|\d D|^2\lesssim L^{-4}_{\rm K3}$, in most of the internal space, in which the correction would be even of $\calo(\alpha^{\prime 3})$. Note that  this approximation would certainly break down in the neighborhood  of possible NS5-branes wrapping the fiber. However, it is also true that our  SUSY-breaking effects are set by the lowest KK scale of the K3 base, and should then be harmless for the consistency of the localized NS5-branes against higher-order corrections -- cf.\ footnote \ref{NS5breaking}.

\subsection{Simple examples}
\label{sec:K3ex}

In order to obtain simple explicit $\caln=0$ examples of the setting provided by the $T^2$ fibration over the {\rm K3} space reviewed above, let us set $N_{\rm NS5}=0$ and take a trivial gauge bundle on  K3. Then, all we have to do is to pick up a primitive harmonic form $\chi$ which has both $(2,0)$ and $(0,2)$ non-vanishing components and such that the condition (\ref{tadcalb}) is satisfied. 

The well known properties of {\rm K3} -- see e.g.\ \cite{aspinwall} for a review --  greatly help in this search. $H^2({\rm K3};\mathbb{R})$ has dimension $b_2=22$ and, picking up a basis $\{e_I\}^{22}_{I=1}$, the inner product matrix
\be\label{metricharm}
\cali_{IJ}=\int_{\rm K3}e_I\wedge e_J
\ee 
has signature $(3,19)$.  In particular, one can choose an integral basis $\{\alpha_I\}^{22}_{I=1}$ of $H^2({\rm K3};\mathbb{Z})$ such that 
\be\label{metricharmcan}
\cali_{IJ}=\left(\begin{array}{ccc} 0 && 1 \\ 1 && 0 \end{array}\right)\oplus \left(\begin{array}{ccc} 0 && 1 \\ 1 && 0 \end{array}\right)\oplus\left(\begin{array}{ccc} 0 && 1 \\ 1 && 0 \end{array}\right)\oplus  (-E_8)\oplus  (-E_8)
\ee
where $E_8$ is the Cartan matrix of the $E_8$ algebra. 
$\Re\hat\Omega_{\rm K3}$, $\Im\hat\Omega_{\rm K3}$ and $\hat J_{\rm K3}$ provide a basis of the self-dual harmonic forms in $H^2({\rm K3};\mathbb{R})$, which is a space-like plane with respect to the metric (\ref{metricharm}). 

Instead of attempting a detailed general discussion of the constraints derived above, we will just provide a couple of simple examples, which should nevertheless give an idea of the qualitative features of more general solutions.  

First, let us take a simple choice for $\hat\Omega_{\rm K3}$:
\be
\hat\Omega_{\rm K3}=\frac{(2\pi)^2\alpha' L^2_{\rm K3}}{\sqrt{\text{Im}\tau}} (\alpha_1+i\alpha_2)
\ee
with $\alpha_1=e_1+e_2$ and $\alpha_2=e_3+e_4$, in terms of the integral basis $\{e_I\}^{22}_{I=1}$ with inner product matrix (\ref{metricharmcan}). Let us then define $\chi$ in terms of four integers $n^a$ and $m^a$, $a=1,2$, as follows
\be
\chi=(n^1-\tau m^1)\alpha_1+(n^2-\tau m^2)\alpha_2
\ee
Taking into account the self-duality of $\alpha_{1,2}$ and the fact that $\int_{\rm K3}\alpha^1\wedge \alpha^1=\int_{\rm K3}\alpha^2\wedge \alpha^2=2$, the condition (\ref{tadcalb})  [with $N_{\rm NS}=Q_{\rm NS5}(E)=0$] reduces to
\be\label{tauL}
24\,{\Im\tau}=L^2_{T^2}\Big(2|n^1-\tau m^1|^2+2|n^2-\tau m^2|^2\Big)
\ee
which, for fixed quantized  numbers $n^a$ and $m^a$, relates $\tau$ and $L^2_{\rm T^2}$. Furthermore, in order to have $\caln=0$ supersymmetry, one needs to impose that 
\be\label{SBcondex}
\frac{n^2-\tau m^2}{n^1-\tau m^1}\neq \pm i
\ee


\bigskip

{\em Example 1}

\smallskip

\noindent A particularly simple example is obtained by setting $n^1=1$ and $n^2=m^1=m^2=0$. In this case, we have only a non-trivial fibration of the $S^1$ described by the coordinate $x^1$ introduced  in (\ref{defeta}), while the second $S^1$ described by $x^2$ is trivially fibered. 
The condition (\ref{tauL}) then gives
\be\label{lengthcond}
L^2_{T^2}=12\,\Im\tau
\ee
while the $H$-field quantization conditions (\ref{Hquantcond}) reduce to
\be\label{Hquant1}
L^2_{T^2}=k_1\,\Im\tau\, ,\quad L^2_{T^2}\,\Re\tau=k_2\,\Im\tau
\ee
with $k_{1,2}\in\mathbb{Z}$, from which we see that 
\be
k_1\Re\tau=k_2
\ee

For example, let us take $\Re\tau=0$. Then, $\tau=iR_2/R_1$ and $L^2_{T^2}=R_1R_2/\alpha^\prime$  and the condition (\ref{Hquant1}) imposes that $k_1=12$ and $k_2=0$, while (\ref{lengthcond}) provides the following constraints 
\be\label{singlebound}
\frac{\alpha^\prime}{R^2_1}\simeq \frac1{12}\simeq 0,083
\ee 
Then $\frac{\alpha^\prime}{R^2_1}$ is relatively small, moderately justifying the supergravity approximation. On the other hand, $R_2$ is obviously non-constrained. Notice that the profile of $e^{D}$ is determined by (\ref{inteqphi}). Since we are assuming that $N_{\rm NS5}=0$, for $L_{\rm K3}\gg 1$ we can reasonably approximate $e^{D}\simeq 1$. Then,  the formula (\ref{apprgm})  for the gravitino mass gives
\be\label{gm11}
m_{3/2}\simeq \frac{g_{\rm s}M_{\rm P}}{2\,L^4_{\rm K3}}\sqrt{\frac{R_1}{\pi R_2}}
\ee

\bigskip

{\em Example 2}

\smallskip

\noindent To obtain a non-trivial fibration also of the second circle, let us take for example $n^1=m^2=1$, $m^1=n^2=0$ and $|\tau|=1$ (with $|\Re\tau|\leq 1/2$), so that $R_1=R_2$. Then  (\ref{tauL}) gives
\bea
L^2_{T^2}=6\,\Im\tau
\eea
while (\ref{Hquantcond}) still takes the form (\ref{Hquant1}). 

Clearly, the general solution of this form  is given by $k_1=6$ and $k_1=0,\pm1,\pm2,\pm3$. By setting $k_2=0$, we get $R^2_1=R^2_2=\alpha^\prime/6$. However, in this case  $\Re\tau=0$ and then $\chi^{0,2}=0$, which implies that the solution is actually $\caln=1$. In the other cases $k_2=\pm1,\pm2,\pm3$, one gets $\caln=0$ solutions. Let us take for example $k_2=1$. In this case
\be
\Re\tau=\frac{1}{6}\, ,\quad\Im\tau=\frac{\sqrt{35}}{6}
\ee
and then we have
\be
\frac{\alpha^\prime}{R^2_1}=\frac{\alpha^\prime}{R^2_2}=\frac1{\sqrt{35}}\simeq 0,17
\ee
which is even less relatively small than (\ref{singlebound}) and, then, even more moderately justifies the supergravity approximation.   
The gravitino mass  (\ref{apprgm}) is now
\be
m_{3/2}\simeq \frac{g_{\rm s}M_{\rm P}}{2\,\sqrt{\pi}\,L^4_{\rm K3}}\times \frac1{6}
\ee
where the suppressing factor with respect to (\ref{gm11}) with $R_1=R_2$  comes from $\|\chi^{0,2}\|\simeq1/6$.


\section{Adding a gaugino condensate}
\label{sec:gaugino}

Up to now we have focused on 4d $\caln=0$ Minkowski vacua where the SUSY-breaking mechanism is due to the torsional geometry of the background. However, in the context of no-scale heterotic string compactifications, the source of supersymmetry breaking has traditionally been identified with the presence of a gaugino condensate generated by non-perturbative effects \cite{din85,drsw85}. It is therefore natural to incorporate a gaugino condensate to the above class of constructions, in order to see which new patterns of supersymmetry breaking it may lead to. In fact, since a gaugino condensate will modify the 4d no-scale scalar potential, one may wonder whether its presence may restore supersymmetry and trigger the decay of the $\caln=0$ vacua discussed above to supersymmetric AdS$_4$ vacua.


\subsection{Gaugino condensate and no-scale SUSY-breaking}
\label{sec:potg}

A simple way to measure the effect of a gaugino condensate in a heterotic compactification is to incorporate the gaugino field up to quartic order into the supergravity action. In particular, one finds that the ten-dimensional string frame bosonic action is modified to
\be\label{10dactiong}
S=\frac{1}{2\kappa^2}\int\d^{10} x\sqrt{-g}\,e^{-2\phi}\big[{\cal R}+4(\d\phi)^2-\frac1{2} T^2+\frac{\alpha^\prime}4 \tr ( R_+^2-  F^2 - 2\bar{\chi}\slashed{D}\chi )\big]
\ee
 where $\chi$ is the 10d gaugino field and we have defined the three-forms
 \be
 T\, =\, H - \oh \Sigma\quad \quad \quad  \Sigma_{MNP}\, =\, \frac{\a'}{4} \tr \bar{\chi} \Gamma_{MNP} \chi
 \label{T}
 \ee

Let us now consider compactifications to four-dimensions, in which $\Sigma$ (and so $T$) has only internal legs: $\Sigma=\frac{1}{3!}\Sigma_{mnp}\d y^m\wedge\d y^n\wedge \d y^p$. As already pointed out in \cite{din85,drsw85,ccdl03} the presence of a non-trivial $\Sigma$ modifies the scalar potential of the compactification. Indeed, following the computations of section \ref{sec:pot}, we see that the potential (\ref{pot2}) is modified to
\be\label{potg}
V^\prime=V(H\rightarrow T) +\frac{\alpha^\prime}{4\kappa^2}\int \text{vol}_M\,e^{4A-2\phi}\bar{\chi}\slashed{D}_J\chi 
\ee
where $V$ is given by (\ref{pot2}) with $H$ substituted by $T$, and 
\be
\slashed{D}_J=\slashed{D}+\frac1{24} e^{-4A+2\phi}[*\d(e^{4A-2\phi}J)]_{mnp}\Gamma^{mnp}\
\ee
 In order to get a 4d Minkowski vacuum in this context, we again need to impose that $V^\prime$ is vanishing and extremized. By separately imposing that $V(H\rightarrow T)$ is extremized,  one is naturally lead to consider configurations of the kind discussed in subsection (\ref{sec:torsusyb}),  up to the replacement $H\rightarrow T$. Namely, one should impose
\be
e^{2\phi}\d(e^{-2\phi}J)=*T\label{sbcond2g}
\ee
instead of (\ref{sbcond2}), and leave $\d A=0$, (\ref{sbcond3}), (\ref{sbtor}) and (\ref{almostHYM}) unchanged. Furthermore the gaugino term
\be\label{gauginokin}
\int \text{vol}_M\,e^{4A-2\phi}\bar{\chi}\slashed{D}_J\chi 
\ee
must also be extremized. This leads to a set of conditions to be satisfied by $\chi$ and $\Sigma$.  

Note that even in the case in which $W_1=0$ supersymmetry is still broken by the gaugino condensate, as one can check by looking directly at the supersymmetry transformations 
\bseq\label{10dsusy2}
\begin{align}
\delta_\eps\psi_M&=\big(\nabla_M-\frac{1}{4}\slashed{T}_M\big)\epsilon-\frac1{16}\Gamma_M\slashed{\Sigma}\epsilon\, ,\label{gravsusy2}\\
\delta_\eps\lambda&= \big(\slashed{\partial}\phi-\frac{1}{2}\,\slashed{T}\big)\epsilon-\frac38\slashed{\Sigma}\epsilon \, ,\label{dilsusy2}\\
\delta_\eps\chi&= \frac12\,\slashed{F}\,\epsilon\, \label{gaugesusy2},
\end{align}
\eseq
In particular,  for compactifications to flat space and non-vanishing gaugino condensate, the external gravitino variation is always non-vanishing: $\delta\psi_\mu=-\frac1{16}\Gamma_\mu\slashed{\Sigma}\epsilon\neq 0$. One may then restore supersymmetry by considering compactifications to AdS$_4$, as analyzed in \cite{fl05}. Such kind of compactifications will be considered in the next subsection.

As discussed in section \ref{sec:calibr}, in the absence of gaugino condensate the background  condition (\ref{susycond2}) can be interpreted in terms of calibrations for gauge bundles and space-time filling NS5-branes. Remarkably, the modification of (\ref{susycond2}) into (\ref{sbcond2g}) is exactly the necessary one in order to preserve such calibration interpretation. This can be seen by going to the dual formulation, briefly reviewed in appendix \ref{app:dual}, where one uses the seven-form $\hat H$ instead of $H$ as fundamental field. Recall from section \ref{sec:calibr} that $\hat H$ is the flux which couples {\em electrically} to NS5-branes and hence the one that appears in the generalized calibration.  Now, as discussed in appendix \ref{app:dual}, in the presence of a gaugino condensate $\hat H$ and $H$ are related by $\hat H=e^{-2\phi}*_{10}T=*_{10}(H-\frac12\Sigma)$. We can then split $\hat H$ as
\be
\hat H=\text{vol}_{X_4}\wedge \tilde H= \text{vol}_{X_4}\wedge (e^{4A-2\phi}*T)
\ee  
where $\text{vol}_{X_4}$ is computed using the unwarped metric. Note that (\ref{sbcond2g}) then insures that $\tilde{H}$ is a closed (as it should be in absence of fundamental strings in the background) and even an exact three-form, and so we can write $\tilde H=\d \tilde B$, where $\tilde B$ is an internal potential two-form.

It is in fact illustrating to express the full potential  (\ref{potg}) in such dual formulation. Indeed, starting from the dual action given in (\ref{dualaction2}),  one arrives to the potential
\bea\label{dualpotg}
\tilde V^\prime&=&V(H\rightarrow -e^{-4A+2\phi}*\tilde H)+\frac{\alpha^\prime}{4\kappa^2}\int \text{vol}_M\,e^{4A-2\phi}\bar{\chi}\slashed{D}_J\chi\cr
&&-\frac{1}{2\kappa^2}\int_{M} (e^{4A-2\phi}J-\tilde B)\wedge[\d H-\frac{\alpha^\prime}{4}\big(\tr R_+\wedge R_++\tr F\wedge F)]
\eea
where the potential $V$ has again the form (\ref{pot2}). We then see that the DWSB ansatz, supplemented by the Bianchi identity (\ref{BI}) and the extremization of (\ref{gauginokin}) is sufficient to get a vacuum, since the last CS-like term in (\ref{dualpotg}) can be seen as being quadratic in vanishing terms because of (\ref{BI}) and (\ref{sbcond2g}). Note that in this formulation $\tilde H$ and $\chi$ are regarded as independent fields and that this gives a simple interpretation of the no-scale structure observed in \cite{drsw85}. Indeed, by starting from a Calabi-Yau compactification, one can allow a non-trivial gaugino condensate $\Sigma\neq 0$ by taking $\chi$ such that $\slashed{D}_{\rm CY}\chi=0$ and still imposing $\tilde H=0$. Of course, by going back to the ordinary formulation, the latter translates into $H=\frac12\Sigma\neq 0$, as originally found in \cite{drsw85}.

Let us stress that so far $\chi$ has not been restricted at all. Of course, $\chi$ should allow for a 4d + 6d splitting $\chi=\chi_{\rm 4D}\otimes\chi_{\rm 6D}+\text{c.c.}$, with $\chi_{\rm 4D}$ playing the role of the condensing gaugino in four-dimensions. Hence $\chi_{\rm 6D}$ (and thus $\Sigma$) cannot be completely arbitrary but should obey certain consistency conditions, like for instance those derived from the potential piece (\ref{gauginokin}) and the other set of equations that must be imposed on the background. In particular note that by imposing (\ref{sbcond2g}), we have
\be
 \slashed D_J|_{\text{(\ref{sbcond2g})}}=\slashed{D}_T\, :=\, \slashed{D}-\frac14\slashed{T}
\ee
where $\slashed D_T$ is the Dirac operator for the gaugino, cf.\ (\ref{dualaction2})
\be\label{gauginoaction}
S_{\rm gaugino}=-\frac{\alpha^\prime}{4\kappa^2}\int \d^{10} x\sqrt{-g}\,e^{-2\phi}\bar{\chi}\slashed{D}_T\chi 
\ee
This means that if we impose the DWSB conditions of section \ref{sec:susybvacua} together with the gaugino equations of motion on our background, then the full potential (\ref{potg}) vanishes, consistently with the requirement of having a four-dimensional  Minkowski vacuum.

As a subset of the above class of vacua one may consider the case in which we have a torsional but complex manifold $M$. This implies that $W_1=0$ and so supersymmetry is not broken at the classical level as in section \ref{sec:susybvacua}, but just by the presence of the gaugino condensate. This generalization to torsional backgrounds of the Calabi-Yau models considered in  \cite{drsw85} has been proposed in \cite{ccdl03} as a way to achieve a richer pattern of moduli stabilization and supersymmetry. As argued there, the fact that $M$ is complex together with the Bianchi identity implies the choice $\chi_{\rm 6D}=\eta$, where $\eta$ comes from the 6d component of the Killing spinor $\eps$ of the compactification, split as in (\ref{fermsplit}). Then $\Sigma=\Sigma^{3,0}+\Sigma^{0,3}$ and (\ref{sbcond2}) can be splitted as
\be
e^{2\phi}\partial(e^{-2\phi}J)=i\,H^{2,1}\quad\quad\quad H^{3,0}=\frac12 \Sigma^{3,0}
\label{splitH}
\ee
so the $(2,1)$ component of $H$ is naturally associated with the compactification scale, while the $(3,0)$ component is associated to the, presumably lower, gaugino condensate scale. Moreover we have that 
\be
\bar{\chi}\slashed{D}_J\chi \sim \Omega \cdot \d J|_{W_1=0}+\text{c.c.}=0
\ee
so that second piece in (\ref{potg}) also vanishes.

The two different scales associated to the components $H^{2,1}$ and $H^{3,0}$ of the $H$-flux suggest that, in principle, below the scale of $H^{2,1}$ one could truncate the potential (\ref{potg}) by imposing the first equation in (\ref{splitH}). In general this would imply freezing the vevs of several compactification moduli in such a way that the first equation in (\ref{splitH}) is satisfied. We would then be left with a truncated potential of the form
\be
V_{\rm no-scale}=\frac{1}{4\kappa^2}\int_M\text{vol}_Me^{4A-2\phi}(H^{3,0}+H^{0,3}-\frac12\Sigma)^2
\ee
which has exactly the same form of the no-scale potential considered in \cite{drsw85}. It is not clear, however, that such no-scale structure will survive at the scale set by $H^{2,1}$, since at this scale we may change the vevs of the complex structure moduli, which in turn change the definition of $H^{3,0}$.

\subsection{Gaugino condensate, supersymmetric AdS$_4$ vacua and calibrations}
 \label{sec:susygaugino}
 
As recalled above, $\Sigma$ enters the supersymmetry transformations (\ref{10dsusy2}) in such a way that it always breaks supersymmetry in compactifications to Minkowski. Indeed, by taking a metric ansatz of the form (\ref{wmetric}) and following the computations in \cite{fl05} one finds that supersymmetry requires a non-vanishing cosmological constant and allows for a possible non-trivial warping, in sharp contrast with the perturbative results of subsection \ref{sec:prel}. More precisely, one defines the AdS$_4$ Killing spinor $\zeta$ as  $\nabla_\mu\zeta=\frac12\,\bar w_0\hat\gamma_\mu\zeta^*$, where $w_0$ is a constant related to the AdS$_4$ radius by
\be
|w_0|^2=\frac{1}{R^2_{\rm AdS}}
\ee  Then, the external gravitino supersymmetry requires that
\be
\label{susyads0}
\Omega \cdot \Sigma \, =\,  8 e^{-A} w_0 \quad \quad \quad \d A\, =\, - \frac{1}{8} *(J  \wedge \Sigma)
\ee
 which indeed reduce to the results of subsection \ref{sec:prel} for $\Sigma = 0$.
 
Moreover, a non-vanishing $\Sigma$ will modify the supersymmetry conditions (\ref{susycond}). Following again the computations in \cite{fl05} it is easy to see that the remaining supersymmetry conditions can be rewritten as
\bseq\label{susyads}
\begin{align}
& e^{-4A + 2\phi} \d \left(e^{4A -2\phi} J \right) \, =\,  * T + 3 e^{-A} \Im \left( \bar{w}_0 \Omega\right) \label{gBPSads}\\
& e^{-3A+2\phi} \d \left( e^{3A -2\phi} \Omega\right) \, = \, - w_0\, e^{-A} J\wedge J \label{dwBPSads}
\end{align}
\eseq
which again reduce to (\ref{susycond}) for $w_0 = \d A = 0$. In fact, (\ref{dwBPSads}) implies that
\be
\d \left( e^{2A -2\phi} J \wedge J\right) \, = \, 0
\label{sBPSads}
\ee
generalizing eq.(\ref{susycond3}) for non-constant warping. Finally, it is easy to see that eqs. (\ref{susyads}) imply that
\be
* T \, = \, -\frac{3}{2}e^{-A} \Im \left( \bar{w}_0 \Omega\right) + \d (3A-\phi)  \wedge J + W_3
\label{mBPSads}
\ee

 Let us now check the consistency of the potential (\ref{potg}) with the above set of supersymmetry conditions. By plugging (\ref{susyads}) into the first term on the r.h.s.\ of (\ref{potg}) one gets\footnote{\label{fn:susyalpha}Here we assume that the deviation from the flat supersymmetric case is at least of order $\alpha^\prime$. Hence we have no contributions to the potential energy coming from the curvature  and $\calo[(\nabla A)^2]$  terms in (\ref{potord1}).}
 \bea\label{potcond1}
 V(H\rightarrow T)|_{\rm SUSY}&=&\frac{3}{\kappa^2}\int \text{vol}_M\,  e^{2A-2\phi}\big\{ 9\big[ \Im(\bar{w}_0\Omega) \big]^2+  |w_0|^2\big[(J\wedge J)^2-(J\wedge J\wedge J)^2\big]\big\}\cr
 &=&\frac{3|w_0|^2}{\kappa^2}\int \text{vol}_M\,  e^{2A-2\phi}
 \eea
On the other hand, by also imposing the gaugino equations of motion derived from (\ref{gauginoaction}), one obtains that the second term in (\ref{potg}) gives
 \bea\label{potcond2}
 \frac{\alpha^\prime}{4\kappa^2}\int \text{vol}_M\,e^{4A-2\phi}\bar{\chi}\slashed{D}_J\chi|_{\rm SUSY}&=& -\frac{3\alpha^\prime}{16\kappa^2}\int \text{vol}_M\,e^{3A-2\phi}\bar{\chi}\Re(\bar{w}_0\slashed{\Omega})\chi=\cr
 &=& -\frac{3}{4\kappa^2}\int \text{vol}_M\,e^{3A-2\phi}\Re(\bar{w}_0\Omega)\cdot \Sigma\cr
 &=&-\frac{6|w_0|^2}{\kappa^2}\int \text{vol}_M\, e^{2A-2\phi} 
 \eea
 where in the last step we have used the first of (\ref{susyads0}).
 Note that the gaugino equations of motion are automatically satisfied if we decompose the ten-dimensional gaugino as $\chi=\chi_{\rm 4D}\otimes\eta+\text{c.c.}$, where $\eta$ defines the Killing spinors written as (\ref{fermsplit}). Combining (\ref{potcond1}) and (\ref{potcond2})  and using (\ref{planck}), we get
 \be
 V^\prime|_{\rm SUSY}=  -\frac{3|w_0|^2}{\kappa^2}\int e^{2A-2\phi} \text{vol}_M=-\frac{3 M^2_{\rm P}}{R^2_{\rm AdS}}
 \ee
 Hence, we reproduce the expected value of the potential energy of an AdS$_4$ compactification with cosmological constant $\Lambda_{\rm AdS}=-3/R^2_{\rm AdS}$. Note that the contribution from the gaugino term (\ref{potcond2}) is crucial to get the correct result. On the other hand, note that in order to evaluate such contribution we have imposed the equations of motion of the gaugino. It seems technically difficult to do it otherwise and so, unlike for purely bosonic backgrounds, we do not have a direct off-shell expression for the scalar potential.
 
Despite this, one can still analyze the supersymmetry conditions of backgrounds with fermion condensates, as well as interpret them in terms of calibrations. In particular, by direct comparison with eqs.~(\ref{susycond}) and the discussion in sections \ref{sec:calibr}, one would still expect the following dictionary between calibrations and BPS objects of the compactification

\TABLE{\renewcommand{\arraystretch}{1.5}
\begin{tabular}{ccccc} 
Calibration & \quad &10d BPS object & \quad &4d BPS object\\ \hline
$e^{4A-2\phi}J$ & \quad & NS5 on $X_4 \times \Pi_2$ & \quad & gauge theory \\
$e^{3A-2\phi}\Omega$ & \quad & NS5 on $X_3 \times \Pi_3$ & \quad & domain wall \\
$e^{2A-2\phi} J \wedge J$ & \quad & NS5 on $X_2 \times \Pi_4$ & \quad & string \\
\end{tabular} }
\noindent
with again $\Pi_p$ a $p$-dimensional submanifold of $M$ and $X_d$ a $d$-dimensional slice of $X_4$.  By extending the results of \cite{km07} to NS5-branes, one can check that (\ref{susyads}) and (\ref{sBPSads}) indeed correspond to the existence of generalized calibrations for NS5-branes in an AdS$_4$ background. More precisely, one can recover the conditions (\ref{susyads}) and (\ref{sBPSads}) from the general expressions (A.28) of \cite{dwsb} by simply taking
\be
\Psi_1 \, =\, i(e^{iJ} - 1) \quad \quad \quad \Psi_2\, =\, -i\Omega \quad \quad \quad F\, =\, e^{-\Phi}T
\ee
and then replacing $e^{-\Phi} \raw e^{-2\phi}$ in order to take into account that we are dealing with NS5-branes. Note that this choice of polyforms $\Psi_1$ and $\Psi_2$ is what we would have taken for type I or type IIB vacua with O9/O5-planes, except for the shift $e^{iJ} \rightarrow e^{iJ} -1$. Performing such shift has the effect of rendering some of the AdS$_4$ BPSness equations trivial, in particular those that would be associated to fundamental strings, which indeed do not appear in the table above.
The energetics of fundamental strings, which naively only depends on the warp factor, is then only affected by the second relation in eq.(\ref{susyads0}), whose interpretation in terms of calibrations is not clear at this point.\footnote{Note that the above shift $e^{iJ} \rightarrow e^{iJ} -1$ is somehow consistent with the tree-level definition of $\mu$-slope and the D-flatness condition that we are imposing for heterotic gauge bundles. Indeed, following section \ref{sec:calibr}, one can see that such condition reads
\be
2 \Im \Psi_1 \wedge e^F\, =-\, J\wedge J \wedge F\, =\, 0
\ee
Having $\Psi_1 = i e^{iJ}$ would add a term $F\wedge F\wedge F$ to the above equation, in contrast to the tree-level DUY equations (\ref{dualHYM}). On the other hand, the DUY equations are expected to be corrected at one-loop.  Adapting the proposal of \cite{blumen05} to our case, one is led to  a one-loop corrected $\Im\Psi_1^{\text{1-loop}}=e^{2\phi}-\frac12 J\wedge J$. Assuming such a modification, it would be natural to assume that also the condition (\ref{sBPSads}) should be corrected into $\d(e^{2A-2\phi}\Im\Psi^{\text{1-loop}}_1)=0$. This would imply that $\d A=0$ as in the flat case. Because of (\ref{susyads0}), this would lead to a  further constraint on $\Sigma$.}


\subsection{$\oh$DWSB AdS$_4$ vacua with gaugino condensate}
\label{sec:ads1/2dwsb}

Having understood in terms of calibrations the conditions for heterotic 4d $\caln=1$ AdS$_4$ vacua with a gaugino condensate, it is now clear how to implement our previous strategy to construct $\caln=0$ AdS$_4$ backgrounds of the same sort. Indeed, recall from section \ref{sec:susybvacua} that the $\caln=0$ vacua considered there were such that the supersymmetry conditions (\ref{sbcond0}) were still satisfied, allowing to define a stable gauge bundle as in section \ref{sec:calibr}. On the other hand, the domain-wall BPSness condition was relaxed to (\ref{violdw}), and half-imposed for $\oh$DWSB backgrounds via (\ref{susycond1/2}).

In the case of AdS$_4$ compactifications, the surviving 1/2 domain-wall BPSness is determined by (\ref{gBPSads}) itself, since the equation of motion $\d(e^{4A-2\phi}*T)=0$  for $T$  implies that
\be\label{1/2dw}
\d [e^{3A -2\phi} \Im(\bar w_0\Omega)] \, = \, 0
\ee
Such result is to be expected, since an AdS$_4$ background can be though as a 4d domain wall supergravity solution, that in the present context comes from a stack of backreacted NS5-branes wrapping an internal three-cycle $\Pi_3$ calibrated by $e^{3A -2\phi} \Im(\bar w_0\Omega)$ -- see e.g.\ \cite{km07}. Now, if such AdS$_4$ background satisfies the equations of motion, a probe NS5-branes on top of the backreacted ones should feel no force from the background. This is guaranteed if such NS5-brane is a BPS object, which in turn means that $e^{3A -2\phi} \Im(\bar w_0\Omega)$ should be a proper calibration and hence the condition (\ref{1/2dw}) should be satisfied.

We are then naturally led to consider $\caln=0$ AdS$_4$ backgrounds where the only source of supersymmetry breaking originates from a background condition of the form
\be\label{gauginoSB}
e^{-3A+2\phi}\d\big[e^{3A-2\phi}\Re(\bar w_0\Omega)\big]=\Re(\bar w_0\, W_1)J\wedge J+\Re(\bar w_0 W_2) \wedge J
\ee
so that supersymmetry is broken if $\Re(\bar w_0\, W_1)\neq -e^{-A}|w_0|^2$ or $W_2 \neq 0$.
This implies that these backgrounds are characterized by the torsion classes of the internal manifold $M$ via
\be\label{45torads}
W_4\,=\, \d(\phi- A)\quad  \quad \quad W_5\,=\, 2\d\phi - 3\d A\quad \quad \quad \Im (\bar w_0 W_1) = \Im (\bar w_0 W_2) = 0
\ee
while $W_3$ is specified by the relation
\be
* T \, = \, -\frac{3}{2}\Im \left((\bar{W}_1+ 2e^{-A} \bar{w}_0) \Omega\right) + \d (3A-\phi)  \wedge J + W_3
\label{msbBPSads}
\ee

A further source of supersymmetry breaking comes from the external gravitino supersymmetry conditions. More precisely, setting 
\be
\Omega \cdot \Sigma \, =\, 8 e^{-A} \sigma_0
\label{defsig}
\ee
we have that the external gravitino supersymmetry is broken if $\sigma_0\neq w_0$. Hence, we have two natural SUSY-breaking scalar parameters for this kind of compactifications
\be
\label{sbparamads}
\cali_1\, =\, W_1 + e^{-A} w_0\quad \quad \quad \cali_2\, =\, \oh e^{-A} (\sig_0 - w_0)
\ee
Note that this family of backgrounds contains all of the supergravity compactifications with non-vanishing $\Sigma$ considered up to date in the literature. In particular, it generalizes the compactifications analyzed in \cite{drsw85,ccdl03,gklm03}, that considered vacua with $W_1 = w_0 = 0$ and so with only the SUSY-breaking parameter $\cali_2$ turned on. Such $\caln=0$ constructions are in some sense orthogonal to the ones considered in section \ref{sec:susybvacua}, since there we had $\sig_0 = w_0 =0$ and so only $\cali_1 \neq 0$. It is therefore natural to wonder to what extent 4d vacua for arbitrary values of both SUSY-breaking parameters turned on can be constructed.

As before, some amount of information can be obtained by analyzing the scalar potential (\ref{potg}).  One can easily see that the first term on the r.h.s.\  of (\ref{potg}) reads
\be\label{SBpotcond1gc}
 V(H\rightarrow T)|_{\rm \frac12DWSB}= \frac{1}{4\kappa^2}\int e^{2A-2\phi} \text{vol}_M (36|w_0|^2+|W_2|^2-24|W_1|^2)
 \ee
 while, by imposing the gaugino equations of motion derived from (\ref{gauginoaction}) and following the same steps as in (\ref{potcond2}),  the second term  on the r.h.s.\  of (\ref{potg}) gives
 \be\label{SBpotcond2gc}
 \frac{\alpha^\prime}{4\kappa^2}\int \text{vol}_M\,e^{4A-2\phi}\bar{\chi}\slashed{D}_J\chi|_{\rm \frac12DWSB}= -\frac{6}{\kappa^2}\int \text{vol}_M\ e^{2A-2\phi}\,\Re(\bar{w}_0 \sig_0) 
 \ee
 Summing up these two terms  one gets 
 \be
 V^\prime|_{\rm \frac12DWSB}=\frac{1}{4\kappa^2}\int e^{2A-2\phi} \text{vol}_M \big[ 36|w_0|^2-24|W_1|^2+|W_2|^2-24\Re(\bar{w}_0 \sig_0)\big]
 \ee
which on-shell should equal $-3|w_0|^2M_P^2$. This is indeed the case for the supersymmetric case, already considered in section \ref{sec:susygaugino}, since there we have that $W_2 = \cali_1 = \cali_2 = 0$ above, which in turn imply that $w_0 = \sig_0 = -e^{A} W_1$. 

If on the other hand we consider the torsional geometries considered in section \ref{sec:susybvacua} and associated to Minkowski DWSB vacua, we need to impose the constraint $|W_2|^2 = 24|W_1|^2$ on the above vacuum energy.  One then concludes that, by consistency, an AdS$_4$ vacuum of this kind needs to satisfy the relation $\sigma_0 = 2 w_0$,\footnote{In fact,  one could also consider the case where $\sigma_0 = 2 w_0 + b i w_0$, $b \in \reals$. However, $\sigma_0$ should enter the holomorphic gravitino mass that can be calculated following section \ref{sec:gravitinomass}.  Since the gravitino mass is proportional to the superpotential and the latter is expected to be aligned with the phase of $w_0$ (as the supersymmetric case shows), we are naturally led to expect that $\sig_0$ has the same phase as $w_0$.} and so supersymmetry is necessarily broken because the first equation in (\ref{susyads0}) is not satisfied. This fact shows that, naively, adding a gaugino condensate on top of the $\caln=0$ torsional geometries of section \ref{sec:susybvacua} is not enough to restore the supersymmetry of the compactification. Indeed, from section \ref{sec:susygaugino} we see that in order to construct supersymmetric AdS$_4$ vacua the torsion class $W_2$ must vanish. By eq.(\ref{sbtor}),  this is not possible for the Minkowski DWSB vacua of section \ref{sec:susybvacua} since there by assumption $W_1 \neq 0$. In particular, if we consider the fibered manifolds of section \ref{sec:oneparSB} we see that $M$ should undergo some kind of topology change in order to flow to a manifold $M'$ with $W_2 = 0$. It is not clear how the presence of $\Sigma$ could trigger such topology change, so one would expect that adding a gaugino condensate on top of the no-scale vacua of section \ref{sec:oneparSB} would most likely take them to a $\oh$DWSB AdS$_4$ background of the kind considered here, not being clear if this would be a vacuum of the theory. Of course, adding further non-perturbative effects produced by, e.g., worldsheet instantons may provide the necessary ingredients to promote our heterotic no-scale vacua to an $\caln=1$ AdS$_4$ vacuum, along the lines of \cite{ckl05}.


\section{Discussion}
\label{sec:conclusions}

In this paper we have addressed the construction of non-supersymmetric heterotic flux vacua from a rather general approach. Such approach is mainly based on the idea of {\em domain-wall supersymmetry breaking} (DWSB), previously developed in \cite{dwsb} in the context of type II flux compactifications. Just as in there, in order to implement the DWSB ansatz we need a microscopic understanding of the scalar potential governing flux compactifications to four dimensions, as well as to rewrite such potential in BPS form in order to analyze which SUSY-breaking backgrounds satisfy the ten-dimensional equations of motion. 

We have performed such analysis in the context of heterotic flux compactifications. We have first derived directly from the 10d heterotic action a scalar potential that, if extremized, guarantees that the 10d equations of motion of the heterotic background are satisfied at tree-level and up to first order in $\alpha'$ corrections. We have then expressed such potential in BPS form, in order to see which flux backgrounds that break supersymmetry at tree-level still satisfy the 10d equations of motion. This has led to identify a class of heterotic SUSY-breaking vacua (dubbed DWSB vacua) satisfying certain geometric conditions, but otherwise flexible enough to generate a large landscape of $\caln=0$ heterotic vacua. This is in sharp contrast with previous approaches to build $\caln=0$ heterotic vacua, like Scherk-Schwarz compactifications that rely on the existence of isometries in the compactification manifold, and so correspond to rather constrained geometries. 

In order to arrive to the above BPS expression we have made the assumption that our compactification manifold $M$ admits an SU(3)-structure, which basically implies that the 10d $\caln=0$ background should correspond to some pattern of 4d spontaneous SUSY-breaking. Such SU(3)-structure not only selects a 10d spinor which, from the 4d viewpoint, corresponds to the generator of an (approximate) supersymmetry. It also allows to relate the BPS bounds of such backgrounds with the concept of calibrated $p$-submanifolds and calibration $p$-forms. More precisely, for heterotic flux compactifications we have that BPS bounds for probe NS5-branes are in one-to-one correspondence with the existence of calibrations. The absence of at least one calibration implies that one class of BPS bounds will not be developed and, moreover, that supersymmetry will be broken at tree-level. When applying these results to the scalar potential analysis, one finds that the class of supersymmetry-breaking vacua above correspond to compactifications without the calibration/BPS bound for NS5-branes showing up as 4d domain-walls. Hence their name DWSB vacua.

Within the class of DWSB heterotic flux vacua, we have considered a subclass of vacua where some domain-wall BPS bounds do exist. This set of vacua (dubbed $\oh$DWSB vacua) is the heterotic analogue of the $\caln=0$ no-scale type IIB/F-theory constructions of \cite{gkp}. By means of the $\oh$DWSB ansatz and the scalar potential we have analyzed the kind of geometries that these vacua should correspond to, obtaining a rather constrained set of fibered manifolds $M$ quite close to torsional geometries recently analyzed in the literature \cite{bsb,bs}. While the torsional ansatz used there was largely motivated by dualities, here we have followed a deductive approach that does not rely on the existence of any dual solution. We should then expect that $\oh$DWSB fibered manifolds allow for a more general and systematic treatment, providing a more complete picture on the landscape of torsional heterotic vacua. 

We have also shown that, in the limit of constant dilaton, $\oh$DWSB compactification manifolds $M$ reduce to half-flat geometries $\tilde{M}$, like the ones considered in \cite{halfflat} in order to describe the effective 4d physics of heterotic flux compactifications. However, the working assumption of \cite{halfflat} is that the half-flat manifold $\tilde{M}$ is a small deviation of a certain Calabi-Yau manifold $\tilde{M}_{\rm CY}$, and so $\tilde{M}$ and $\tilde{M}_{\rm CY}$ share the same set of light fields. As our explicit construction does not rely on such assumption, one can use it to check to what extent such Calabi-Yau-inspired truncation is justified. 

Note that above the derivation of a four-dimensional effective theory is based on the rather crude approximation that the dilaton is constant on the compactification manifold. It would be however interesting to derive in a more rigorous way which kind of 4d effective theory can be obtained from this kind of compactifications, and to which extent the effects of a varying dilaton can change the four-dimensional physics. In particular, it would be interesting to compute the spectrum of soft masses of this kind of compactifications, as well as how they are affected by a varying dilaton. We have already provided a glimpse of such effect by computing the 4d gravitino and gaugino masses in general and in the limit of constant dilaton, but it should be easy to extend such analysis to the computation of soft masses for 4d chiral multiplets, at least for the explicit examples constructed in section \ref{sec:K3ex}, in ref.\cite{bsb} and similar vacua based on twisted tori. As pointed out, the effective theory description of such SUSY-breaking vacua should be simplified in the case of anisotropic fibrations, where the SUSY-breaking scale will be suppressed with respect to the Kaluza-Klein scale and solving the Bianchi identity and the BPS conditions for the curvature tensore becomes more tractable. While anisotropic compactifications have been proposed as in interesting playground to build heterotic GUT models \cite{ht04}, it is a priori not clear that breaking supersymmetry at tree-level could generate a TeV scale for the soft masses of such models, even for the case of a strongly varying dilaton. In this sense, let us point out that our analysis can also be applied to the construction of type I flux vacua, where such potential problem can be solved by going to the limit of very weak string coupling. Indeed, such suppression is already manifest in the computation of Type I soft masses for simple elliptic fibrations in the limit of constant dilaton \cite{cm07,cm09}.

In fact, since our contact with effective four-dimensional theories has only been made for the subclass of $\oh$DWSB compactifications, an natural question is how this effective theory generalizes to the full set of DWSB vacua. In particular, one may wonder if the no-scale structure of $\oh$DWSB vacua is also present there. If not, there is a chance that (almost) all the moduli of DWSB compactifications are lifted at tree-level. Note that in order to answer this question one does not necessarily need to have a very good understanding of the four-dimensional effective theory. Indeed, since DWSB vacua are defined in terms of first order differential equations for the SU(3)-invariant forms $J$ and $\Omega$, one may implement the approach of \cite{bt05,lucaeff09}, which is based on the existence of integrable structures such as the ones associated to $J$ and $\Omega$, to compute the amount of moduli in these compactifications. 

As is well-known, whatever the effective field theory is, the dilaton should not be stabilized at tree-level, and one should consider backgrounds with gaugino condensates in order to lift such modulus. Remarkably, heterotic backgrounds with fermion condensates seem to share many of the interesting properties of purely bosonic vacua analyzed in this paper. Indeed, we have shown that the supersymmetry conditions for gaugino condensate heterotic vacua can also be understood in terms of calibrations. Hence, one could in principle also extend the DWSB ansatz to this richer class of backgrounds. In practice, however, in order to compute the 4d vacuum energy one has to impose the gaugino equations of motion, and so an off-shell expression of the scalar potential is not available. While this prevents a systematic analysis of $\caln=0$ vacua with gaugino condensate, the classification of these backgrounds in terms of calibrations should hopefully constitute a useful framework for future investigations on these kinds of issues. 

Indeed, an obvious application of our results would be to adapt the type IIB scenarios in \cite{kklt,LVS} to the heterotic context. Recall that the basic ingredients of such constructions involve an $\caln=0$ no-scale vacuum and a mechanism breaking such no-scale structure, such as $\a'$-corrections and/or non-perturbative effects. All these ingredients have a worldsheet and/or a ten-dimensional supergravity description in the present heterotic context, which may allow for a more rigorous understanding of the de Sitter landscape in string theory.

At any rate, we hope that the developments in this paper help developing a broader picture of the set of $\caln=0$ vacua in string theory vacua, allowing to derive interesting results on the above and related issues.


\bigskip

\bigskip

\centerline{\bf Acknowledgments}

\bigskip

It is a pleasure to thank P.~G.~C\'amara, M.~Haack, S.~Groot Nibbelink, A.~Sagnotti, M.~Trapletti, D.~Tsimpis and P.~K.~S.~Vaudrevange for useful discussions. This work  is supported in part by the Cluster of Excellence ``Origin and Structure of the Universe'' in M\"unchen, Germany.

\bigskip
\bigskip

\newpage

\appendix


\section{Fermionic conventions and SU(3)-structure}
\label{app:fermconv}

In ten dimensions, we use $M,N,\ldots$ as curved indices and we underline flat indices $\ul{M},\ul{N},\ldots$.
We use a real representation of the ten-dimensional gamma matrices $\Gamma^M$, in which the Majorana spinors are real. The ten-dimensional chiral operator is
\be
\Gamma_{(10)}:= \Gamma^{\ul{0\ldots 9}}\ .
\ee
The heterotic ten-dimensional supersymmetry generator $\epsilon$ is Majorana-Weyl, meaning that it is real and it satisfies the chirality condition $\Gamma_{(10)}\epsilon =\epsilon$.

For any $p$-form $\rho$, then we use the notation 
\be
\slashed{\rho}:= \frac{1}{p!}\,\rho_{M_1\ldots M_p}\,\Gamma^{M_1\ldots M_p}\, .
\ee
where $\Gamma^{M_1\ldots M_p}\equiv \Gamma^{[M_1}\cdots\Gamma^{M_p]}$.

In a compactification of the form $X_{10}=X_4\times M$, we split the ten-dimensional gamma matrices $\Gamma^M$  in terms of four- and six-dimensional gamma matrices $\hat\gamma^\mu$ (associated with the unwarped $X_4$ metric) and $\gamma^m$ in the following way
\be
\Gamma^{\mu}\, =\, e^{-A}\hat\gamma^{\mu}\otimes \bbone\quad \quad \quad\Gamma^{{m}}\, =\,\gamma_{(4)}\otimes \gamma^{{m}} 
\ee
where $\gamma_{(4)}=i\hat\gamma^{\ul{0123}}$ is the standard four-dimensional chiral operator. The six-dimensional chiral operator is in turn $\gamma_{(6)}=-i\gamma^{\ul{123456}}$ and so we have that $\Gamma_{(10)}=\gamma_{(4)}\otimes\gamma_{(6)}$. 

The ten-dimensional supersymmetry generator $\epsilon$ can be accordingly  decomposed as
\be\label{fermsplit}
\epsilon =\, \zeta\otimes\eta+\ \text{c.c.}\,
\ee
with $\gamma_{(4)}\zeta=\zeta$ and $\gamma_{(6)}\eta=\eta$. If $X_4$ is Minkowski, then $\zeta$ is a constant chiral spinor.  If $X_4$ is AdS$_4$, then $\zeta$ is the Killing spinor defined by $\nabla_\mu\zeta=\frac12\,\bar w_0\hat\gamma_\mu\zeta^*$, where $w_0$ has arbitrary phase and is related to the AdS$_4$ radius $R$ by $|w_0|=1/R$. 

The internal spinor $\eta$ can be used to construct a real two-form $J$ and a complex three-form $\Omega$ as follows
\bea\label{su3forms}
J_{mn}=\frac{i}{\|\eta\|^2}\,\eta^\dagger\gamma_{mn}\eta\, ,\quad \Omega_{mnp}=\frac{1}{\|\eta\|^2}\,\eta^T\gamma_{mnp}\eta\, ,
\eea
where $\|\eta\|^2\equiv \eta^\dagger\eta$. Furthermore
\be
I^m{}_n=\frac{i}{\|\eta\|^2}\,\eta^\dagger\gamma^m{}_{n}\eta=g^{mk}J_{km}
\ee
is an almost complex structure with respect to which the metric is hermitian, with associated projector onto $(1,0)$ vectors
\bea
P^m{}_n=\frac12(\delta^m{}_n-iI^m{}_n)\, .
\eea
Then $J$ is the associated K\"ahler (or fundamental) (1,1)-form and $\Omega$ is a globally defined (3,0)-form. 

The compatibility conditions between $\Omega$ and $J$ imply that their exterior derivatives take the following general form \cite{chiossi02}
\bea\label{tc}
\d J &=&-\frac32\Im(\overline{W}_1\Omega)+W_4\wedge J+W_3\cr
\d\Omega &=& W_1 J\wedge J+W_2\wedge J+\overline{W}_5\wedge \Omega
\eea
where $W_1$ is a complex scalar, $W_2$ is $(1,1)$ and primitive, $W_3$ is real $(2,1)+(1,2)$ and primitive, $W_4$ is a real one-form, $W_5$ is a real one-form. $W_1,\ldots, W_5$ are called {\em torsion classes}. Roughly speaking, the torsion classes measure the failure of an SU(3)-structure to define a  Calabi-Yau metric, which has $W_1=\ldots=W_5=0$.


\section{Supersymmetry breaking and relation between spinorial and tensorial formalism}\label{app:spinsb}

In this appendix we discuss the relation between the most general violation of the standard Killing spinors equations obtained by setting $0=\delta\psi_M=\delta\lambda=\delta\chi$ and the violation of the equivalent supersymmetry conditions in SU(3)-structure form. We use the spinorial decomposition described in appendix \ref{app:fermconv} and split the ten-dimensional gravitino $\psi_M$ into external components $\psi_\mu$ and internal components $\psi_m$. We will discuss both the cases with and without a non-vanishing gaugino bilinear.

\subsection{Classical supersymmetry breaking and spinors}
We first discuss external gravitino and gaugino, which are straightforward, and later discuss the internal gravitino and dilatino.

The external components of (\ref{gravsusy}) takes the form
\be
\delta\psi_{\mu}=\frac12\,e^A\hat{\gamma}_{\mu}\zeta\otimes (\slashed{\partial}A\eta+e^{-A}w_0\eta^*)+{\rm c.c.}
\ee
Since $\slashed{\partial}A\eta$ and $\eta^*$ are orthogonal, it is clear that the violation of the condition $\delta\psi_\mu=0$ is in one-to-one correspondence with the constancy of the warping and the vanishing of the cosmological constant. 

On the other hand, the gaugino transformation (\ref{gaugesusy}) takes the form
\be
\delta\chi=\frac12\zeta\otimes \slashed{F}\eta+\text{c.c.\ }\, 
\ee
We can write
\be
\slashed{F}\eta=-i(J\cdot F)\,\eta+\frac12(\iota_m\Omega\cdot F)\,\gamma^m\eta^*\, 
\ee
The two terms on the r.h.s.\ are orthogonal and their vanishing corresponds to the two conditions that $F$ is primitive and $F^{0,2}=0$ respectively.

Let us now turn to the internal gravitino and dilatino. The corresponding variations (\ref{gravsusy}) and (\ref{dilsusy})  split as follows
\bseq\label{intgravdil}
\begin{align}
&\delta\psi_m=\zeta\otimes\Big(\nabla_m-\frac{1}{4}\slashed{H}_m\Big)\eta+\text{c.c}.~\, \label{intgrav}\\
&\delta\lambda=\zeta\otimes\Big(\slashed{\partial}\phi-\frac{1}{2}\slashed{H}\Big)\eta+\text{c.c.}~\ \label{intdil}
\end{align}
\eseq
We introduce the decompositions
\bseq\label{gravdildec}
\begin{align}
&\Big(\nabla_m-\frac{1}{4}\slashed{H}_m\Big)\eta= i\,p_m\eta+q_{mn}\gamma^n\eta^* \, \label{intgravdec}\\
&\Big(\slashed{\partial}\phi-\frac{1}{2}\slashed{H}\Big)\eta= u_m\gamma^m\eta+r\eta^*\ \label{intdildec}
\end{align}
\eseq
where $p_m$ is real and  $q_{mn}$ and $u_m$ are restricted by the projector conditions $\bar P^k{}_n q_{mk}=P^k{}_nu_k=0$. Then,  we can translate the most general violation of $\delta\psi_m=\delta\lambda=0$ in terms of the non-vanishing of the parameters $p_m$, $q_{mn}$, $u_m$ and $r$. These are in turn related to the exterior derivatives of the SU(3)-structure tensors $J$ and $\Omega$ as follows
\bseq\label{su3violation}
\begin{align}
&e^{2\phi}\d\big(e^{-2\phi}J\wedge J\big)=-4\,\text{Re}\, u \wedge J\wedge J-8\text{Re}\left(s^*\wedge \Omega\right)\, \\
&e^{2\phi}\text{d}\big(e^{-2\phi}{\Omega}\big)=2(i\,p-u)\wedge\Omega-rJ\wedge J+8is\wedge J\, \\
&e^{2\phi}\text{d}\big(e^{-2\phi}J\big)-\ast H=2\text{Im}\big(r^*{\Omega}\big)-4\,\text{Re}\,u\wedge J-2\text{Im}\big(t^{*n}\wedge\iota_n\Omega\big)\, 
\end{align}
\eseq
Here $s=\frac{1}{2}q_{mn}\,\d y^{m}\wedge \d y^n$, $t^n=q_m{}^n \d y^m$, $u=u_m\d y^m$ and $p=p_m\d y^m$.  Clearly,  by setting $p_m=q_{mn}=u_m=r=0$ one gets the conditions (\ref{susycond}).

\bigskip

Let us now discuss how the general SUSY-breaking ansatz (\ref{sbcond0})-(\ref{addcond}) restricts the form of the SUSY-breaking parameters $p_m$, $q_{mn}$, $u_m$ and $r$. First of all, (\ref{sbcond3}) and the non-primitive component of (\ref{sbcond2}) imply that $u=0$ and $s^{(2,0)}=0$. Then, by imposing (\ref{addcond}) one gets $p=0$. Furthermore, from (\ref{sbcond2}) one obtains $r=g^{mn}q_{mn}$ and $q^{(2,0)}=0$. Thus, we see that the SUSY-breaking condition (\ref{violdw}) takes the form
\be\label{susybreakapp}
e^{2\phi}\text{d}\big(e^{-2\phi}{\Omega}\big)=-rJ\wedge J+8is\wedge J
\ee
where $s$ is (1,1) and can be decomposed in primitive and non-primitive part as follows: $s=-\frac{i}6\, r\, J +s_{\rm P}$. Comparing with the first condition in (\ref{sbtor}), we see that 
\be
r=3W_1\quad ,\quad s_{\rm P}=-\frac{i}{8}\,W_2
\ee


\subsection{Supersymmetry breaking in presence of a gaugino condensate}
\label{app:gaugino}

Here we discuss how the supersymmetry breaking equations alter if one allows for a gaugino condensate. The supersymmetry variations of the gravitino and dilatino (\ref{gravsusy}) and (\ref{dilsusy}) get changed into
\bseq\label{gauginosusy}
\begin{align}
\delta\psi_M&=\Big(\nabla_M-\frac{1}{4}\slashed{H}_M+\frac{1}{16}\slashed{\Sigma}\Gamma_M\Big)\epsilon ,\label{gagravsusy}\\
\delta\lambda&=\Big(\slashed{\partial}\phi-\frac{1}{2}\slashed{H}-\frac{1}{8}\slashed{\Sigma}\Big)\epsilon ,\label{gadilsusy}
\end{align}
\eseq
while the variation of the gaugino (\ref{gaugesusy}) remains unchanged. For the external component of (\ref{gagravsusy}) one obtains then
\begin{equation}
\delta\psi_{\mu}=\frac{1}{2}e^A\hat{\gamma}_{\mu}\zeta\otimes\Big(\slashed{\partial}A\eta+e^{-A}w_0\eta^*-\frac{1}{8}\slashed{\Sigma}\eta\Big)+c.c.
\end{equation}
This shows that after a gaugino condensate is added the condition $\delta\psi_{\mu}=0$ no longer forces the cosmological constant to be zero or the warp factor to be constant. Allowing for additional violation of $\delta\psi_{\mu}=0$ yields
\begin{equation}
\slashed{\Sigma}\eta=8\slashed{\partial}A\eta-16v_m\gamma^m\eta+8e^{-A}w_0\eta^*-16h\eta^*=8\slashed{\partial}A\eta-16v_m\gamma^m\eta+8e^{-A}\sigma_0\eta^*
\end{equation}
where  $\sigma_0=w_0-2e^{A}h$ and $v_m$ is restricted by $P^k{}_nv_k=0$. Below, we will impose that $v_m=0$, so that $\delta\psi_{\mu}\propto \zeta\otimes \eta^*+\text{c.c.}$, which is a natural assumption if we want to interpret the SUSY-breaking in $\caln=1$ four-dimensional terms. 

The internal component of the gravitino variations and the dilatino variation read
\bseq
\begin{align}
\delta\psi_m&=\zeta\otimes\Big(\nabla_m-\frac{1}{4}\slashed{H}_m+\frac{1}{16}\slashed{\Sigma}\Gamma_{m}\Big)\eta+c.c\\
\delta\lambda&=\zeta\otimes\Big(\slashed{\partial}\phi-\frac{1}{2}\slashed{H}-\frac{1}{8}\slashed{\Sigma}\Big)\eta+c.c
\end{align}
\eseq
which can be decomposed as
\bseq
\begin{align}
\Big(\nabla_m-\frac{1}{4}\slashed{H}_m+\frac{1}{16}\slashed{\Sigma}\Gamma_{m}\Big)\eta&=i\tilde p_m\eta+\tilde q_{mn}\gamma^n\eta^*\\
\Big(\slashed{\partial}\phi-\frac{1}{2}\slashed{H}-\frac{1}{8}\slashed{\Sigma}\Big)\eta&=\tilde u_m\gamma^m\eta+\tilde r\eta^*
\end{align}
\eseq
where $\tilde p_m$ is real and  $\tilde q_{mn}$ and $\tilde u_m$ are restricted by the projector conditions $\bar P^k{}_n \tilde q_{mk}=P^k{}_n\tilde u_k=0$. Hence, the violation of $\delta\psi_m=\delta\lambda=0$ can be expressed by the parameters $\tilde p_m$, $\tilde q_{mn}$, $\tilde u_m$ and $\tilde r$. The exterior derivatives of the SU$(3)$-structure tensors $J$ and $\Omega$ read then
\bseq
\begin{align}
e^{-2A+2\phi}\text{d}\left(e^{2A-2\phi}J\wedge J\right)&=4\text{Re}\left(v-\tilde u\right)\wedge J\wedge J-8\text{Re}\left(\tilde s^*\wedge\Omega\right)\\
e^{-3A+2\phi}\text{d}\left(e^{3A-2\phi}\Omega\right)&=(2i\tilde p-2u+5v)\wedge\Omega\nonumber\\ &\quad\quad-\left(\tilde r+e^{-A}w_0-2h\right)J\wedge J+8i\tilde s J\\
e^{-4A+2\phi}\text{d}\left(e^{4A-2\phi}J\right)-\ast T&=\text{Im}\left(\left[2\tilde r^*+3e^{-A}\bar{w}_0-6h^*\right]\Omega\right) + \nonumber\\ &\quad\quad\text{Re}\left(14v-4\tilde u'\right)\wedge J - 2\text{Im}\left(\tilde t^{*n}\wedge\iota_n\Omega\right)
\end{align}
\eseq
Here we used $\tilde s=\frac{1}{2}\tilde q_{mn}\d y^m\wedge\d y^n$, $\tilde t^n=\tilde q_m{}^n\d y^m$, $\tilde u=\tilde u_m\d y^m$ and $\tilde p=\tilde p_m\d y^m$. After setting  $\tilde p_m=\tilde q_{mn}=\tilde u_m=v_m=\tilde r=h=0$ one obtains (\ref{susyads}) and (\ref{sBPSads}).

\bigskip
The parameters $\tilde p_m$, $\tilde q_{mn}$, $\tilde u_m$, $v_m$, $\tilde r$ and $h$ get severely restricted by our SUSY-breaking ansatz. First, we impose $v=0$, for the reason discussed above. Then, by imposing (\ref{gBPSads}) and (\ref{sBPSads}) one obtains $\tilde u=0$, $\tilde s^{2,0}=0$, $\tilde r-3h=g^{mn}q_{mn}$ and $\tilde q^{2,0}=0$. Furthermore, from (\ref{addcond}) one gets $\tilde p=0$. Hence, the remaining SUSY-breaking condition is
\begin{equation}
e^{-3A+2\phi}\text{d}\left(e^{3A-2\phi}\Omega\right)=-\left(\tilde r+e^{-A}\sigma_0\right)J\wedge J+8i\tilde s\wedge J=-\hat{r}J\wedge J+8i\tilde s\wedge J
\end{equation}
which is of the same form as (\ref{susybreakapp}). Thus, $\tilde s=-\frac{i}{6}\hat{r}J+s_P$ and the comparison to (\ref{gauginoSB}) gives
\begin{equation}
\hat{r}=3W_1~,~~~s_P=-\frac{i}{8}W_2
\end{equation}


\section{Dual formulation of heterotic supergravity}
\label{app:dual}

The dual formulation of the heterotic theory is expressed in terms of the seven-form flux  $\hat H=e^{-2\phi}*H$. In this formulation the six-form potential $\hat B$, $\d \hat B=\hat H$, plays the role of the fundamental field and couples electrically to the 5-brane charge of the background and the BI (\ref{BI}) arises as the equation of motion of $\hat B$.  For this reason, this frame is the natural one to describe the coupling of NS5-branes. 

The complete dual formulation up to order $\alpha^\prime$ can be found in \cite{bdr2} and the dualization procedure relating the two formulations is discussed in detail in \cite{bdr3}.  Here we just focus on the bosonic sector. Start from the new action
\bea\label{extaction}
S^\prime=S-\frac{1}{2\kappa^2}\int_{X_{10}} \hat B\wedge \big[  \d H+\frac{\alpha^\prime}{4}(\tr F\wedge F-\tr R_+\wedge R_+) \big]\ , 
\eea
where $S$ as in (\ref{10daction}), but where $H$ and $\hat B$ should be considered as elementary independent fields. By varying with respect to $\hat B$ one gets the Bianchi identity (\ref{BI}) and then, integrating out $\hat B$ just produces the original action  (\ref{10daction}). 

On the other hand, by extremizing $S^\prime$ with respect to $H$, one gets
\bea\label{invrel}
\hat H=e^{-2\phi}*H\, 
\eea
By plugging (\ref{invrel}) into (\ref{extaction}) and keeping only $\calo(\alpha^\prime)$ terms, we arrive at the dual action
\bea\label{dualaction}
\hat S&=&\frac{1}{2\kappa^2}\int\d^{10} x\sqrt{-g}\,e^{-2\phi}\big[{\cal R}+4(\d\phi)^2-\frac1{2}\, e^{4\phi}\, \hat H^2+\frac{\alpha^\prime}4(\tr R_+^2- \tr F^2)\big]\cr
&& -\frac{\alpha^\prime}{8\kappa^2}\int_{X_{10}} \hat B\wedge(\tr F\wedge F-\tr R_+\wedge R_+)\, .  
\eea
The supersymmetry transformations in the dual formulation are as in (\ref{10dsusy}), up to terms which vanish on-shell at order $\alpha^\prime$. 

\bigskip

Let us also consider the duality transformation in presence of a non-vanishing gaugino.
It is useful to introduce a three-form $\Sigma$ defined by
\be
 \Sigma_{MNP}\, =\, \frac{\a'}{4} \tr \bar{\chi} \Gamma_{MNP} \chi
\ee
In the ordinary formulation which uses the 3-form $H$ as fundamental, the relevant terms in the action are now as in (\ref{10dactiong}).
By performing the duality transformation described above we now get
\be
\hat H=e^{-2\phi}*T\quad,\quad \text{with}\quad T=H-\frac12 \Sigma
\ee
and the dual action with non-vanishing gaugino terms now reads
\be\label{dualaction2}
\hat S^\prime=\hat S -\frac{\alpha^\prime}{4\kappa^2}\int \d^{10} x\sqrt{-g}\,e^{-2\phi}\bar{\chi}(\slashed{D}-\frac14\slashed{T})\chi 
\ee
Now the supersymmetry transformations are modified at $\calo(\alpha^\prime)$ by the presence of $\Sigma\neq 0$ and take the form (\ref{10dsusy2}).




\newpage

\begin{thebibliography}{99}


\bibitem{Scherk:1978ta}
  J.~Scherk and J.~H.~Schwarz,
  {\em ``Spontaneous Breaking Of Supersymmetry Through Dimensional Reduction,''}
  Phys.\ Lett.\  B {\bf 82}, 60 (1979).
  
\bibitem{din85}
  J.~P.~Derendinger, L.~E.~Ib\'a\~nez and H.~P.~Nilles,
  {\em ``On The Low-Energy D = 4, N=1 Supergravity Theory Extracted From The D = 10, N=1 Superstring,''}
  Phys.\ Lett.\  B {\bf 155}, 65 (1985).

\bibitem{drsw85}
  M.~Dine, R.~Rohm, N.~Seiberg and E.~Witten,
  {\em ``Gluino Condensation In Superstring Models,''}
  Phys.\ Lett.\  B {\bf 156}, 55 (1985).
  
\bibitem{Font:1990nt}
  A.~Font, L.~E.~Ib\'a\~nez, D.~L\"ust and F.~Quevedo,
  {\em ``Supersymmetry breaking from duality invariant gaugino condensation,''}
  Phys.\ Lett.\  B {\bf 245}, 401 (1990).
  
\bibitem{Ferrara:1990ei}
  S.~Ferrara, N.~Magnoli, T.~R.~Taylor and G.~Veneziano,
  {\em ``Duality and supersymmetry breaking in string theory,''}
  Phys.\ Lett.\  B {\bf 245}, 409 (1990).
  
\bibitem{Nilles:1990jv}
  H.~P.~Nilles and M.~Olechowski,
  {\em ``Gaugino condensation and duality invariance,''}
  Phys.\ Lett.\  B {\bf 248}, 268 (1990).
  
\bibitem{Ibanez:1992hc}
  L.~E.~Ib\'a\~nez and D.~L\"ust,
  {\em ``Duality anomaly cancellation, minimal string unification and the effective low-energy Lagrangian of 4-D strings,''}
  Nucl.\ Phys.\  B {\bf 382}, 305 (1992)
  [arXiv:\hepth{9202046}].
  
\bibitem{Kaplunovsky:1993rd}
  V.~S.~Kaplunovsky and J.~Louis,
  {\em ``Model independent analysis of soft terms in effective supergravity and in string theory,''}
  Phys.\ Lett.\  B {\bf 306}, 269 (1993)
  [arXiv:\hepth{9303040}].
  
 \bibitem{Grana:2005jc}
  M.~Gra\~na,
  {\em ``Flux compactifications in string theory: A comprehensive review,''}
  Phys.\ Rept.\  {\bf 423}, 91 (2006)
  [arXiv:\hepth{0509003}].

\bibitem{Douglas:2006es}
  M.~R.~Douglas and S.~Kachru,
  {\em ``Flux compactification,''}
  Rev.\ Mod.\ Phys.\  {\bf 79}, 733 (2007)
  [arXiv:\hepth{0610102}].

\bibitem{Blumenhagen:2006ci}
  R.~Blumenhagen, B.~K\"ors, D.~L\"ust and S.~Stieberger,
  {\em ``Four-dimensional String Compactifications with D-Branes, Orientifolds and Fluxes,''}
  Phys.\ Rept.\  {\bf 445}, 1 (2007)
  [arXiv:\hepth{0610327}].


\bibitem{chiossi02}
S.~Chiossi and S.~Salamon,
{\em ``Intrinsic torsion of SU(3) and G${}_2$ structures,"}
arXiv: \Math{DG}{0202282}.

\bibitem{lucal}
  L.~Martucci and P.~Smyth,
  {\em ``Supersymmetric D-branes and calibrations on general N = 1 backgrounds,''}
  JHEP {\bf 0511} (2005) 048
  [arXiv:\hepth{0507099}].
  
  
\bibitem{gmpt05}
  M.~Gra\~na, R.~Minasian, M.~Petrini and A.~Tomasiello,
  {\em ``Generalized structures of N=1 vacua,''}
  JHEP {\bf 0511} (2005) 020
  [arXiv:\hepth{0505212}].
  

\bibitem{dwsb}
  D.~L\"ust, F.~Marchesano, L.~Martucci and D.~Tsimpis,
  {\em ``Generalized non-supersymmetric flux vacua,''}
  JHEP {\bf 0811} (2008) 021
  [\arXivid{0807.4540} [hep-th]].

\bibitem{noscale}
  E.~Cremmer, S.~Ferrara, C.~Kounnas and D.~V.~Nanopoulos,
  {\em ``Naturally Vanishing Cosmological Constant In N=1 Supergravity,''}
  Phys.\ Lett.\  B {\bf 133} (1983) 61.
  \\
  A.~B.~Lahanas and D.~V.~Nanopoulos,
  {\em ``The Road to No Scale Supergravity,''}
  Phys.\ Rept.\  {\bf 145}, 1 (1987).

\bibitem{gkp}
  S.~B.~Giddings, S.~Kachru and J.~Polchinski,
  {\em ``Hierarchies from fluxes in string compactifications,''}
  Phys.\ Rev.\  D {\bf 66} (2002) 106006
  [arXiv:\hepth{0105097}].
  
  
\bibitem{Hull}
  C.~M.~Hull,
  {\em ``Superstring Compactifications With Torsion And Space-Time Supersymmetry,''}
  Print-86-0251 (CAMBRIDGE)
  
  \bibitem{stromtorsion}
  A.~Strominger,
  {\em ``Superstrings with Torsion,''}
  Nucl.\ Phys.\  B {\bf 274} (1986) 253.
  
  
\bibitem{Lust:1986ix}
  D.~L\"ust,
  {\em ``Compactification Of Ten-Dimensional Superstring Theories Over Ricci Flat}
  Coset Spaces,''
  Nucl.\ Phys.\  B {\bf 276}, 220 (1986).



\bibitem{bpsaction}
  G.~Lopes Cardoso, G.~Curio, G.~Dall'Agata and D.~L\"ust,
   {\em ``BPS action and superpotential for heterotic string compactifications  with
  fluxes,''}
  JHEP {\bf 0310} (2003) 004
  [arXiv:\hepth{0306088}].
  
\bibitem{kklt}
   S.~Kachru, R.~Kallosh, A.~D.~Linde and S.~P.~Trivedi,
  {\em ``De Sitter vacua in string theory,''}
  Phys.\ Rev.\  D {\bf 68}, 046005 (2003)
  [arXiv:\hepth{0301240}].
  
\bibitem{LVS}
  V.~Balasubramanian, P.~Berglund, J.~P.~Conlon and F.~Quevedo,
  {\em ``Systematics of Moduli Stabilisation in Calabi-Yau Flux Compactifications,''}
  JHEP {\bf 0503}, 007 (2005)
  [arXiv:\hepth{0502058}].

\bibitem{hetF}  
  H.~Jockers, P.~Mayr and J.~Walcher,
  {\em ``On N=1 4d Effective Couplings for F-theory and Heterotic Vacua,''}
  \arXivid{0912.3265}.

\bibitem{bdr}
  E.~A.~Bergshoeff and M.~de Roo,
  {\em ``The Quartic Effective Action Of The Heterotic String And Supersymmetry,''}
  Nucl.\ Phys.\  B {\bf 328} (1989) 439.

\bibitem{bedulli06}
L.~Bedulli and L.~Vezzoni, {\em ``The Ricci tensor of SU(3)-manifolds,"} J.~Geom.~Phys.~57 
(2007), n. 4, 1125 [arXiv:\Math{DG}{0606786}].

\bibitem{cassani08}
  D.~Cassani,
  {\em ``Reducing democratic type II supergravity on SU(3) x SU(3) structures,''}
  JHEP {\bf 0806} (2008) 027
  [\arXivid{0804.0595} [hep-th]].


\bibitem{bsb}
  K.~Becker, C.~Bertinato, Y.~C.~Chung and G.~Guo,
  ``Supersymmetry breaking, heterotic strings and fluxes,''
  Nucl.\ Phys.\  B {\bf 823} (2009) 428
  [\arXivid{0904.2932} [hep-th]].
  

  
  
 \bibitem{michelson82}
M.~L.~Michelson, 
{\em ``On the existence of special metrics in complex geometry,"}
 Acta Math.~ {\bf 149}, 261 (1982)


\bibitem{sentorsion}
  A.~Sen,
  {\em ``(2,0) Supersymmetry and Space-Time Supersymmetry in the Heterotic String
  Theory,''}
  Nucl.\ Phys.\  B {\bf 278} (1986) 289.
 
  
\bibitem{cardoso02}
  G.~Lopes Cardoso, G.~Curio, G.~Dall'Agata, D.~L\"ust, P.~Manousselis and G.~Zoupanos,
  {\em ``Non-Kaehler string backgrounds and their five torsion classes,''}
  Nucl.\ Phys.\  B {\bf 652} (2003) 5
  [arXiv:\hepth{0211118}].


\bibitem{bryant}
R.~L.~Bryant {\em ``On the geometry of almost complex 6-manifolds,"} Asian J. Math. {\bf 10} (2006), no. 3, 561--605.  


\bibitem{gauntlett1}
  J.~P.~Gauntlett, N.~Kim, D.~Martelli and D.~Waldram,
  {\em ``Fivebranes wrapped on SLAG three-cycles and related geometry,''}
  JHEP {\bf 0111} (2001) 018
  [arXiv:\hepth{0110034}].
  \\
  J.~P.~Gauntlett, D.~Martelli and D.~Waldram,
  {\em ``Superstrings with intrinsic torsion,''}
  Phys.\ Rev.\  D {\bf 69} (2004) 086002
  [arXiv:\hepth{0302158}].

\bibitem{lawson}
  R.~Harvey and H.~B.~Lawson,
  {\em ``Calibrated geometries,''}
  Acta Math.\  {\bf 148} (1982) 47.

\bibitem{papa0}
  J.~Gutowski and G.~Papadopoulos,
  {\em ``AdS calibrations,''}
  Phys.\ Lett.\  B {\bf 462} (1999) 81
  [arXiv:\hepth{9902034}].
  \\
  J.~Gutowski, G.~Papadopoulos and P.~K.~Townsend,
  {\em ``Supersymmetry and generalized calibrations,''}
  Phys.\ Rev.\  D {\bf 60} (1999) 106006
  [arXiv:\hepth{9905156}].


\bibitem{lucal2}
  J.~Evslin and L.~Martucci,
  {\em ``D-brane networks in flux vacua, generalized cycles and calibrations,''}
  JHEP {\bf 0707} (2007) 040
  [arXiv:\hepth{0703129}].


\bibitem{km07}
  P.~Koerber and L.~Martucci,
  {\em ``D-branes on AdS flux compactifications,''}
  JHEP {\bf 0801}, 047 (2008)
  [arXiv:0710.5530 [hep-th]].
  

\bibitem{witten86}
  E.~Witten,
  {\em ``New Issues In Manifolds Of SU(3) Holonomy,''}
  Nucl.\ Phys.\  B {\bf 268} (1986) 79.

  
\bibitem{DUY}
S.~K.~Donaldson, {\em "Anti-self dual Yang-Mills connections over complex algebraic surfaces and stable vector bundles,"} Proc.~Lond~Math.~Soc.~{\bf 50} (1985) 1. \\
K.~Uhlenbeck and S.~T.~Yau, {\em "On the existence of Hermitian-Yang-Mills connections in stable vector bundles,"} 
Comm. Pure Appl. Math. {\bf 39} (1986), no. S, suppl., S257ÐS293.


\bibitem{liyau}
  J.~Li and S.~T.~Yau,
{\em  ``Hermitian Yang-Mills connection on non-K\"ahler manifolds,"}
in {\em Mathematical aspects of string theory}, World Scientific Publ.~, S.~T.~ Yau, editor; London 560 (1987).


\bibitem{HL2}
  F.~R.~Harvey and H.~B.~Lawson,
  {\em ``An Introduction To Potential Theory In Calibrated Geometry,''}
  Am.\ J.\ Math.\  {\bf 131} (2009) 893.



\bibitem{gurrieri07}
  S.~Gurrieri, A.~Lukas and A.~Micu,
  {\em ``Heterotic String Compactifications on Half-flat Manifolds II,''}
  JHEP {\bf 0712} (2007) 081
  [\arXivid{0709.1932} [hep-th]].

\bibitem{louis08}
  I.~Benmachiche, J.~Louis and D.~Martinez-Pedrera,
  {\em ``The effective action of the heterotic string compactified on manifolds with SU(3) structure,''}
  Class.\ Quant.\ Grav.\  {\bf 25} (2008) 135006
  [\arXivid{0802.0410} [hep-th]].

\bibitem{halfflat}
  S.~Gurrieri, A.~Lukas and A.~Micu,
  {\em ``Heterotic on half-flat,''}
  Phys.\ Rev.\  D {\bf 70} (2004) 126009
  [arXiv:\hepth{0408121}].
  \\
  B.~de Carlos, S.~Gurrieri, A.~Lukas and A.~Micu,
  {\em ``Moduli stabilisation in heterotic string compactifications,''}
  JHEP {\bf 0603} (2006) 005
  [arXiv:\hepth{0507173}].
  \\
  A.~Micu,
 {\em  ``A Note on Moduli Stabilisation in Heterotic Models in the Presence of Matter Fields,''}
  Phys.\ Lett.\  B {\bf 674} (2009) 139
  [\arXivid{0812.2172} [hep-th]];
  {\em ``Moduli Stabilisation in Heterotic Models with Standard Embedding,''}
  JHEP {\bf 1001} (2010) 011
  [\arXivid{0911.2311} [hep-th]].



\bibitem{beckersup}
  K.~Becker, M.~Becker, K.~Dasgupta and S.~Prokushkin,
  {\em ``Properties of heterotic vacua from superpotentials,''}
  Nucl.\ Phys.\  B {\bf 666} (2003) 144
  [arXiv:\hepth{0304001}].


\bibitem{gvw}
  S.~Gukov, C.~Vafa and E.~Witten,
  {\em ``CFT's from Calabi-Yau four-folds,''}
  Nucl.\ Phys.\  B {\bf 584} (2000) 69
  [Erratum-ibid.\  B {\bf 608} (2001) 477]
  [arXiv:\hepth{9906070}].


\bibitem{tentofour}
  P.~Koerber and L.~Martucci,
 {\em  ``From ten to four and back again: how to generalize the geometry,''}
  JHEP {\bf 0708} (2007) 059
  [\arXivid{0707.1038} [hep-th]].


\bibitem{drs}
  K.~Dasgupta, G.~Rajesh and S.~Sethi,
  {\em ``M theory, orientifolds and G-flux,''}
  JHEP {\bf 9908} (1999) 023
  [arXiv:\hepth{9908088}].


\bibitem{elliptic}
 J.~X.~Fu and S.~T.~Yau,
  {\em ``The theory of superstring with flux on non-Kaehler manifolds and the
  complex Monge-Ampere equation,''}
  J.\ Diff.\ Geom.\  {\bf 78} (2009) 369
  [arXiv:\hepth{0604063}]. 
  K.~Becker, M.~Becker, J.~X.~Fu, L.~S.~Tseng and S.~T.~Yau,
  {\em ``Anomaly cancellation and smooth non-Kaehler solutions in heterotic string theory,''}
  Nucl.\ Phys.\  B {\bf 751} (2006) 108
  [arXiv:\hepth{0604137}].


\bibitem{bs}
  K.~Becker and S.~Sethi,
  {\em ``Torsional Heterotic Geometries,''}
  Nucl.\ Phys.\  B {\bf 820} (2009) 1
  [\arXivid{0903.3769} [hep-th]].



\bibitem{aspinwall}
  P.~S.~Aspinwall,
  {\em ``K3 surfaces and string duality,''}
  arXiv:\hepth{9611137}.



\bibitem{bdr2}
  E.~A.~Bergshoeff and M.~de Roo,
  {\em ``The string effective action in the dual formulation of D = 10 supergravity,''}
  Phys.\ Lett.\  B {\bf 247} (1990) 530.




\bibitem{ccdl03}
  G.~Lopes Cardoso, G.~Curio, G.~Dall'Agata and D.~L\"ust,
  {\em ``Heterotic string theory on non-Kaehler manifolds with H-flux and  gaugino condensate,''}
  Fortsch.\ Phys.\  {\bf 52}, 483 (2004)
  [arXiv:\hepth{0310021}].
  
  
\bibitem{fl05}
  A.~R.~Frey and M.~Lippert,
  {\em ``AdS strings with torsion: Non-complex heterotic compactifications,''}
  Phys.\ Rev.\  D {\bf 72}, 126001 (2005)
  [arXiv:\hepth{0507202}].
  
\bibitem{blumen05}
  R.~Blumenhagen, G.~Honecker and T.~Weigand,
  {\em ``Loop-corrected compactifications of the heterotic string with line bundles,''}
  JHEP {\bf 0506} (2005) 020
  [arXiv:\hepth{0504232}].


\bibitem{gklm03}
  S.~Gukov, S.~Kachru, X.~Liu and L.~McAllister,
  {\em ``Heterotic moduli stabilization with fractional Chern-Simons invariants,''}
  Phys.\ Rev.\  D {\bf 69}, 086008 (2004)
  [arXiv:\hepth{0310159}].
  
  
 
\bibitem{ckl05}
  G.~Curio, A.~Krause and D.~L\"ust,
  {\em ``Moduli stabilization in the heterotic / IIB discretuum,''}
  Fortsch.\ Phys.\  {\bf 54}, 225 (2006)
  [arXiv:\hepth{0502168}].



\bibitem{ht04}
  A.~Hebecker and M.~Trapletti,
  {\em ``Gauge unification in highly anisotropic string compactifications,''}
  Nucl.\ Phys.\  B {\bf 713}, 173 (2005)
  [arXiv:\hepth{0411131}].
  


\bibitem{cm07}
  P.~G.~C\'amara and M.~Gra\~na,
  {\em ``No-scale supersymmetry breaking vacua and soft terms with torsion,''}
  JHEP {\bf 0802}, 017 (2008)
  [\arXivid{0710.4577} [hep-th]].

\bibitem{cm09}
  P.~G.~C\'amara and F.~Marchesano,
  {\em ``Open string wavefunctions in flux compactifications,''}
  JHEP {\bf 0910}, 017 (2009)
  [\arXivid{0906.3033} [hep-th]].

\bibitem{bt05}
  K.~Becker and L.~S.~Tseng,
  {\em ``Heterotic Flux Compactifications and Their Moduli,''}
  Nucl.\ Phys.\  B {\bf 741}, 162 (2006)
  [arXiv:\hepth{0509131}].
  \\
  M.~Becker, L.~S.~Tseng and S.~T.~Yau,
  {\em ``Moduli space of torsional manifolds,''}
  Nucl.\ Phys.\  B {\bf 786} (2007) 119
  [arXiv:\hepth{0612290}].

\bibitem{lucaeff09}
  L.~Martucci,
  {\em ``On moduli and effective theory of N=1 warped flux compactifications,''}
  JHEP {\bf 0905} (2009) 027
  [\arXivid{0902.4031} [hep-th]].


  
 \bibitem{bdr3}
  E.~A.~Bergshoeff and M.~de Roo,
  ``Duality transformations of string effective actions,''
  Phys.\ Lett.\  B {\bf 249} (1990) 27.
 

\end{thebibliography}
\end{document}